\documentclass{article}
\usepackage{amsmath,amsfonts,setspace,cancel,url,hyperref,verbatim,slashed,amsthm,graphicx,color,cite}
\usepackage[a4paper,margin=1in]{geometry}
\newcommand{\be}{\begin{equation}}
\newcommand{\ee}{\end{equation}}
\newcommand{\gM}{\mathcal{M}}

\newcommand{\hmu}{{\hat{\mu}}}
\newcommand{\hnu}{{\hat{\nu}}}

\newcommand{\hrho}{\hat{\rho}}

\newcommand{\p}{\bullet}
\newcommand{\pl}{\!\!\p\!}

\newcommand{\hpartial}{\hat{\partial}}
\newcommand{\hd}{\hpartial}
\newcommand{\D}{D}

\newcommand{\Aa}{A}
\newcommand{\Ab}{B}
\newcommand{\Ac}{C}

\newcommand{\Fa}{\mathcal{F}}
\newcommand{\Fb}{\mathcal{H}}

\numberwithin{equation}{section}

\newcommand{\G}{\mathrm{SL}(2) \times \mathbb{R}^+}
\newcommand{\Gthree}{\mathrm{SL}(3) \times \mathrm{SL}(2)}
\newcommand{\Gfour}{\mathrm{SL}(5)}
\newcommand{\Gfive}{\mathrm{SO}(5,5)}
\newcommand{\Gsix}{\mathrm{E}_6}
\newcommand{\Gseven}{\mathrm{E}_7}
\newcommand{\Geight}{\mathrm{E}_8}

\newcommand{\Hthree}{\mathrm{SO}(3) \times \mathrm{SO}(2)}
\newcommand{\Hfour}{\mathrm{SO}(5)}
\newcommand{\Hfive}{\mathrm{SO}(5)\times \mathrm{SO}(5)}
\newcommand{\Hsix}{\mathrm{USp}(8)}
\newcommand{\Hseven}{\mathrm{SU}(8)}
\newcommand{\Height}{\mathrm{SO}(16)}

\newcommand{\gV}{{V}}
\newcommand{\ED}{{\rm E}_{D(D)}}

\newcommand{\mi}{\hat i}
\newcommand{\mj}{\hat j}
\newcommand{\mk}{\hat k}
\newcommand{\ml}{\hat l}
\newcommand{\mm}{\hat m}
\newcommand{\mn}{\hat n}
\renewcommand{\mp}{\hat p} 
\newcommand{\mq}{\hat q}
\newcommand{\mr}{\hat r}

\begin{document}

\begin{titlepage}
\begin{center}

\hfill Imperial-TP-2018-ASA-01

\vskip 1.5cm

{\Large \bf  The Exceptional Sigma Model}

\vskip 1cm

{\bf Alex S. Arvanitakis${}^1$, Chris D. A. Blair${}^2$} \\

\vskip 25pt

{\em $^1$ \hskip -.1truecm
\em The Blackett Laboratory,\\
Imperial College London, \\
Prince Consort Road London SW7 2AZ, U.K.\vskip 5pt}

${}^2$ Theoretische Natuurkunde, Vrije Universiteit Brussel, and the International Solvay Institutes,\\
Pleinlaan 2, B-1050 Brussels, Belgium
\vskip 5pt 

{email: {\tt a.arvanitakis@imperial.ac.uk } , {\tt cblair@vub.ac.be}} \\

\end{center}

\vskip 0.5cm

\begin{center} {\bf ABSTRACT}\\[3ex]\end{center}

We detail the construction of the \emph{exceptional sigma model}, which describes a string propagating in the ``extended spacetime'' of exceptional field theory. This is to U-duality as the doubled sigma model is to T-duality.
Symmetry specifies the Weyl-invariant Lagrangian uniquely and we show how it reduces to the correct 10-dimensional string Lagrangians. We also consider the inclusion of a Fradkin-Tseytlin (or generalised dilaton) coupling as well as a reformulation with dynamical tension.

\noindent

\end{titlepage}

\newpage
\setcounter{page}{1}

\tableofcontents

\section{Introduction}
\label{intro} 

\subsection{Background}
Duality underlies the modern understanding that the different superstring theories and the eleven-dimensional M-theory should not be viewed as being disjoint but as particular descriptions of a greater theory. 
The appearance of duality groups, including the $O(d,d)$ of T-duality and $\ED$ of U-duality, warrant our close attention.
At least conceptually, one might wonder if there is some alternate formulation in which such $O(d,d)$ or $\ED$ symmetries do not appear only on reduction, but are present from the outset. 

Steps in this direction can indeed be taken. These approaches include the
doubled sigma model (or doubled worldsheet)
\cite{Duff:1989tf, Tseytlin:1990nb, Tseytlin:1990va, Hull:2004in, Hull:2006va}, double field theory (DFT)
\cite{Siegel:1993th, Siegel:1993xq, Hull:2009mi}, and exceptional field theory (EFT)
\cite{
Berman:2010is, Berman:2011jh,Berman:2011cg, Berman:2012vc, Hohm:2013vpa, Hohm:2013uia, Hohm:2014fxa, Hohm:2015xna, Abzalov:2015ega, Musaev:2015ces, Berman:2015rcc}. 
These theories provide reformulations of the string worldsheet action and the low energy supergravities in which extra ``dual'' coordinates are introduced, in order to realise the $O(d,d)$ or $\ED$ symmetry ``geometrically'', acting on an enlarged spacetime described by the usual coordinates and the duals. 
In this paper, as outlined in \cite{Arvanitakis:2017hwb}, we provide a construction of an ``exceptional sigma model'': a two-dimensional sigma model which describes a string coupled to the background fields of this enlarged spacetime, with a (formally) manifest  exceptional symmetry related to the U-duality groups of M-theory. 

The ideas here can be traced back to the approaches of 
\cite{Duff:1989tf, Tseytlin:1990nb, Tseytlin:1990va, Hull:2004in, Hull:2006va},
where the string sigma model was formulated in a T-duality symmetric manner.
There, a dual coordinate is added for each T-dualisable target space coordinate.
The geometry of the doubled background is encoded in fields which are in representations of $O(d,d)$ and which group together components of the different spacetime fields (e.g. the metric and B-field appear together in a so-called generalised metric). The number of on-shell degrees of freedom is still $d$ rather than $2d$, and this ``reduction'' is achieved in various ways depending on the exact model under consideration. In all cases one can think of the $2d$ scalars as chiral, with some sort of chirality constraint implemented differently in different models.
In this paper we will follow the approach of Hull \cite{Hull:2004in, Hull:2006va} and eliminate the dual coordinates by a gauging procedure.

It is not only the sigma model that can be ``doubled'', but also the low energy effective theory describing the background fields.
This leads to double field theory
\cite{Siegel:1993th, Siegel:1993xq,Hull:2009mi}. The DFT equations of motion following from the action of \cite{Hohm:2010pp} can be obtained by requiring the vanishing of the conformal anomaly of the doubled worldsheet 
\cite{Berman:2007xn, Berman:2007yf, Copland:2011wx}, just as one obtains the usual string background field equations. 

Importantly, the local symmetries of supergravity --- diffeomorphisms and $p$-form gauge transformations --- can also be written $O(d,d)$-covariantly: the combination of diffeomorphisms and B-field gauge transformations yields ``generalised diffeomorphisms'' with gauge parameter $\Lambda^M$ ($M$ here is an $O(d,d)$ index) acting on generalised tensors, just as ordinarily diffeomorphisms can be viewed as infinitesimal $\mathrm{GL}(d)$ transformations. Consistency of the algebra of these symmetry transformations leads to a constraint on the coordinate dependence, implying that fields and gauge parameters can depend only on at most half the doubled coordinates. All these constructions are formally invariant under $O(d,d)$.\footnote{This is $O(d,d;\mathbb{R})$ in both cases classically.}

In order to make contact with the usual worldsheet or supergravity actions, we have to identify half the doubled coordinates as  ``physical'' ones. Of course this identification generically breaks the $O(d,d)$ symmetry; the freedom to make alternative choices of which $d$-dimensional set of coordinates is physical is T-duality \cite{Hull:2004in}. For backgrounds with $d$ isometries in the physical directions, one can freely choose any set of $d$ coordinates to be physical, the different choices are related by honest Buscher \cite{Buscher:1987sk, Buscher:1987qj} dualities, and indeed one now has an unbroken $O(d,d)$ symmetry. 

This provides the link between the $O(d,d)$ manifest theories of the doubled sigma model and DFT with the $O(d,d)$ T-duality symmetry of compactified string theory or supergravity. 
T-duality is of course a perturbative duality of string theory. 
U-duality, on the other hand, is a non-perturbative symmetry,
with M-theory compactified on a $D$-torus leading to an exceptional duality symmetry group $\ED$ (the split real form of the exceptional series). 

Although we do not have access to a non-perturbative version of string theory (or the full M-theory), U-duality is also realised in (compactifications of) the low energy supergravity actions.
In this context, we can construct the generalisation of double field theory to the exceptional case, as was carried out in 
\cite{Berman:2010is, Berman:2011jh, Hohm:2013vpa, Hohm:2013uia, Hohm:2014fxa, Hohm:2015xna, Abzalov:2015ega, Musaev:2015ces, Berman:2015rcc}. 
Again, we introduce extra ``dual'' coordinates so that the total set of coordinates appear in a representation of $\ED$, as do the fields. 
We now have $\ED$ generalised diffeomorphisms \cite{Berman:2011cg, Berman:2012vc}, and a section condition
implying that we cannot depend on all the coordinates. 
Different solutions \cite{Hohm:2013vpa,Blair:2013gqa} to this condition (usually termed ``sections'') lead to us having no more than eleven or ten coordinates -- in the former case, the theory reduces to eleven-dimensional supergravity while in the latter case it reduces to the two type II supergravities. Note that the type IIA section is always contained trivially within the M-theory one, while the type IIB section is an inequivalent solution (i.e. it cannot be transformed into a IIA section by $\ED$).

Now, it is an immediate\footnote{That is to say, it predates the birth of one of the authors and the second birthday of the other.} question as to whether there is some analogous description of the underlying brane actions which realises $\ED$ symmetries just as the doubled sigma model does for $O(d,d)$. This was first investigated for M2 branes in \cite{Duff:1990hn}, by studying worldvolume duality relations. However, as finally established a quarter-century later, this approach is limited in scope \cite{Duff:2015jka} -- and only completely works for certain target space dimensions for which the number of dual coordinates introduced equals the number of physical coordinates. Fundamentally, the underlying problem is that U-duality mixes branes of different worldvolume dimensions (whereas T-duality exchanges winding and momentum modes of the fundamental string itself).

Our approach to this problem will be to restrict to fixed worldvolume dimension, and construct the action that naturally couples to the $\ED$ covariant variables describing the background in the EFT formalism.
This idea was used in \cite{Blair:2017gwn} to construct an EFT particle action. This action can be viewed in three ways: 
\begin{itemize}
\item as an action for a massless particle in an ``extended spacetime'', with extra worldline scalars corresponding to dual directions,
\item as an action for massless particle-like states in $10$ or $11$ dimensions on integrating out these dual coordinates,
\item as an action for massive particles corresponding to wrapped branes in $n$-dimensions, on further reduction.
\end{itemize}
This EFT particle couples geometrically to the EFT metric degrees of freedom and electrically to the $\ED$ multiplet $A_\mu{}^M$, which are one-forms from the point of view of the non-dualisable directions $X^\mu$ and generalised vectors carrying the index $M$ running over the representation of $\ED$ corresponding to the extended coordinates $Y^M$ (for instance, for $\Gsix$, this is the 27-dimensional fundamental representation). This representation is often denoted by $R_1$. The fields of EFT include a tensor hierarchy $R_1,R_2,\dots$ of such generalised form fields \cite{Berman:2012vc}.

Our approach in this paper will be similar in spirit to the above. It will provide a generalisation of the ``Hull-style'' doubled string \cite{Hull:2004in, Hull:2006va}, while making use of ideas grounded in the local symmetries of DFT in \cite{Lee:2013hma} which were adapted to EFT in \cite{Blair:2017gwn}. 
The EFT particle of the latter paper could be viewed as a massless particle in the extended spacetime, and our action will have a similar re-interpretation in a quasi-tensionless reformulation. This directly generalises the type IIB $\mathrm{SL}(2)$ covariant string of \cite{Townsend:1997kr, Cederwall:1997ts} to larger groups (and to type IIA as well). 
 Due to the analogy to the double string we will be calling this the exceptional sigma model, or more simply the exceptional string.

Before summarising our main result, let us note that the worldvolume duality approach was used in \cite{Berman:2010is} to motivate the construction of the $\Gfour$ EFT, and alongside the further development of EFT there have been some efforts to re-approach the problem of brane actions from the perspective of exceptional geometry, including \cite{Hatsuda:2012vm, Hatsuda:2013dya, 
Linch:2015fya, Linch:2015qva, Linch:2015fca,
Sakatani:2016sko, Sakatani:2017vbd}.
Other work includes the study of brane dynamics in the closely related approach based on $\mathrm{E}_{11}$, as in \cite{West:2004iz, West:2018lfn}, and the formulation of a $\Gthree$ covariant membrane action in \cite{Bengtsson:2004nj}.

\subsection{Main result}

In this paper we present the Lagrangian of the two-dimensional sigma model with an $\ED$ EFT background as target space, valid for $D=2,3\dots 6$ (for these values of $D$, the structures of EFT are relatively homogeneous and similar to that of DFT. For $D=7$, various differences begin to appear in the essential features of EFT, and most relevantly for us no generalised two-form appears in the action of EFT). 
Perhaps the simplest form of the action of the $\ED$ exceptional sigma model is:
\be
\begin{split}
S_{EWS}  = - \frac{1}{2}  \int d^2 \sigma 
\Big(
&
T(\gM, q)
 \Big(
\sqrt{-\gamma} \gamma^{\alpha \beta} g_{\mu\nu} 
\partial_\alpha X^\mu \partial_\beta X^\nu 
+ \frac{1}{2} \sqrt{-\gamma} \gamma^{\alpha \beta} \gM_{MN} D_\alpha Y^M D_\beta Y^N \Big)
\\ & 
\qquad + 
q_{MN} \epsilon^{\alpha \beta} \left( 
B_{\mu \nu}{}^{MN} \partial_\alpha X^\mu \partial_\beta X^\nu 
+ \partial_\alpha  X^\mu  A_\mu{}^M  D_\beta Y^N
+ \partial_\alpha Y^M  \gV_\beta^N 
\right) \Big) \,.
\end{split} 
\label{ESintro}
\ee
with
\be
D_\alpha Y^M = \partial_\alpha Y^M + \partial_\alpha X^\mu A_\mu{}^M + \gV_\alpha{}^M
\ee
where the ``tension'' is:
\be
T(\gM,q) = 
\sqrt{ \frac{q_{MN} q_{PQ} \gM^{MP} \gM^{NQ}}{2(D-1)}}\,.
\ee
This string couples to an EFT background: $\{g_{\mu\nu},\gM_{MN},A_\mu{}^M,B_{\mu\nu}{}^{MN},\dots\}$; respectively, the external and generalised metrics, the ``1-form'' field $A_\mu{}^M\in R_1$, as well as a ``2-form'' field $B_{\mu\nu}{}^{MN}$ in the representation $R_2$ of the tensor hierarchy, which couples electrically to the worldsheet much like the Kalb-Ramond $B$-field does to the usual string.
The role of the $O(d,d)$ structure $\eta$ is now played by the \emph{constant charge parameter $q_{MN}\in \bar R_2$}. We will see how consistency --- i.e. covariance under (generalised) diffeomorphisms and gauge-invariance of the $B$-field coupling --- implies that $q$ is constrained by 
\be
q_{MN} Y^{NK}{}_{PQ} \partial_K = q_{PQ} \partial_M 
\label{magicintro}
\ee
(equivalently, $q \otimes \partial |_{\bar R_3} = 0$).

Crucially, the exceptional string Lagrangian is compatible with the symmetries of EFT. As explained in section \ref{esm} this places stringent requirements on its form. This Lagrangian also correctly reduces to the usual 10-dimensional string Lagrangian (when appropriate), which we confirm in section \ref{chargered}.

\section{The doubled sigma model}
\label{dsm} 

\subsection{Action for the doubled string}

The part of the fundamental string (F1) action which couples to the 10-dimensional metric  $\hat g_{\hmu\hnu}$ and B-field $\hat B_{\hmu \hnu}$ is: 
\be
S = - \frac{T_{F1}}{2} \int d^2\sigma
( \sqrt{-\gamma} \gamma^{\alpha \beta} \hat g_{\hmu\hnu} + \epsilon^{\alpha \beta} \hat B_{\hmu \hnu} ) \partial_\alpha X^\hmu \partial_\beta X^\hnu \,,
\label{F1action} 
\ee
where we take $\epsilon^{01} = -1$.
We split the coordinates $X^\hmu = (X^\mu, Y^i)$, with $i=1,\dots,d$, and insert the following decompositions of the backgrounds fields into the action \eqref{F1action}:
\be
\begin{array}{cll}
\hat g_{\mu\nu} & =& g_{\mu\nu} + \phi_{ij} A_\mu{}^i A_\nu{}^j \,,\\
\hat g_{\mu i} & = & \phi_{ij} A_\mu{}^j \,,\\
\hat g_{ij} & = & \phi_{ij} \,,
\end{array} 
\quad\quad
\begin{array}{cll} 
\hat B_{\mu\nu} & = & B_{\mu\nu} - A_{[\mu}{}^j A_{\nu]j} + B_{ij} A_\mu{}^i A_\nu{}^j\,, \\
 \hat B_{\mu i} & = & A_{\mu i} + A_\mu{}^j B_{ji}\,, \\
\hat B_{ij} & = & B_{ij}\,.
\end{array} 
\ee
Letting $D_\alpha Y^i \equiv \partial_\alpha Y^i + \partial_\alpha X^\mu A_\mu{}^i$ one can write the action as 
\be
\begin{split}
S = - \frac{T_{F1}}{2} \int d^2\sigma
&\Big(
 \sqrt{-\gamma} \gamma^{\alpha \beta}
 \left( g_{\mu\nu} \partial_\alpha X^\mu \partial_\beta X^\nu 
 + \phi_{ij} D_\alpha Y^i D_\beta Y^j
 \right) 
\\ & 
\quad 
+ \epsilon^{\alpha \beta} (  ( B_{\mu\nu} + A_\mu{}^i A_{\nu i} ) \partial_\alpha X^\mu \partial_\beta X^\nu  
+ 2  A_{\mu i} \partial_\alpha X^\mu D_\beta Y^i
+ B_{ij} D_\alpha Y^i D_\beta Y^j )
\Big)\,.
\end{split} 
\label{splitF1action}
\ee
Now, double the $Y^i$, introducing duals $\tilde Y_i$ which together are written as $Y^M = ( Y^i, \tilde Y_i)$. 
A doubled action equivalent to \eqref{splitF1action} is \cite{Hull:2004in, Hull:2006va}
\be
\begin{split}
S_{DWS}  = - \frac{T_{F1}}{2} \int d^2 \sigma
& \Big(
\sqrt{-\gamma} \gamma^{\alpha \beta}
\left( 
 g_{\mu\nu} \partial_\alpha X^\mu \partial_\beta X^\nu 
 + \frac{1}{2}\gM_{MN} D_\alpha Y^M D_\beta Y^N 
\right) 
\\ & \quad + \epsilon^{\alpha \beta} 
\left( B_{\mu\nu} \partial_\alpha X^\mu \partial_\beta X^\nu 
+  \eta_{MN} A_\mu{}^M \partial_\alpha X^\mu D_\beta Y^N 
 +  \eta_{MN} \partial_\alpha Y^M \gV_\beta{}^M
 \right) 
\Big) \,.
\end{split} 
\label{DWS}
\ee
Here, the background fields have been combined into an $O(d,d)$ vector and an $O(d,d)$ generalised metric: 
\be
A_\mu{}^M = \begin{pmatrix}  A_\mu{}^i \\ A_{\mu i} \end{pmatrix} \quad,\quad
\gM_{MN} = \begin{pmatrix} 
\phi_{ij} - B_{ik} \phi^{kl} B_{lj} & B_{ik} \phi^{kj} \\
- \phi^{ik} B_{kj} & \phi^{ij} 
\end{pmatrix} \,,
\ee
and $\eta_{MN}$ is the $O(d,d)$ metric 
\be
\eta_{MN} = \begin{pmatrix} 0 & I \\ I & 0 \end{pmatrix} \,.
\ee
We have introduced an extra worldline one-form (and $O(d,d)$ vector) $V_\alpha{}^M$, and defined
\be
D_\alpha Y^M \equiv \partial_\alpha Y^M + \partial_\alpha X^\mu A_\mu{}^M + \gV_\alpha{}^M\,.
\ee
The one-form $V_\alpha{}^M$ plays several roles. In the original formulation \cite{Hull:2004in, Hull:2006va} it is introduced in order to implement the self-duality constraint
\be
\partial_\alpha Y^M + \partial_\alpha X^\mu A_\mu{}^M = \eta^{MN} \gM_{NP} \sqrt{-\gamma} \epsilon_{\alpha \beta}\gamma^{\beta \gamma} ( \partial_\gamma Y^P + \partial_\gamma X^\mu A_\mu{}^P )
\ee
by gauging a shift symmetry in the dual directions. (Note that the consistency of this constraint is guaranteed by the relationship $\gM_{MN} \eta^{NP} \gM_{PQ} \eta^{QK} = \delta_M{}^K$.) More recently, while also gauging away the dual directions, it has been pointed out that $V_\alpha{}^M$ must transform under gauge transformations of the background fields in order to ensure covariance on the doubled worldsheet \cite{Lee:2013hma}. 

In the approach of double field theory (DFT), one formally allows the background fields to depend on the doubled coordinates $Y^M$ subject to the section condition, $\eta^{MN} \partial_M \otimes \partial_N = 0$. This is solved by $\partial_i \neq 0$, $\tilde \partial^i = 0$, i.e. by having no dependence on half the coordinates. The gauge field $V_\alpha{}^M$ is required to obey $V_\alpha{}^M \partial_M = 0$.
Then in this canonical ``choice of section'' we have $\gV_\alpha{}^i = 0$ and $\gV_{\alpha i} \neq 0$. 
Integrating out $\gV_{\alpha i}$ gives exactly the action \eqref{splitF1action}, plus a total derivative. 

To see this, we note that the terms involving $Y^M$ in \eqref{DWS} can be decomposed as follows:
\be
\begin{split} 
S_{DWS}  \supset 
 - \frac{T_{F1}}{2}\int d^2 \sigma 
& \Big(
\frac{1}{2} \sqrt{-\gamma} \gamma^{\alpha \beta} \phi_{ij} D_\alpha Y^i D_\beta Y^j - \epsilon^{\alpha \beta} \partial_\alpha Y^i \partial_\beta \tilde Y_j 
+ \epsilon^{\alpha \beta} A_{\mu i} \partial_\alpha X^\mu ( 2 D_\beta Y^i - A_\nu{}^i \partial_\beta X^\nu )
\\ & 
\quad 
+ \frac{1}{2} \sqrt{-\gamma} \gamma^{\alpha \beta} \phi^{ij} 
( D_\alpha \tilde Y_i - B_{ik} D_\alpha Y^k)
( D_\beta \tilde Y_j - B_{jl} D_\beta Y^l)
- \epsilon^{\alpha \beta} D_\alpha \tilde Y_i D_\beta Y^i
\Big)\,.
\end{split} 
\ee
Integrating out $\gV_{\alpha i}$ is equivalent to eliminating $D_\alpha \tilde Y_i$ from the second line, which amounts to replacing it by
\be
+ \frac{1}{2} \sqrt{-\gamma} \gamma^{\alpha \beta} \phi_{ij} D_\alpha Y^i D_\beta Y^j 
+ \epsilon^{\alpha \beta} B_{ij} D_\alpha Y^i D_\beta Y^j
\ee
Then we get exactly \eqref{splitF1action} with the additional term
\be
S \supset -\frac{T_{F1}}{2} \int d^2\sigma ( -\epsilon^{\alpha \beta} \partial_\alpha Y^i \partial_\beta \tilde Y_i )
\ee
This reduction matches exactly that in \cite{Lee:2013hma}, while in \cite{Hull:2006va} this term is removed by adding to $S_{DWS}$ a ``topological term'' required for invariance under large gauge transformations. This term is 
\be
S_{top} =  \frac{T_{F1}}{2} \int d^2 \sigma \epsilon^{\alpha \beta} 
\frac{1}{2} \Omega_{MN} \partial_\alpha Y^M \partial_\beta Y^N
\quad,\quad
\Omega_{MN} = \begin{pmatrix} 0 & 1 \\ -1& 0 \end{pmatrix} \,.
\label{topterm}
\ee

\subsection{Gauge transformations}

Let us check how the doubled sigma model respects the gauge symmetries of double field theory. 
Note that we are using the formulation with only a partial doubling of the spacetime coordinates, as described in \cite{Hohm:2013nja}. 
We have three types of local symmetries: external diffeomorphisms, generalised diffeomorphisms and generalised gauge transformations. 
Let us focus only on the latter two here. We start with the generalised gauge transformations of the gauge fields $A_\mu{}^M$ and $B_{\mu\nu}$:
\be
\begin{split}
 \delta_\lambda A_\mu{}^M  = - \partial^M \lambda_\mu \,,
 \qquad
 \delta_\lambda B_{\mu\nu}  = 2 \partial_{[\mu} \lambda_{\nu]} - A_{[\mu}{}^M \partial_M \lambda_{\nu]} \,.
 \label{DFTgauge}
\end{split} 
\ee
It is convenient to specify a ``covariant'' transformation of $B_{\mu\nu}$ by 
\be
\Delta B_{\mu\nu} \equiv \delta B_{\mu\nu} + A_{[\mu}{}^M \delta A_{\nu] M}
\ee
for which $\Delta_\lambda B_{\mu\nu} = 2 D_{[\mu} \lambda_{\nu]}$ where $D_\mu = \partial_\mu - \mathcal{L}_{A_{\mu}}$ is the covariantisation of the partial derivative $\partial_\mu$ under generalised diffeomorphisms (defined below).

The doubled sigma model action \eqref{DWS} should be invariant under these tranformations. 
This requires that in addition we need $\gV_\alpha^M$ to transform as
\be
\delta_\lambda V_\alpha{}^M = \partial_\alpha X^\mu \partial^M \lambda_\mu\,.
\ee
Then, the quantity $D_\alpha Y^M$ is automatically invariant, and the action transforms into
\be
\delta_\lambda S = -\frac{T_{F1}}{2} \int d^2\sigma 
\epsilon^{\alpha \beta} 
\left(
2 \partial_\alpha \lambda_\mu \partial_\beta X^\mu  - \partial_\alpha X^\mu \gV_\beta^M \partial_M \lambda_\mu   
\right) 
\ee
which is a total derivative using the condition $V_\alpha^M \partial_M = 0$.

Next we consider generalised diffeomorphisms.
The name reflects the fact that these act via a generalised Lie derivative $\mathcal{L}_\Lambda$ on generalised tensors, where on a generalised vector $U^M$ we have
\be
\mathcal{L}_\Lambda U^M = \Lambda^N \partial_N U^M - U^N \partial_N \Lambda^M + U_N \partial^M \Lambda^N \,.
\label{DFTgd}
\ee
For instance, the generalised metric transforms as a rank two tensor under these transformations. Acting on the gauge fields $A_\mu{}^M$ and $B_{\mu\nu}$, however, one can think of generalised diffeomorphisms as more like traditional gauge transformations than diffeomorphisms. We have
\be
\begin{split} 
\delta_\Lambda A_\mu{}^M  = D_\mu \Lambda^M = \partial_\mu \Lambda^M - \mathcal{L}_{A_\mu} \Lambda^M \,, \qquad
\Delta_\Lambda B_{\mu\nu}  = \Lambda_M \mathcal{F}_{\mu\nu}{}^M \,,
\end{split} 
\label{DFTgdAB}
\ee
with the field strength for $A_\mu{}^M$ -- this field strength is \emph{covariant} under generalised diffeomorphisms, transforming as a generalised vector -- given by
\be
\mathcal{F}_{\mu\nu}{}^M = 2 \partial_{[\mu} A_{\nu]}{}^M - 2 A_{[\mu}{}^N \partial_N A_{\nu]}{}^M + A_{[\mu}{}^N \partial^M A_{\nu] N} + \partial^M B_{\mu\nu} \,.
\label{DFTF}
\ee
Now, generalised diffeomorphisms are not a \emph{symmetry} on the worldsheet; the reason being that the position of a brane embedded in a target space is not invariant under target space diffeomorphisms.
The correct way that these transformations should appear on the worldsheet (or any worldvolume action) is the following.
We should transform the doubled coordinates and require that this induces the correct transformation rules for the background fields in the worldsheet action, in the sense that these induced transformations on the background fields correspond to generalised diffeomorphisms.
Let us denote by $\bar \delta_\Lambda$ the following transformation which amounts solely to shifting $Y^M$ by $\Lambda^M$: 
\be
\begin{split} 
\bar\delta_\Lambda Y^M & = \Lambda^M \,, \\
\bar\delta_\Lambda X^\mu & = 0 \,,\\
\bar \delta_\Lambda \mathcal{O}( Y ) & =  \Lambda^P \partial_P \mathcal{O}(Y) \,,
\end{split}
\label{deltabar}
\ee
where $\mathcal{O}(Y)$ signifies any background field which depends on $Y$. Letting $S[X,Y,\gV; g,\gM, A, B]$ denote the action for the worldsheet fields $X^\mu,Y^M$ and $\gV_\alpha^M$ coupled by the background fields 
$g_{\mu\nu}$, $\gM_{MN}$, $A_\mu{}^M$ and $B_{\mu\nu}$
(which may depend on $X^\mu$ and $Y^M$), the covariance condition is that:
\be
\bar \delta_\Lambda \mathcal{S}[ X,Y,\gV; g, \gM, A, B ]
=
 \mathcal{S}[ X,Y,\gV; \delta_\Lambda g, \delta_\Lambda \gM, \delta_\Lambda A, \delta_\Lambda B ]
\ee
which leads to a symmetry if $\Lambda$ is a generalised Killing vector, i.e. a generalised diffeomorphism which annihilates the background fields. 
This requires the following transformation of $\gV^M$ under generalised diffeomorphisms, as originally worked out (for the fully doubled case) in \cite{Lee:2013hma}:
\be
\bar\delta_{\Lambda}  \gV_\alpha^M  = 
 -\partial^M \Lambda_N ( \partial_\alpha Y^N + \gV_\alpha^N) 
+ \partial^M ( A_\mu{}^N \Lambda_N ) \partial_\alpha X^\mu
\label{bardeltaV} 
\ee 
for which the action transforms as required up to a total derivative term arising from the Wess-Zumino part: 
\be
\frac{T_{F1}}{2} \int d^2\sigma \epsilon^{\alpha \beta} \partial_\alpha \Lambda^M \partial_\beta Y_M\,.
\ee
The strategy for the construction of the exceptional sigma model will be to begin with the EFT generalisations of the transformations \eqref{DFTgauge}, \eqref{DFTgd}, \eqref{DFTgdAB} for the background fields, and check what sigma model action is compatible with these. There we will also check the covariance requirement under external diffeomorphisms, which allow us to completely fix all relative coefficients in the action (our approach will be quite general and so also applies to the doubled sigma model, hence one can use the results of section \ref{esm} to confirm the covariance of the latter under external diffeomorphisms, which we have not discussed here).

\subsection{Tensor hierarchy and generalisations}

We now finish our review of the doubled sigma model by pointing out the ingredients that generalise naturally (if surprisingly) to the exceptional case.
Let us focus on the Wess-Zumino term
\be
S_{WZ} = - \frac{T_{F1}}{2} \int d^2 \sigma
\epsilon^{\alpha \beta} 
\left( B_{\mu\nu} \partial_\alpha X^\mu \partial_\beta X^\nu 
+  \eta_{MN} A_\mu{}^M \partial_\alpha X^\mu D_\beta Y^N 
 +  \eta_{MN} \partial_\alpha Y^M \gV_\beta{}^M
 \right)  \,.
 \label{DWSWZ}
\ee
The leading term involves the DFT generalised two-form $B_{\mu\nu}$, which is what we expect a string to couple to. 
Both $B_{\mu\nu}$ and the generalised one-form $A_\mu{}^M$ transform under generalised diffeomorphisms and gauge transformations, while $B_{\mu\nu}$ appears in the field strength \eqref{DFTF} for $A_\mu{}^M$. 
The pair $(A_\mu{}^M, B_{\mu\nu})$ are the ``tensor hierarchy'' of $O(d,d)$ DFT.
Similar tensor hierarchies appear in EFT, and reflect the fact that the $O(d,d)$ or $\ED$ covariant fields incorporate components from the same supergravity fields, hence their gauge transformations, field strengths and Bianchi identities are linked in a systematic fashion. 

Part of the systematisation is that we can always associated the generalised $p$-form fields to a sequence of representations $R_p$ of $O(d,d)$ or $\ED$.
The DFT one-form is in the representation $R_1 = \bf{2d}$ of $O(d,d)$ while the DFT two-form is in the representation $R_2 = \mathbf{1}$.
Given objects $A_1, A_2 \in R_1$, there is a map $\p : R_1 \otimes R_1 \rightarrow R_2$, which for DFT is given by $ A_1 \p A_2 = \eta_{MN} A_1{}^M A_2{}^N$.
The above Wess-Zumino term clearly involves:
\be
 B_{\mu\nu} \partial_\alpha X^\mu \partial_\beta X^\nu 
+  \partial_\alpha X^\mu A_\mu \p  D_\beta Y 
 + \partial_\alpha Y \p \gV_\beta \,.
\ee
A natural conjecture would be that the same formula should hold for the groups and representations of EFT. 
In general though, $R_2$ will not be the trivial representation. 
In order to obtain a quantity that can be integrated, we will need to introduce a charge $q \in \bar R_2$, and define
\be
S_{WZ} = - \frac{1}{2} \int d^2 \sigma
\epsilon^{\alpha \beta} 
q \cdot 
\left(  B_{\mu\nu} \partial_\alpha X^\mu \partial_\beta X^\nu 
+  \partial_\alpha X^\mu A_\mu \p  D_\beta Y 
 + \partial_\alpha Y \p \gV_\beta 
 \right)  \,,
\label{guessWZ}
\ee
where $q \cdot B \in \mathbf{1}$. This charge will encode the tension of the string action. Clearly for the doubled sigma model, we just have $q = T_{F1}$. 

Remarkably, the guess \eqref{guessWZ} turns out to be correct, as long as the charge $q$ obeys a constraint \eqref{magicintro}. This constraint comes about when one checks the gauge invariance of the Wess-Zumino term.

\section{Exceptional field theory}
\label{eft}

In this section, we will introduce the core elements of exceptional field theory, focusing on the fields to which the exceptional sigma model couples, and their symmetries.
After presenting the general details, we will focus on the group $\Gsix$, for which some extra details have to be filled in.

\subsection{Field content and symmetries}

Exceptional field theory is a reformulation of supergravity 
with a formal manifest $\ED$ symmetry, realised on an extended set of coordinates $(X^\mu , Y^M)$.
The coordinates $Y^M$ lie in a representation of $\ED$ denoted by $R_1$.
The fields of the theory are also assigned to various $\ED$ representations, and include the following: an external metric, $g_{\mu\nu}$, in the trivial representation, a generalised metric, $\gM_{MN}$, in the coset $\ED / H_D$ where $H_D$ is the maximal compact subgroup of $\ED$, and a tensor hierarchy of generalised form fields, $A_\mu{}^M \in R_1$, $B_{\mu\nu} \in R_2$, $C_{\mu\nu\rho} \in R_3$, $\dots$. 

The exceptional string will couple to the generalised two-form $B_{\mu\nu}$, as well as to the external and generalised metric, and the generalised one-form. 
The list of the representations $R_1$ and $R_2$ for the groups $\ED$, $D=2, \dots 8$ is displayed in table \ref{DGH}.

Note that in practice, not all the fields appearing in the tensor hierarchy are needed in formulating the dynamics of EFT, as in many cases they will contain dualisations of the physical degrees of freedom. 
Additionally, we may also have to include additional ``constrained compensator fields'' which also drop out of the dynamics, but are important in formulating the complete gauge invariant set of field strengths. Such compensator fields appear when $R_p = \bar R_1$, and are related to the appearance of components of the dual graviton. These extra fields are necessary to construct the field strength of the of the generalised one-form in $\Gseven$ \cite{Hohm:2013uia}, and are needed already at the level of the generalised Lie derivative for $\Geight$ \cite{Hohm:2014fxa}. In this paper, they will become relevant when considering the $\Gsix$ exceptional sigma model, appearing in the field strength for the generalised two-form to which the exceptional string couples.

\begin{table}[h]\centering
\begin{tabular}{|c|c|c|c|c|c|}
\hline
$n = 10-d$ & $d$ &  $G = E_{D,D},\, D=d+1$ & $H$ & $R_1$ & $R_2$  \\ \hline
9 & 1 & $\G$ & $\mathrm{SO}(2)$ & $\mathbf{2}_1 \oplus \mathbf{1}_{-1}$  & $ \mathbf{2}_0 $    \\
8 & 2 & $\Gthree$ & $\Hthree$  & $ (\mathbf{3,2})$ & $(\mathbf{\bar 3,1})$  \\
7 & 3 &  $\Gfour$ & $\Hfour$ & $ \mathbf{10}$& $\mathbf{\overline{5}}$  \\
6 & 4 & $\Gfive$ & $\Hfive$  & $\mathbf{16}$ & $\mathbf{10}$  \\
5 & 5 & $\Gsix$ & $\Hsix$  & $\mathbf{27}$& $\mathbf{\overline{27}}$ \\
4 & 6 &  $\Gseven$ & $\Hseven$  & $\mathbf{56}$ & $\ast$  \\
3 & 7 & $\Geight$ & $\Height$ & $\mathbf{248}$ & $\ast$   \\
\hline
\end{tabular}
\caption{Generalised diffeomorphism groups and tensor hierarchy representations for EFT. $n$ is the external dimension while $d$ is the internal dimension from the point of view of type II sections.}
\label{DGH}
\end{table}

The local symmetries of exceptional field theory will be important for us. There are two types of diffeomorphism symmetry: external diffeomorphisms, with parameters $\xi^\mu$ in the trivial representation of $\ED$, and generalised diffeomorphisms, with parameters $\Lambda^M$. In addition, there is a set of generalised gauge transformations of the tensor hierarchy fields. 

Generalised diffeomorphisms can be defined using the generalised Lie derivative, which acting on a generalised vector $U \in R_1$, of weight $\lambda_U$, is given by:
\be
\begin{split} 
\mathcal{L}_\Lambda U^M & = \Lambda^N \partial_N U^M - U^N \partial_N \Lambda^M + Y^{MN}{}_{PQ} \partial_N \Lambda^P U^Q + ( \lambda_U + \omega ) \partial_N \Lambda^N U^M \\
 & = \Lambda^N \partial_N U^M - \alpha P_{adj}^M{}_Q{}^N{}_P \partial_N \Lambda^P U^Q + \lambda_U \partial_N \Lambda^N \,, 
\end{split} 
\ee
where the intrinsic weight is
\be
\omega = \begin{cases} 0 & \mathrm{DFT} \\ 
 - \frac{1}{n-2} & \mathrm{EFT} \,.
 \end{cases} 
\ee
The generalised diffeomorphism parameter $\Lambda$ itself carries weight $-\omega$. 
Here $P_{adj}^M{}_Q{}^N{}_P$ is the projector onto the adjoint representation in the tensor product $R_1 \otimes \bar{R_1}$ and $\alpha$ is a group-dependent constant recorded in \cite{Berman:2012vc}. The $Y$-tensor $Y^{MN}{}_{PQ}$ is formed from group invariants. 
From $D=2$ to $D=6$, the Y-tensor is symmetric on upper and lower indices, and the section condition, restricting the coordinate dependence of the theory, is 
\be
Y^{MN}{}_{PQ} \partial_M \otimes \partial_N = 0 
\ee
or $\partial \otimes \partial |_{R_2} = 0$, and is required for consistency.

Due to these properties of the $Y$-tensor we will restrict our attention to $D \leq 6$, for which one can largely treat EFT (and DFT) in a general manner.

The generalised metric transforms as a generalised tensor of weight 0 under generalised diffeomorphisms.
The external metric transforms as a scalar of weight $-2 \omega$. 
The remaining fields, in the tensor hierarchy do not transform as tensors but rather as gauge fields. Note that we still assign weight $-p\omega$ to the field in the representation $R_p$. 
To formulate their transformations, we introduce some general notation following \cite{Cederwall:2013naa, Hohm:2015xna, Wang:2015hca}. 

There are two useful operations which map between fields of weight $-p\omega$ in representations $R_p$ of the tensor hierarchy. 
There is a nilpotent derivative operator:
\be
\hat \partial: R_{p+1} \rightarrow R_{p} 
\ee
which is automatically covariant (under generalised diffeomorphisms) for $p=1, \dots n-4$ \cite{Cederwall:2013naa}. 
There is also a map
\be
\p : R_{p} \otimes R_q \rightarrow R_{p+q} \,,
\ee
which is taken to be symmetric for $p\neq q$, and is defined for $p+q \leq n-2$.

If we consider just the fields $A \in R_1$ and $B \in R_2$, which are most relevant for our exceptional sigma model, we can express these operations using the $Y$-tensor directly.
First, note that representation $R_2$ always appears in the symmetric part of the tensor product of $R_1$ with itself.
Therefore it is convenient to denote fields in $R_2$ as carrying a (projected) pair of symmetrised $R_1$ indices, thus we write $B^{MN}$.
We define
\be
( \hat \partial B)^M = Y^{MN}{}_{PQ} \partial_N B^{PQ} \,,
\label{hatpartialB}
\ee
and
\be
(A_1 \p A_2 )^{MN} =
\begin{cases} 
 \frac{1}{2d}  Y^{MN}{}_{PQ} A_1^P A_2^Q & O(d,d) \\ 
 \frac{1}{2(D-1)}  Y^{MN}{}_{PQ} A_1^P A_2^Q & E_{D(D)} 
\end{cases} \,.
\ee
Given the definition of $\hat \partial B$ for $B \in R_2$ in the conventions of some EFT, then equation \eqref{hatpartialB} effectively defines our convention for the relationship between $B \in R_2$ and $B^{MN}$. 
We summarise the precise definitions in appendix \ref{EFTconv}. 
Observe also that $\hat \partial B$ defines a trivial generalised diffeomorphism parameter, that is $\mathcal{L}_{\hat \partial B} = 0$ acting on anything.

Now, let us write down the gauge transformations and field strengths associated to the first few fields of the tensor hierarchy. 
First, define the covariant external partial derivative $D_\mu = \partial_\mu - \mathcal{L}_{A_\mu}$ in terms of the generalised Lie derivative $\mathcal{L}$ (which shows that the generalised one-form $A_\mu{}^M$ provides a gauge field for generalised diffeomorphisms).
For the fields in $R_p$, $p > 1$, define ``covariant'' variations 
\be
\begin{split}
 \Delta\Ab_{\mu\nu} = \delta\Ab_{\mu\nu} + \Aa_{[\mu}\pl\delta\Aa_{\nu]} \,, \qquad
\Delta\Ac_{\mu\nu\rho} = \delta\Ac_{\mu\nu\rho} - 3\delta\Aa_{[\mu}\pl\Ab_{\nu\rho]} + \Aa_{[\mu}\pl(\Aa_\nu\pl\delta\Aa_{\rho]}) \,.
\end{split}
\label{DeltaGauge}
\ee
Although we are only really interested in the generalised forms $A\in R_1$ and $B \in R_2$, we have here to include the generalised three-form $C \in R_3$. The exceptional string will not couple to this field, but the nature of the tensor hierarchy means that it still appears in the field strength of $B \in R_2$. 
Then, in terms of generalised diffeomorphisms parametrised by $\Lambda \in R_1$, and gauge transformations $\lambda_\mu \in R_2$, $\Theta_{\mu\nu} \in R_2$, $\Omega_{\mu\nu\rho} \in R_3$, we have
\be
\begin{split} 
 \delta \Aa_{\mu} &= D_{\mu} \Lambda - \hat{\partial}\, \lambda_\mu \,, \\
 \Delta \Ab_{\mu\nu} &= \Lambda\, \pl \Fa_{\mu\nu} + 2 D_{[\mu} \lambda_{\nu]} - \hat{\partial} \Theta_{\mu\nu} \,, \\
\Delta \Ac_{\mu\nu\rho} &= \Lambda\, \pl \Fb_{\mu\nu\rho} + 3 \Fa_{[\mu\nu} \pl \lambda_{\rho]} + 3 D_{[\mu} \Theta_{\nu\rho]} - \hat{\partial} \Omega _{\mu\nu\rho} \,. \\
\end{split} 
\label{deltaGauge}
\ee
Letting $[A_\mu, A_\nu]_E = \frac{1}{2} ( \mathcal{L}_{A_\mu} A_\nu - \mathcal{L}_{A_\nu} A_\mu)$ be the analogue of the Lie bracket, the field strengths for $A_\mu$ and $B_{\mu\nu}$ are:
\be
\begin{split} 
\Fa_{\mu\nu} &= 2\partial_{[\mu} \Aa_{\nu]} - [\Aa_\mu,\Aa_\nu]_E + \hd\Ab_{\mu\nu} \,, \\
\Fb_{\mu\nu\rho} &= 3\D_{[\mu}\Ab_{\nu\rho]} - 3\partial_{[\mu}\Aa_{\nu}\pl\Aa_{\rho]} + \Aa_{[\mu}\pl[\Aa_\nu,\Aa_{\rho]}]_E + \hd\Ac_{\mu\nu\rho} \,, \\
\end{split} 
\label{FAB}
\ee
and their variations are given by:
\be
\begin{split}
 \delta\Fa_{\mu\nu} &= 2\D_{[\mu}\delta\Aa_{\nu]} + \hd\Delta\Ab_{\mu\nu} \,, \\
 \delta\Fb_{\mu\nu\rho} &= 3\D_{[\mu}\Delta\Ab_{\nu\rho]} - 3\delta\Aa_{[\mu}\pl\Fa_{\nu\rho]} + \hd\Delta\Ac_{\mu\nu\rho} \,. \\
\end{split}
\ee
Under \eqref{deltaGauge} the field strengths transform as generalised tensors (of weight $-p\omega$) under generalised diffeomorphisms parametrised by $\Lambda$.

In addition, one has transformations under external diffeomorphisms with parameter $\xi^\mu$. In this paper we will need to use:
\be
\label{deltaexternaldiff}
\begin{split} 
\delta_\xi A_\mu  = \xi^\rho \Fa_{\rho \mu} + \gM^{MN} g_{\mu\nu} \partial_N \xi^\nu \,,\qquad
\Delta_\xi B_{\mu\nu}  = \xi^\rho \Fb_{\mu \nu \rho} \,\,.
\end{split}
\ee
In the EFT construction, requiring invariance of the (bosonic part of the) action under such transformations uniquely fixes the relative coefficients of every term. Remarkably, we will find below that the same holds true on the worldsheet: a subtle interplay between the kinetic and Wess-Zumino terms is needed to ensure covariance under external diffeomorphisms. 

The general expressions here may have to be modified in some groups. 
For $D=7$, we have already mentioned that the field strength of the one-form $A_\mu{}^M$ involves a second two-form field which is necessary for gauge invariance \cite{Hohm:2013uia}, with the generalised one-form transforming under additional ``constrained'' gauge transformations that are necessary in order to shift away components which represent ``dual graviton'' degrees of freedom.
For $D=6$, a similar situation arises at the level of the generalised two-form $B_{\mu\nu}$,
which as described in \cite{Hohm:2013vpa} has a similar additional symmetry.
The full details of the field strength associated to $B_{\mu\nu}$, and the precise form of its extra gauge transformations were not specified in \cite{Hohm:2013vpa}.
Our check of the symmetries of the exceptional sigma model require us to understand the full field strength. To do this, we now look at the example of $\Gsix$ more closely, both to clarify this situation and to make our main example be one in which the group $\ED$ is genuinely exceptional.

\subsection{Example: the $\Gsix$ EFT}

\subsubsection*{General details}

For $\Gsix$, the representation $R_1$, in which the coordinates $Y^M$ appear, is the fundamental 27-dimensional representation.
The representation $R_2$ is the conjugate representation to the fundamental, while the representation $R_3$ is the adjoint.

The group $\Gsix$ has two cubic symmetric invariant tensors, $d^{MNP}$ and $d_{MNP}$. 
These are normalised such that $d^{MPQ} d_{NPQ} = \delta^M_N$, and obey a cubic identity:
\be
\begin{split} 
\frac{1}{3}d_{SPT}  \left( d^{PQK} d^{NRT} + d^{PQN} d^{RKT}+ d^{PQR} d^{KNT}
\right)  = \frac{1}{30} \left( \delta_S^Q d^{KNR} 
+ \delta_S^K d^{NRQ} 
+ \delta_S^N d^{RQK} 
+ \delta_S^R d^{NKQ} 
\right) \,.
\end{split}
\label{d3}
\ee 
The Y-tensor for $\Gsix$ is 
\be
Y^{MN}{}_{PQ} = 10 d^{MNK} d_{PQ K} \,,
\ee 
and the section condition is
\be
d^{MNP} \partial_N \otimes \partial_P = 0 \,.
\label{e6sec} 
\ee
Note that the adjoint projector, $P_{adj}^M{}_N{}^P{}_Q = t_\alpha{}^M{}_N t^{\alpha K}{}_L$, (where $t_\alpha{}^M{}_N$ are the 78 adjoint generators valued in the fundamental) is:
\be
P_{adj}^M{}_P{}^N{}_Q  =  - \frac{5}{3} d_{RPQ} d^{RMN} + \frac{1}{18} \delta^M_P \delta^N_Q + \frac{1}{6} \delta^M_Q \delta^N_P \,,
\label{padj}
\ee
so that
\be
 \mathcal{L}_\Lambda U^M =  \Lambda^N \partial_N U^M - 6 P_{adj}^M{}_Q{}^N{}_P \partial_N \Lambda^P U^Q + \lambda_U \partial_N \Lambda^N U^M\,.
\ee
The operations $\p$ and $\hat \partial$ relevant to the fields in $R_1$ and $R_2$ are
\be
(A_1 \p A_2)_M = d_{MNP} A_1^N A_2^P 
\quad,\quad
( \hat \partial B)^M = 10 d^{MNP} \partial_N B_P\,.
\ee
One therefore has 
\be
B^{MN} = d^{MNP} B_P \,.
\ee
Finally, note that the generalised metric is $\gM_{MN}$, and we have $\gM_{MN} \gM_{PQ} \gM_{KL} d^{NQL} = d_{MPK}$ (expressing the fact $\gM$ is a group element).

\subsubsection*{Sections}

For the components of the $\Gsix$ invariant tensors, we follow the conventions of \cite{Hohm:2013vpa}.
Under the decomposition $\Gsix \rightarrow \mathrm{SL}(6)$ corresponding to obtaining an M-theory, we have $A^M = ( A^i, A_{ij} , A^{\bar i} )$ with $i, \bar i  = 1,\dots,6$ and $A_{ij} = A_{[ij]}$ (as in \cite{Hohm:2013vpa} we do not include an explicit factor of $1/2$ in contractions). Then,
\be
\begin{array}{cccc}
d_{MNP} & \rightarrow & d_{i \bar j}{}^{kl} = \frac{1}{\sqrt{5}} \delta^{kl}_{[ij]} 
& d^{ij,kl,mn} = \frac{1}{4\sqrt{5}} \epsilon^{ijklmn} \\
d^{MNP} & \rightarrow & d^{i \bar j}{}_{kl} = \frac{1}{\sqrt{5}} \delta^{ij}_{[kl]} 
& 
d_{ij,kl,mn} =\frac{1}{4 \sqrt{5}} \epsilon_{ijklmn} \,.
\end{array} 
\label{dm}
\ee
The section condition \eqref{e6sec} is then solved by $\partial_i \neq 0$, $\partial^{ij} = 0 = \partial_{\bar i}$.

Alternatively, under $\Gsix \rightarrow \mathrm{SL}(5) \times \mathrm{SL}(2)$ (corresponding to a IIB section), we have $A^M = ( A^i, A_{i a} , A^{ij} , A_a)$, with now $i,j=1,\dots,5$ and $a,b=1,2$, and
\be
\begin{array}{ccccc}
d_{MNP} & \rightarrow & d_i{}^{ja, b} = \frac{1}{\sqrt{10}} \delta^j_i \epsilon^{a b} &
 d_{ij}{}^{ka,lb} = \frac{1}{\sqrt{5}} \delta^{kl}_{[ij]} \epsilon^{a b} & d_{ij,kl,m} = \frac{1}{2 \sqrt{10}} \epsilon_{ijklm} \\ 
d^{MNP} & \rightarrow & d^i{}_{ja, b} = \frac{1}{\sqrt{10}} \delta^i_j \epsilon_{a b} &
 d^{ij}{}_{ka,lb} = \frac{1}{\sqrt{5}} \delta^{ij}_{[kl]} \epsilon_{a b} & d^{ij,kl,m} = \frac{1}{2 \sqrt{10}} \epsilon^{ijklm} \\ 
\end{array}
\label{db}
\ee
The section condition \eqref{e6sec} is then solved by $\partial_i \neq 0$, $\partial^{i a} = 0 =\partial_{ij} = \partial^a$.

\subsubsection*{Continuing the tensor hierarchy}

The fields that appear in the action of the $\Gsix$ EFT are $g_{\mu\nu}, \gM_{MN}, A_\mu{}^M$ and $B_{\mu\nu M}$. These are also the fields to which the exceptional string will couple. However, the generalised two-form is not dynamical: there is no kinetic term involving its field strength $\Fb_{\mu\nu\rho}$, and its equation of motion leads to: 
\be
d^{MNK} \partial_K \left(
\sqrt{g} \gM_{NL} \Fa^{\mu\nu L} 
+ \frac{\sqrt{10}}{6} \epsilon^{\mu\nu\rho\sigma\kappa} \Fb_{\rho \sigma \kappa N} 
\right) 
= 0  \,,
\label{E6duality} 
\ee
which is interpreted as a duality relation relating components of $\Fa_{\mu\nu}{}^M$ to components of $\Fb_{\mu\nu\rho M}$. 

As this is the only place in the dynamics of the $\Gsix$ EFT that $\Fb_{\mu\nu\rho M}$ appears, in \cite{Hohm:2013vpa} this field strength was only determined up to pieces which vanished under the action of $\hat \partial$ (i.e. under $d^{MNK} \partial_K$ as above).
This is consistent with the observation that the standard formulae for the tensor hierarchy field strengths, \eqref{FAB}, do not apply anymore, as for $\Gsix$ the derivative $\hat \partial: R_3 \rightarrow R_2$ is no longer automatically covariant under generalised diffeomorphisms. In general, this happens when one reaches a form-field representation $R_p$ which coincides with $\bar R_1$.
As we have said, for $\Gseven$, problems arise already for $A_\mu \in R_1$, and these can be circumvented by introducing a second ``constrained'' two-form \cite{Hohm:2013uia}. Here, we detail the analogous construction that applies for $\Gsix$, following the clues provided in \cite{Musaev:2014lna}. 

For $\Gsix$, the representation $R_3$ is the adjoint. 
We introduce a three-form $C_{\mu\nu\rho}{}^\alpha \in R_3$, and a second three-form $\tilde C_{\mu\nu\rho M} \in \mathbf{\bar{27}}$, which is constained to obey the same constraints as the derivatives $\partial_M$. That is, $d^{MNP} \tilde C_{\mu\nu\rho M} \tilde C_{\sigma \kappa \lambda N }  = d^{MNP} \tilde C_{\mu\nu \rho M} \partial_N = 0$.
We introduce a derivative map, which we may as well persist in calling $\hat\partial$,
\be
(\hat \partial C_{\mu\nu\rho} )_M = 6 t_\alpha{}^N{}_M \partial_N C_{\mu\nu\rho}{}^\alpha
\ee
and define the field strength by 
\be
\Fb_{\mu\nu\rho} = 3\D_{[\mu}\Ab_{\nu\rho]} - 3\partial_{[\mu}\Aa_{\nu}\pl\Aa_{\rho]} + \Aa_{[\mu}\pl[\Aa_\nu,\Aa_{\rho]}]_E + \hd\Ac_{\mu\nu\rho}
+ \tilde C_{\mu\nu\rho}
 \,, \\
\ee
where now
\be
\delta\Fb_{\mu\nu\rho} = 3\D_{[\mu}\Delta\Ab_{\nu\rho]} - 3\delta\Aa_{[\mu}\pl\Fa_{\nu\rho]} + \hd\Delta\Ac_{\mu\nu\rho} + \Delta \tilde C_{\mu\nu\rho} \,. \\
\ee
The covariant variations for the three-forms are given by
\be
\Delta C_{\mu\nu\rho}{}^\alpha
 = \delta C_{\mu\nu\rho}{}^\alpha 
 - 3(t^\alpha)^P{}_Q \delta A_{[\mu}{}^Q B_{\nu \rho] P }
+  (t^\alpha)^P{}_Q d_{PKL}A_{[\mu}{}^Q  A_\nu{}^K \delta A_{\rho]}{}^L \,,
\ee
which in fact conforms to the usual structure of \eqref{DeltaGauge} if one says
$( A \p  B)_\alpha = t_\alpha{}^P{}_Q A^Q B_P$,
while
\be
\begin{split} 
\Delta \tilde C_{\mu\nu\rho M}  &
 =  \delta \tilde C_{\mu\nu\rho M} 
+ \partial_M ( \delta A_{[\mu}{}^K B_{\nu \rho] K} ) 
- 3 \partial_M \delta A_{[\mu}{}^K B_{\nu \rho ]K} 
\\ & \qquad
-\frac{1}{3}d_{PKL} \partial_M ( A_{[\mu}{}^P A_\nu{}^K \delta A_{\rho]}{}^L )
+ d_{PKL} \partial_M A_{[\mu}{}^P A_\nu{}^K \delta A_{\rho]}{}^L  \,.
\end{split}
\ee
The usual gauge transformations \eqref{DeltaGauge}
must be accompanied by a transformation
\be
\begin{split} 
  \Delta \tilde C_{\mu \nu \rho M}  =
  \partial_M \Lambda^N \mathcal{H}_{\mu\nu\rho N} - \frac{1}{3} \partial_M ( \Lambda^N \mathcal{H}_{\mu\nu\rho N}) 
 + 3 \partial_M \mathcal{F}_{[\mu\nu}{}^K \lambda_{\rho] K} 
 - \partial_M ( \mathcal{F}_{[\mu\nu}{}^K \lambda_{\rho] K} )
\\ +  18 \partial_M \partial_N A_{[\mu}{}^P \Theta_{\nu \rho]}{}^\alpha t_\alpha{}^N{}_P
\end{split} 
\ee
of the constrained three-form: in addition, one has a gauge transformation of this object given by
\be
\begin{split} 
\Delta B_{\mu\nu M}  = -\Theta_{\mu\nu M}\,, \qquad
\Delta\tilde  C_{\mu\nu\rho M}  = 3 D_{[\mu} \Theta_{\nu \rho ]M} \,,
\end{split}
\ee
where $\Theta_{\mu\nu M}$ is constrained in the same manner as $\tilde C_{\mu\nu\rho M}$.

Verifying that these gauge transformations work requires the use of the Bianchi identity for $A_\mu$, $3 D_{[\mu } \Fa_{\nu \rho]}{} = \hat \partial \Fb_{\mu\nu\rho}$, and showing that 
\be
t_\alpha{}^N{}_P d^{QKP} \partial_K \partial_N = 0 
\label{tdsec}
\ee
which can be done using the section condition, the relationship $P^M{}_N{}^K{}_L = t_\alpha{}^M t^{\alpha K}{}_L$, and the cubic identity \eqref{d3}.
Note that
\be
\label{E6undeterminedvanishing}
d^{MNK} \partial_K \left(
 6 t_\alpha{}^P{}_N \partial_P C_{\mu\nu\rho}{}^\alpha
 + \tilde C_{\mu\nu\rho N} 
\right)  = 0 \,,
\ee
with the first term vanishing due to \eqref{tdsec} and the second vanishing due to the constrained nature of $\tilde C_{\mu\nu\rho M}$. The terms inside the bracket are the ``undetermined terms'' $\mathcal O_{\mu\nu\,M}$ of \cite{Hohm:2013vpa}. 
Equation \eqref{E6undeterminedvanishing} ensures that the three-form potentials do not appear in the $\Gsix$ EFT action of \cite{Hohm:2013vpa}, nor in the duality relation \eqref{E6duality}. In \cite{Baguet:2015xha}, this leads to an ambiguity in the ``integrated'' form of this duality relation between certain components of $\Fa_{\mu\nu}$ and $\Fb_{\mu\nu\rho}$. 
Here, as we have access to the fully covariant field strengths, we instead assume the duality relation holds without the derivative:
\be\sqrt{g} \gM_{NL} \Fa^{\mu\nu L} 
+ \frac{\sqrt{10}}{6} \epsilon^{\mu\nu\rho\sigma\kappa} \Fb_{\rho \sigma \kappa N} = 0 
\label{intduality}
\ee 
and can use the gauge freedoms $\Theta_{\mu\nu \alpha}$ and $\Theta_{\mu\nu M}$ associated to the three-forms to gauge these away, recovering the same duality relations between components used in \cite{Baguet:2015xha}.
The expression \eqref{intduality} is natural to take as the complete ``integrated'' form of the duality relation, as the objects appearing in it are proper generalised tensors.

\section{The exceptional sigma model}
\label{esm}

We will now present and construct the action for the exceptional sigma model.

\subsection{Action and symmetries} 

Denote the worldsheet coordinates $\sigma^\alpha$, the worldsheet metric by $\gamma_{\alpha\beta}$ and the worldsheet Levi-Civita symbol by $\epsilon^{\alpha \beta}$.
The worldsheet fields that appear are the extended spacetime coordinates $(X^\mu,Y^M)$, and the worldsheet one-form $V_\alpha^M$, which is constrained by the requirement 
\be
V_\alpha{}^M \partial_M = 0 \,.
\ee
These worldsheet fields are coupled by the background fields $(g_{\mu\nu}, \gM_{MN}, A_\mu{}^M, B_{\mu\nu})$, which depend on the coordinates $(X^\mu,Y^M)$ subject to the section condition. 
We further introduce a charge $q \in \bar R_2$.
Then the action for the exceptional sigma model is
\be
\begin{split}
S_{EWS}  = - \frac{1}{2}  \int d^2 \sigma 
\Big(
&
T(\gM, q)
 \Big(
\sqrt{-\gamma} \gamma^{\alpha \beta} g_{\mu\nu} 
\partial_\alpha X^\mu \partial_\beta X^\nu 
+ \frac{1}{2} \sqrt{-\gamma} \gamma^{\alpha \beta} \gM_{MN} D_\alpha Y^M D_\beta Y^N \Big)
\\ & 
\qquad + 
q \cdot \epsilon^{\alpha \beta} \left( 
B_{\mu \nu} \partial_\alpha X^\mu \partial_\beta X^\nu 
+ \partial_\alpha  X^\mu ( A_\mu \p D_\beta Y )
+ (\partial_\alpha Y \p \gV_\beta)
\right)\Big)
\end{split} 
\label{ES}
\ee
where the ``tension'' is:
\be
T(\gM,q) = 
\sqrt{ q \cdot ( \mathcal{M}^{-1}  \mathcal{M}^{-1} )|_{R_2 \otimes R_2} \cdot q }\,,
\ee
where 
the notation means that
given the product $\gM^{MN} \gM^{PQ}$ 
we project the index pairs $MP$ and $NQ$ separately into $R_2$ before contracting each with one $q \in \bar R_2$.
This projection is of course automatic if we express $q$ using $R_1$ indices as $q_{MN}$ (as opposed to having $q$ carry a $\bar R_2$ index, in which case we think of the product of generalised metrics as being projected instead).
In this case, 
\be
\begin{split}
S_{EWS}  = - \frac{1}{2}  \int d^2 \sigma 
\Big(
&
T(\gM, q)
 \Big(
\sqrt{-\gamma} \gamma^{\alpha \beta} g_{\mu\nu} 
\partial_\alpha X^\mu \partial_\beta X^\nu 
+ \frac{1}{2} \sqrt{-\gamma} \gamma^{\alpha \beta} \gM_{MN} D_\alpha Y^M D_\beta Y^N \Big)
\\ & 
\qquad + 
q_{MN} \epsilon^{\alpha \beta} \left( 
B_{\mu \nu}{}^{MN} \partial_\alpha X^\mu \partial_\beta X^\nu 
+ \partial_\alpha  X^\mu  A_\mu{}^M  D_\beta Y^N
+ \partial_\alpha Y^M  \gV_\beta^N 
\right)\Big)\,,
\end{split} 
\label{actionnice}
\ee
with
\be
T ( \gM, q)
= \sqrt{ \frac{  q_{MN} q_{PQ} \gM^{MP} \gM^{NQ} }{2d} } \,.
\label{Tm} 
\ee
For convenience, we will just write $T \equiv T(\gM,q)$ in the rest of the paper.

We may summarise the symmetries of this action:

\begin{itemize}
\item Gauge symmetries: the usual worldsheet diffeomorphisms $\sigma^\alpha \to (\sigma'(\sigma))^\alpha$ acting on the worldsheet metric $\gamma_{\alpha \beta}$, scalars $X^\mu, Y^M$ and the $R_1$-valued 1-forms $V_\alpha^M$, Weyl transformations rescaling $\gamma$, and finally the following less usual \emph{shift symmetries}\footnote{For the doubled string these were used in \cite{Hull:2004in, Hull:2006va}, and reflect the fact the section condition reduces the dependence on the coordinates. In \cite{Park:2013mpa, Lee:2013hma} this is referred to as the coordinate gauge symmetry of DFT or EFT, corresponding here to $\varepsilon^M=Y^{MN}{}_{PQ} ( \varphi\partial_N \varphi')^{PQ}$ for $\varphi, \varphi'$ arbitrary $\ED$ tensors.} for which the $V_\alpha^M$ are gauge fields:  (we denote worldsheet variations by $\bar\delta$; unadorned variations act on the EFT background)
\be
\label{shiftsymmetry}
\bar\delta_\varepsilon Y^A=\varepsilon^A\,,\qquad \bar\delta_\varepsilon V_\alpha^A=-\partial_\alpha \varepsilon^A
\ee
where the $Y^A\in Y^M$ are dual coordinates in a given section of the EFT background (i.e. $\partial_A=0$). We thus have $V^M_\alpha \partial_M=0$. The $V$ equation of motion is equivalent to the twisted self-duality constraint
\be
\label{selfduality}
T(\gM,q) DY^M= \gM^{MN} q_{NP} (\star_\gamma DY^P)
\ee
generalising that of the doubled string, with the constant $q_{MN}$ replacing the $O(d,d)$ structure $\eta_{MN}$.

\item Background gauge symmetry: the EFT gauge transformations $\delta_\lambda B_{\mu\nu}{}^{MN}=2 \partial_{[\mu}\lambda_{\nu]}{}^{MN}+\dots$, accompanied by the transformation $\bar \delta_\lambda V_\alpha^M$ \eqref{oneformgaugetransformsAandV}, are invariances of the exceptional sigma model action. In other words the electric $B$-field coupling is gauge-invariant. This is true only if the constant charge parameter $q_{MN}\in \bar R_2$ is \emph{constrained} by \eqref{magicintro} which can be more suggestively rewritten $\mathcal L_\Lambda q_{MN}=0$.
Its surviving components depend on the number of generalised Killing vectors of the EFT background and the type of section condition; for a generic background in a IIB section $q_{MN}$ has two independent surviving components which form an $\mathrm{SL}(2)$ doublet and determine the couplings to the IIB supergravity NS-NS and R-R 2-forms, while for ``IIA'' sections $q_{MN}$ has one independent surviving component which is simply identified with the type IIA string tension. This interpretation is justified when we use the EFT-to-supergravity dictionary in section \ref{reductionsection} to relate the EFT background to a type IIB or eleven-dimensional supergravity background, which serves to identify the precise relation of the exceptional string to the usual string theory strings.

\item Global symmetries: for each generalised Killing vector $\Lambda^M(X,Y)$ --- i.e. $\Lambda$ such that $\delta_\Lambda F=0$ for any EFT background field $F\in\{g_{\mu\nu},\gM_{MN},A_\mu{}^M,B_{\mu\nu}{}^{MN},\dots\}$ --- the variation 
\be
\bar\delta_\lambda Y^M=\Lambda^M
\ee
along with the $\bar\delta_\lambda V^M_\alpha$ of \eqref{wsgd} leaves the action invariant. Similarly for each external Killing vector $\xi^\mu(X,Y)$ with $\delta_\xi F=0$ we have a global symmetry of the sigma model acting as $\bar \delta_\xi X^\mu=\xi^\mu$. In other words,
\begin{center}
\emph{(generalised) Killing vectors induce infinitesimal global symmetries.}
\end{center}
This is ensured by the stronger requirement that $\bar\delta_{\Lambda,\xi} S_{EWS}$ induces the usual transformation of the background fields under infinitesimal generalised and external diffeomorphisms  
$\Lambda^M,\xi^\mu$. This is a covariance condition expressing the fact pullbacks of generalised tensors are geometric. The analogous property in Riemannian geometry is trivially true: consider $\bar\delta_\xi X^\mu=\xi^\mu(X)$ acting on $g_{\mu\nu}(X)\partial_\alpha X^\mu\partial_\beta X^\nu$, then a short calculation gives
\be
\bar\delta_\xi\big(g_{\mu\nu}(X)\partial_\alpha X^\mu\partial_\beta X^\nu\big)=L_\xi g_{\mu\nu}(X) \partial_\alpha X^\mu\partial_\beta X^\nu\,.
\ee
Finally, there is a formal $\ED$ invariance acting on the fields and coordinates in the obvious manner. As usual in a formalism with the same philosophy as DFT or EFT, a choice of section on which the background fields depend breaks this. 
\end{itemize}

\subsection{Fixing the action: generalised diffeomorphisms and gauge transformations}

We claim that the above action can be fixed essentially from scratch, based on a few reasonable assumptions. 
We begin by deciding we are searching for a Weyl-invariant Polyakov-style string action, quadratic in worldsheet derivatives of the extended spacetime coordinates $(X^\mu, Y^M)$, depending polynomially on the EFT background fields and in particular coupling electrically to the EFT two-form $B_{\mu\nu}{}^{MN}$.

\subsubsection*{Kinetic terms} 

First, let us think about pullbacks of exceptionally geometric quantities. 
As was pointed out in \cite{Park:2013mpa}, the pullback of the ``generalised line element'', which becomes $\gM_{MN} \partial_\alpha Y^M \partial_\beta Y^N$ on the worldsheet, 
is not a good object as it does not respect the generalised diffeomorphism symmetry correctly.
Under the shift $\bar\delta_\Lambda Y^M = \Lambda^M(X,Y)$, we do not have $\bar \delta_\Lambda ( \gM_{MN} \partial_\alpha Y^M \partial_\beta Y^N ) = \mathcal L_\Lambda \gM_{MN} \partial_\alpha Y^M \partial_\beta Y^N$.
Indeed, for $\Lambda^M = \Lambda^M(X)$ the generalised diffeomorphism reduces to a gauge transformation of $A_\mu{}^M$, and the pullback $\gM_{MN} \partial_\alpha Y^M \partial_\beta Y^N$ is evidently not gauge invariant as it should be. This necessitates the combination $\partial_\alpha Y^M \rightarrow \partial_\alpha Y^M + \partial_\alpha X^\mu A_\mu{}^M$, and to further take care of $Y$-dependent transformations we introduce $\gV_\alpha^M$ and define
\be
D_\alpha Y^M = \partial_\alpha Y^M + \partial_\alpha X^\mu A_\mu{}^M + \gV_\alpha^M \,,
\ee
The extra worldsheet one-form $\gV_\alpha^M$ ensures the correct covariance under $Y$-dependent generalised diffeomorphisms and is also needed for invariance under generalised gauge transformations, with
\be
\label{oneformgaugetransformsAandV}
\begin{split}
\delta_\lambda A_\mu{}^M & = - Y^{MN}{}_{PQ} \partial_N \lambda_\mu{}^{PQ} \,,\\
\bar\delta_\lambda \gV_\alpha^M & = Y^{MN}{}_{PQ} \partial_N \lambda_\mu{}^{PQ} \partial_\alpha X^\mu \,.
\end{split}  
\ee
and also for invariance under the ``coordinate gauge symmetry'' of \cite{Park:2013mpa, Lee:2013hma} which is a consequence of the section condition. 

We fix the transformation of $\gV_\alpha^M$ under generalised diffeomorphisms by postulating that, for $U_M$ a generalised covector carrying special weight $+\omega$, that
\be
\bar \delta_\Lambda ( U_M D_\alpha Y^M ) = ( \mathcal L_\Lambda U_M ) D_\alpha Y^M + U_M \partial_\alpha X^\mu \delta_\Lambda A_\mu{}^M \,.
\ee
(This is because $Y^M$ (and hence $\partial_\alpha Y^M$), $A_\mu{}^M$ and $\Lambda^M$ all have weight $-\omega$. The transformation of $\gV_\alpha^M$ is also consistent with taking its weight to be $-\omega$.) This equation expresses the fact the pullback of $U_M$ to the worldsheet is ``geometric''; its analogue in ordinary geometry is identically true. It implies that under generalised diffeomorphisms, one has
\be
\begin{split}
\bar\delta_\Lambda Y^M & = \Lambda^M \,,\\
\bar\delta_\Lambda X^\mu & = 0\,,\\
\bar\delta_\Lambda \gV_\alpha^M & = - Y^{MN}{}_{PQ} ( \partial_N \Lambda^P D_\alpha Y^Q + \partial_N A_\mu{}^P \partial_\alpha X^\mu \Lambda^Q )\,,\\
\bar\delta_\Lambda \mathcal{O} & = \Lambda^P \partial_P \mathcal{O} \,,
\end{split} 
\label{wsgd} 
\ee
where $\mathcal{O}$ is any function of the extended coordinates defined on the worldsheet.
As a result,
\be
\bar \delta_\Lambda D_\alpha Y^M = \left( \partial_P \Lambda^M - Y^{MN}{}_{KP} \partial_N \Lambda^K \right) D_\alpha Y^P 
+ \delta_\Lambda A_\mu{}^M \partial_\alpha X^\mu \,.
\label{wsgdDY}
\ee
Now, the generalised metric $\gM_{MN}$ has weight zero. With the above transformations, this means that: 
\be
\begin{split}
\bar \delta_\Lambda ( \gM_{MN} D_\alpha Y^M D_\beta Y^N) & = 
(\delta_\Lambda \gM_{MN} )D_\alpha Y^M D_\beta Y^N + 2 \gM_{MN} D_{(\alpha} Y^M \delta_\Lambda A_\mu{}^N \partial_{\beta)} X^\mu 
\\ & \qquad
+ 2 \omega \partial_K \Lambda^K \gM_{MN} D_\alpha Y^M D_\beta Y^N \,.
\end{split}
\ee
To cancel the final term, we need to introduce some object of weight $+2\omega$. Note that one cannot cancel these terms by modifying the transformation rule of $\gV_\alpha^M$ as this breaks the condition $\gV_\alpha^M \partial_M = 0$. We introduce a charge $q_{MN}$ in the $\bar R_2$ representation, carrying weight $+2\omega$. This weight assignment is natural as $B_{\mu\nu}{}^{MN} \in R_2$ has weight $-2\omega$.
We require that the generalised Lie derivative of $q_{MN}$ be zero, which leads to the constraint
\be
q_{MN} \partial_P = q_{NK} Y^{KL}{}_{MN} \partial_L 
\label{magic} \,, 
\ee
which will appear again later as being necessary for gauge invariance of the Wess-Zumino term. 
If we define the tension $T$ as in \eqref{Tm} (the numerical factor is in principle arbitrary at this point, and will be fixed later when we examine external diffeomorphisms), then this provides a scalar of weight $+2\omega$, and it follows that 
\be
\begin{split}
\bar \delta_\Lambda ( T \gM_{MN} D_\alpha Y^M D_\beta Y^N) & = 
( \delta_\Lambda T \gM_{MN} ) D_\alpha Y^M D_\beta Y^N + 2 T\gM_{MN} D_{(\alpha} Y^M \delta_\Lambda A_\mu{}^N \partial_{\beta)} X^\mu \,,
\end{split}
\ee
which is the desired transformation rule.

Similarly, one finds that $T g_{\mu\nu} \partial_\alpha X^\mu \partial_\beta X^\nu$ behaves correctly, as $g_{\mu\nu}$ is a scalar of weight $-2\omega$ under generalised diffeomorphisms. 

We conclude that the only gauge invariant kinetic terms, quadratic in the derivatives of the worldsheet scalars, and transforming in the appropriate manner under $\bar\delta_\Lambda Y^M = \Lambda^M$ are $T \gM_{MN} D_\alpha Y^M D_\beta Y^N$ and $T g_{\mu\nu} \partial_\alpha X^\mu \partial_\beta X^\nu$, assuming we exclude terms nonpolynomial in the EFT background fields, such as $B_{\mu\nu}g^{\nu\rho} B_{\rho\sigma} \partial_\alpha X^\mu \partial_\beta X^\sigma$ and the like.

\subsubsection*{Wess-Zumino terms} 

Our starting assumption is that the coupling to the generalised two-form involves 
\be
\epsilon^{\alpha \beta} q_{MN} B_{\mu\nu}{}^{MN} \partial_\alpha X^\mu \partial_\beta X^\nu\,.
\ee
We need to search for the gauge invariant completion of this.
Assuming that the external and generalised metrics do not appear, we write down the following general (up to total derivatives) guess for a quadratic Wess-Zumino term:
\be
\mathcal{L}_{WZ} = \epsilon^{\alpha \beta} q_{MN} \left( 
B_{\mu\nu}{}^{MN} \partial_\alpha X^\mu \partial_\beta X^\nu
+ \alpha \partial_\alpha X^\mu A_\mu{}^M D_\beta Y^N 
+ \beta \partial_\alpha X^\mu A_\mu{}^M  \gV_\beta{}^N 
+ \gamma \gV_\alpha^M D_\beta Y^N 
\right) \,,
\label{WZguess}
\ee
where we now want to determine the numerical coefficients $\alpha,\beta,\gamma$.
Under the gauge transformation $\lambda_\mu{}^{MN}$, one finds 
\be
\begin{split}
\delta_\lambda \mathcal{L}_{WZ}  =
\epsilon^{\alpha \beta} q_{MN} &\Big(
\partial_\alpha X^\mu\left( 2 \partial_\mu \lambda_\nu{}^{MN}\partial_\beta X^\nu +(\gamma-\alpha)Y^{M R}{}_{PQ}\partial_R \lambda_\mu{}^{PQ}  \partial_\beta Y^N\right)
\\ &  \qquad+
\partial_\alpha X^\mu\partial_\beta X^\nu\left(-2 \mathcal L_{A_\mu} \lambda_\nu^{MN} + 
(-1-\alpha-\beta +\gamma\big)
Y^{M R}{}_{PQ}\partial_R \lambda_\mu{}^{PQ} A_\nu{}^N 
\right)
\\ &  \qquad+
(\gamma-\alpha-\beta)\partial_\alpha X^\mu Y^{M R}{}_{PQ}\partial_R \lambda_\mu{}^{PQ} V_\beta^N
\Big) \,.
\end{split} 
\label{deltalwz1}
\ee
We expect this should equal the total derivative $2 \epsilon^{\alpha \beta} q_{MN}  \partial_\alpha \lambda_\nu{}^{MN}\partial_\beta X^\nu$.
This requires:
\be
2 q_{PQ} \partial_N = ( \alpha-\gamma) q_{MN} Y^{MK}{}_{PQ} \partial_K \,,
\ee
from terms in the first line, while for the second line to vanish we need 
\be
2 q_{PQ} \partial_N = (1 + \beta + \alpha-\gamma) q_{MN} Y^{MK}{}_{PQ} \partial_K\,,
\ee
from the terms involving derivatives of $\lambda_\mu{}^{PQ}$, and also
\be
 q_{Q(P} \partial_{N)} = q_{M(N} Y^{MK}{}_{P)Q} \partial_K\,,
\ee
from terms involving derivatives of $A_\mu{}^N$. Thus we need the condition \eqref{magic}, and to fix $\alpha-\gamma=2$, $\beta = - 1$.
The final line in \eqref{deltalwz1} vanishes using \eqref{magic} and $\gV_\alpha{}^M \partial_M = 0$.

We then fix the final coefficient in
\be
\mathcal{L}_{WZ} = \epsilon^{\alpha \beta} q_{MN} \left( 
B_{\mu\nu}{}^{MN} \partial_\alpha X^\mu \partial_\beta X^\nu
+ \alpha \partial_\alpha X^\mu A_\mu{}^M D_\beta Y^N 
- \partial_\alpha X^\mu A_\mu{}^M  \gV_\beta{}^N 
+ (\alpha-2)\gV_\alpha^M D_\beta Y^N 
\right) \,,
\label{WZguess2}
\ee
using covariance under generalised diffeomorphisms. We require the pullback of $B_{\mu\nu}$ to be geometric in the sense
\be
\begin{split}
\bar\delta_\Lambda
\mathcal{L}_{WZ} & \stackrel{!}{=} \epsilon^{\alpha \beta} q_{MN} \Big( 
\delta_\Lambda B_{\mu\nu}{}^{MN} \partial_\alpha X^\mu \partial_\beta X^\nu
+  \partial_\alpha X^\mu \delta_\Lambda A_\mu{}^M 
( \alpha D_\beta Y^N - \gV_\beta^N ) 
\\  & \qquad\qquad \qquad \qquad 
+ 
( \alpha \partial_\alpha X^\mu A_\mu{}^M + (\alpha-2) \gV_\alpha^M ) \delta_\Lambda A_\mu{}^M \partial_\beta X^\mu 
\Big) \,,
\label{WZwant1}
\end{split} 
\ee
where $\delta_\Lambda A_\mu{}^M = D_\mu \Lambda^M$ and $\delta_\Lambda B_{\mu\nu}{}^{MN} = \frac{1}{2d} Y^{MN}{}_{PQ} \left( \Lambda^P \Fa_{\mu\nu}{}^Q - A_{[\mu}{}^P D_{\nu]} \Lambda^Q\right)$.
The required variation \eqref{WZwant1} can be simplified to:
\be
\begin{split}
\bar\delta_\Lambda
\mathcal{L}_{WZ} & \stackrel{!}{=} \epsilon^{\alpha \beta} q_{MN} \Big( 
\Big[
\Lambda^P \partial_P B_{\mu\nu}{}^{MN} 
+ 2 \Lambda^M \partial_\mu A_\nu{}^N - A_\mu{}^M \partial_\nu \Lambda^N 
\\  & \qquad\qquad \qquad \qquad 
- \Lambda^M A_\mu{}^P \partial_P A_\nu{}^N + A_\mu{}^M A_\nu{}^P \partial_P \Lambda^N
\Big]
 \partial_\alpha X^\mu \partial_\beta X^\nu
 \\  & \qquad\qquad \qquad \qquad 
 + 
\partial_\alpha X^\mu ( \partial_\mu \Lambda^M - A_\mu{}^P \partial_P \Lambda^M + \Lambda^P \partial_P A_\mu{}^M ) 
( \alpha D_\beta Y^N - \gV_\beta^N ) 
\\  & \qquad\qquad \qquad \qquad 
- \partial_\alpha X^\mu \partial_P A_\mu{}^N \Lambda^N ( \alpha D_\beta Y^N - \gV_\beta^N ) 
\Big) \,.
\label{WZwant2}
\end{split} 
\ee
Now, the direct variation $\bar\delta_\Lambda$ \eqref{wsgd} can never produce the derivatives $\partial_\mu A_\nu{}^M$ or $\partial_\mu \Lambda^M$. 
Thus, in \eqref{WZwant2} we should replace $\partial_\alpha X^\mu \partial_\mu = \partial_\alpha - \partial_\alpha Y^M \partial_M$ and then remove the $\partial_\alpha$ derivative acting on the background field by integration by parts. In order not to generate an unwanted -- and uncancellable -- term involving $\partial_\alpha V_\beta{}^N$ from the third line when doing so, we have to take $\alpha = 1$.
With the coefficient fixed to this value, a straightforward calculation shows that the direct variation $\bar \delta_\Lambda \mathcal{L}_{WZ}$ indeed leads to equation \eqref{WZwant1} up to the total derivative terms we have just indicated.

A nice property of the final Wess-Zumino term,
\be
\begin{split} 
\mathcal{L}_{WZ} & = \epsilon^{\alpha \beta} q_{MN} \left( 
B_{\mu\nu}{}^{MN} \partial_\alpha X^\mu \partial_\beta X^\nu
+ \partial_\alpha X^\mu A_\mu{}^M D_\beta Y^N 
- \partial_\alpha X^\mu A_\mu{}^M  \gV_\beta{}^N 
-\gV_\alpha^M D_\beta Y^N 
\right) 
\\ & 
= \epsilon^{\alpha \beta} q_{MN} \left( 
B_{\mu\nu}{}^{MN} \partial_\alpha X^\mu \partial_\beta X^\nu
+ \partial_\alpha X^\mu A_\mu{}^M D_\beta Y^N 
+ \partial_\alpha Y^M \gV_\beta{}^N \right) \,,
\end{split} 
\label{WZfinal}
\ee
is that if one varies the background fields only, one obtains the covariant variation \eqref{DeltaGauge} of $B_{\mu\nu}$ automatically:
\be
\delta_{(A,B)} \mathcal{L}_{WZ} = 
\epsilon^{\alpha \beta}  q_{MN} \left( \Delta B_{\mu\nu}{}^{MN} \partial_\alpha X^\mu \partial_\beta X^\nu + \partial_\alpha X^\mu \delta A_\mu{}^M  D_\beta Y^N \right) \,.
\label{genvarWZ}
\ee
Finally, we must consider the transformation of $B_{\mu\nu}{}^{MN}$ under the gauge transformations $\Theta_{\mu\nu} \in R_3$, under which we have
$\Delta B_{\mu\nu} = - \hat \partial \Theta_{\mu\nu}$. 
For $C \in R_3$, we need 
\be
q_{MN} (\hat \partial C)^{MN} = 0 \,.
\label{qhatC}
\ee
We have not considered the general structure of $\hat\partial: R_3 \rightarrow R_2$, however the results of \cite{Aldazabal:2013via} indicate that for $D<6$ we can write this as
\be
\hat \partial C^{MN}=\tilde Y^{MNT}{}_{QRS} \partial_T C^{QRS}\,,\qquad \tilde Y^{MNT}{}_{QRS}= Y^{MN}{}_{PQ}Y^{PT}{}_{RS}-\delta^Q_T Y^{MN}{}_{RS}
\ee
which vanishes when hit with $q_{MN}$ using \eqref{magic}. 

We should consider $D=6$ separately. There, with $q_{MN} = d_{MNP} q^P$, we find instead that we need 
\be
 \label{gaugemalarkey}
 q^N \left(
 6 t_\alpha{}^P{}_N \partial_P \Theta_{\mu\nu}{}^\alpha
 + \tilde \Omega_{\mu\nu N}\right) =0
\ee
which is in fact true by a calculation similar to the one leading to \eqref{E6undeterminedvanishing} as long as $q$ satisfies \eqref{magic}.

\subsection{Fixing the action: external diffeomorphisms} 

To finish what we have started, we consider external diffeomorphisms. 

\subsubsection*{Kinetic terms}

The action of external diffeomorphisms on the worldsheet coordinates and hence on the background fields viewed as functions on the worldsheet is:
\be
\begin{split} 
\bar{\delta}_\xi X^\mu & = \xi^\mu \,,\\
\bar{\delta}_\xi Y^M & = - \xi^\mu A_\mu{}^M \,, \\
\bar{\delta}_\xi \mathcal{O} & = \xi^\mu ( \partial_\mu - A_\mu{}^M \partial_M ) \mathcal{O} \,.
\end{split}
\label{wvolext}
\ee
Observe that we include a transformation of the $Y^M$ which takes the form of a field dependent generalised diffeomorphism. This is necessary due to the form of the covariant derivative $D_\mu=\partial_\mu - A_\mu{}^P\partial_P+\dots$ acting on the background.
We take:
\be
\bar\delta_\xi \gV_\alpha^M = \xi^\mu Y^{MN}{}_{PQ} \left(
\partial_N A_\mu{}^P D_\alpha Y^Q 
+ \partial_N A_{[\nu}{}^P A_{\mu]}{}^Q \partial_\alpha X^\nu
+ \partial_N B_{\mu\nu}{}^{PQ} \partial_\alpha X^\nu 
\right)
+  \delta'_\xi \gV_\alpha^M \,,
\ee
where $ \delta'_\xi \gV_\alpha^M$ indicates possible further terms which depend on the worldsheet metric and Levi-Civita symbol. We set these to zero for now and will reconsider them at the very end.

Then one has
\be
\begin{split} 
\bar \delta_\xi \partial_\alpha X^\mu & = \partial_\alpha X^\rho D_\rho \xi^\mu + D_\alpha Y^M \partial_M \xi^\mu \\
\bar \delta_\xi D_\alpha Y^M & = \delta_\xi A_\mu{}^M \partial_\alpha X^\mu - \gM^{MN} g_{\mu\nu} \partial_N \xi^\mu \partial_\alpha X^\nu  \\ & \qquad\qquad\qquad\qquad
- \xi^\mu \partial_N A_\mu{}^M D_\alpha Y^N + \xi^\mu Y^{MN}{}_{PQ} \partial_N A_\mu{}^P D_\alpha Y^Q 
\end{split}
\label{bardeltaxi}
\ee
where 
\be
\delta_\xi A_\mu{}^M = \xi^\nu \Fa_{\nu \mu}{}^M + \gM^{MN} g_{\mu\nu} \partial_N \xi^\nu 
\ee
is the transformation of $A_\mu{}^M$ under external diffeomorphisms.
Now, the transformations of the metrics are:
\be
\begin{split} 
\delta_\xi g_{\mu\nu}  = \xi^\rho D_\rho g_{\mu\nu} + 2 D_{(\mu} \xi^\rho g_{\nu) \rho} \,,\qquad
\delta_\xi \gM_{MN}  = \xi^\rho D_\rho \gM_{MN}\,.
\end{split} 
\ee
It is straightforward to calculate:
\be 
\bar \delta_\xi \left( g_{\mu\nu} \partial_\alpha X^\mu \partial_\beta X^\nu \right) 
 = (\delta_\xi g_{\mu \nu}) \partial_\alpha X^\mu \partial_\beta X^\nu 
 + 2 g_{\mu\nu} \partial_M \xi^\mu D_{(\alpha} Y^M \partial_{\beta)} X^\nu 
 - 2 \omega \xi^\rho \partial_K A_\rho{}^K g_{\mu\nu} \partial_\alpha X^\mu \partial_\beta X^\nu  \,.
\ee
The weight term can be dealt with as before using $T$, as we have
\be
\bar \delta_\xi T = \delta_\xi T + 2 \omega T \xi^\rho \partial_K A_\rho{}^K \,,
\ee
where $\delta_\xi T = \xi^\rho D_\rho T$. From \eqref{bardeltaxi}, we see that the extra term involving the derivative $\partial_M \xi^\mu$ can be cancelled against the second term in the transformation of $D_\alpha Y^M$, if we combine $T g_{\mu\nu} \partial_\alpha X^\mu \partial_\beta X^\nu$ and $T \gM_{MN} D_\alpha Y^M D_\beta Y^N$. Indeed, we find 
\begin{align}
\nonumber
\bar\delta_\xi ( T g_{\mu\nu} \partial_\alpha X^\mu \partial_\beta X^\nu + T \gM_{MN} D_\alpha Y^M D_\beta Y^N ) 
& = 
( \delta_\xi  T g_{\mu\nu} ) \partial_\alpha X^\mu \partial_\beta X^\nu 
+ ( \delta_\xi T \gM_{MN}) D_\alpha Y^M D_\beta Y^N 
\\ 
& + 2 T \gM_{MN} \delta_\xi A_\mu{}^M \partial_{(\alpha} X^\mu D_{\beta)} Y^N 
\end{align}
which is the required behaviour. Note that this establishes that it is the pullback of the whole ``generalised line element'' $g_{\mu\nu} dX^\mu d X^\nu + \gM_{MN} DY^M D Y^N$ to the worldvolume that respects the external diffeomorphism symmetry (up to weight terms which can be cancelled via something like $T$ as done here or against a worldvolume metric as in the EFT particle case \cite{Blair:2017gwn}).

However, for the string, this is not the end of the story.

\subsubsection*{Wess-Zumino terms}

The desired variation of the Wess-Zumino term is 
\be
\bar \delta_\xi \mathcal{L}_{WZ} \stackrel{!}{=}
\epsilon^{\alpha \beta}  q_{MN} \left( \Delta_\xi B_{\mu\nu}{}^{MN} \partial_\alpha X^\mu \partial_\beta X^\nu + \partial_\alpha X^\mu \delta_\xi A_\mu{}^M  D_\beta Y^N \right) \,,
\label{genvarWZ}
\ee
with
\be
\begin{split} 
\delta_\xi A_\mu{}^M  = \xi^\rho \Fa_{\rho \mu}{}^M + \gM^{MN} g_{\mu\nu} \partial_N \xi^\nu \,, \qquad
\Delta_\xi B_{\mu\nu}{}^{MN}  = \xi^\rho \Fb_{\mu \nu \rho}{}^{MN} \,.
\end{split}
\ee
A subtlety here is that for $\Gsix$ we assume that the spacetime background is on-shell, so that we can use the duality relation \eqref{intduality} to write $\Fb_{\mu\nu\rho}$ here.
Now, we have (up to $D \leq 5$, with modifications for $D=6$ as explained in the previous section):
\be
\begin{split}
\Fa_{\mu\nu}{}^M & = 2 \partial_{[\mu} A_{\nu]}{}^M - 2 A_{[\mu}{}^N \partial_N A_{\nu]}{}^M - Y^{MN}{}_{PQ} \partial_N A_{[\mu}{}^P A_{\nu]}{}^Q + Y^{MN}{}_{PQ} \partial_N B_{\mu \nu}{}^{PQ} \\
\Fb_{\mu\nu\rho}{}^{MN} & = 3 \partial_{[\mu} B_{\nu\rho]}{}^{MN} - 3 \mathcal{L}_{A_{[\mu}} B_{\nu \rho]}{}^{MN} 
- 3 \frac{1}{2d} Y^{MN}{}_{PQ} \partial_{[\mu} A_\nu{}^P A_{\rho]}{}^Q  + ( \hat \partial C_{\mu\nu\rho} )^{MN} 
\\ & \qquad + 2 \frac{1}{2d} Y^{MN}{}_{PQ} A_{[\mu}{}^P A_\nu{}^K \partial_K A_{\rho]}{}^Q 
+  \frac{1}{2d} Y^{MN}{}_{PQ}Y^{QR}{}_{KL} A_{[\mu}{}^P \partial_R A_{\nu}{}^K A_{\rho]}{}^L 
\end{split} 
\ee
Then varying $A$ and $B$ we should find:
\be
\begin{split}
\bar \delta \mathcal{L}_{WZ} & \stackrel{!}{=} q_{MN} \epsilon^{\alpha \beta} 
\Big(
\xi^\rho \partial_\alpha X^\mu \partial_\beta X^\nu 
\left[
 3 \partial_{[\mu} B_{\nu\rho]}{}^{MN} - 3 \mathcal{L}_{A_{[\mu}} B_{\nu \rho]}{}^{MN} 
\right] 
\\ & \qquad\qquad\quad
+ \,\xi^\mu \partial_\alpha X^\nu D_\beta Y^N Y^{MR}{}_{KL} \partial_R B_{\mu\nu}{}^{KL} 
\\ & \qquad\qquad\quad
+ \xi^\rho \partial_\alpha X^\mu \partial_\beta X^\nu  (\hat \partial C_{\mu\nu\rho})^{MN}
\\ & \qquad\qquad\quad
+ \partial_\alpha X^\nu D_\beta Y^N \gM^{MK}  g_{\mu\nu} \partial_K \xi^\mu
\\ & \qquad\qquad\quad
+ \xi^\rho \partial_\alpha X^\mu \partial_\beta X^\nu  
\Big(
-3  \partial_{[\mu} A_\nu{}^M A_{\rho]}{}^N + 2 A_{[\mu}{}^M A_\nu{}^K \partial_K A_{\rho]}{}^N  
\\ &  \qquad\qquad\quad \qquad\qquad\quad \qquad\qquad\quad \qquad\qquad\quad
+ Y^{NR}{}_{KL} A_{[\mu}{}^M \partial_R A_{\nu}{}^K A_{\rho]}{}^L 
\Big)
\\ & \qquad\qquad\quad
- \xi^\mu \partial_\alpha X^\nu D_\beta Y^N
\left(
2 \partial_{[\mu} A_{\nu]}{}^M - 2 A_{[\mu}{}^K \partial_K A_{\nu]}{}^M - Y^{MK}{}_{RS} \partial_K A_{[\mu}{}^R A_{\nu]}{}^S
\right) 
\Big) 
\end{split} 
\label{target}
\ee
Let us consider this. The third line will never be generated by $\bar \delta_\xi$ acting on any of the coordinates or fields. 
It vanishes for $D \leq 5$ by \eqref{qhatC}; for $D=6$ the situation is similar to the discussion there, as instead of just $q_{MN} (\hat \partial C_{\mu\nu\rho})^{MN} = 0$ we require
 \be
 \label{y}
 q^N \left(
 6 t_\alpha{}^P{}_N \partial_P C_{\mu\nu\rho}{}^\alpha
 + \tilde C_{\mu\nu\rho N}\right) =0\,,
 \ee
which is the same calculation as needed for the gauge invariance requirement \eqref{gaugemalarkey} 

The fourth line involves the external metric $g_{\mu\nu}$. It will also never be generated by $\bar \delta_\xi$. One finds that this is the only problematic term: all the rest can be obtained from $\bar\delta_\xi$ up to total derivatives. Thus, 
\be
\begin{split} 
\bar\delta_\xi \mathcal{L}_{WZ} & = \epsilon^{\alpha \beta}  q_{MN} \left( \Delta_\xi B_{\mu\nu}{}^{MN} \partial_\alpha X^\mu \partial_\beta X^\nu + \partial_\alpha X^\mu \delta_\xi A_\mu{}^M  D_\beta Y^N \right) 
\\ & \qquad 
- \epsilon^{\alpha \beta}  q_{MN} \partial_\alpha X^\nu D_\beta Y^N \gM^{MK}  g_{\mu\nu} \partial_K \xi^\mu \,.
\end{split}
\ee
The way to cancel this term is to combine it with contributions from the kinetic terms, alongside the extra transformation $\delta'_\xi V_\alpha^M$ we have heretofore neglected.

\subsubsection*{Combining kinetic and Wess-Zumino}

Define $\hat \Delta_\xi$ to be the anomalous variation given by the difference between $\bar\delta_\xi$ and the expected variation. 
We trial the full Lagrangian
\be
\mathcal{L} = 
x T \sqrt{-\gamma} \gamma^{\alpha \beta} g_{\mu\nu} \partial_\alpha X^\mu \partial_\beta X^\nu
+ y T \sqrt{-\gamma} \gamma^{\alpha \beta}  \gM_{MN} D_\alpha Y^M D_\beta Y^N
+ \mathcal{L}_{WZ} \,,
\ee
where $x,y$ are numerical constants to be determined.
The total anomalous variation of this is $\hat \Delta_\xi \mathcal{L} \equiv \bar\delta_\xi \mathcal{L}(X,Y,\gV;g,\gM,A,B) - \mathcal{L}(X,Y,\gV; \delta_\xi g, \delta_\xi \gM, \delta_\xi A, \delta_\xi B)$,
where we have individually that 
\be
\begin{split}
\hat \Delta_\xi ( T \sqrt{-\gamma} \gamma^{\alpha \beta} g_{\mu\nu} \partial_\alpha X^\mu \partial_\beta X^\nu)  & = 
2 T \sqrt{-\gamma} \gamma^{\alpha \beta} g_{\mu\nu} \partial_M \xi^\mu D_{\alpha} Y^M \partial_{\beta} X^\nu 
\,\\
\hat \Delta_\xi( T \sqrt{-\gamma} \gamma^{\alpha \beta}  \gM_{MN} D_\alpha Y^M D_\beta Y^N ) & = 
-2 T \sqrt{-\gamma} \gamma^{\alpha \beta} g_{\mu\nu} \partial_M \xi^\mu D_{\alpha} Y^M \partial_{\beta} X^\nu 
 \,\\
\hat \Delta_\xi \mathcal{L}_{WZ} & = - \epsilon^{\alpha \beta}  q_{MN} \partial_\alpha X^\nu D_\beta Y^N \gM^{MK}  g_{\mu\nu} \partial_K \xi^\mu \,.
\end{split} 
\label{anomvars}
\ee
Hence, 
\be
\begin{split}
\hat \Delta_\xi \mathcal{L} 
& =  2 (x-y) T \sqrt{-\gamma} \gamma^{\alpha \beta} g_{\mu\nu} \partial_M \xi^\mu D_{\alpha} Y^M \partial_{\beta} X^\nu
+ \epsilon^{\alpha \beta}  q_{MN}  \gM^{NK}g_{\mu\nu}  \partial_K \xi^\mu    D_\alpha Y^M  \partial_\beta X^\nu
\\ & 
\qquad + \delta'_\xi V_\alpha{}^M \left( 
2 y  T \sqrt{-\gamma} \gamma^{\alpha \beta}  \gM_{MN} D_\beta Y^N - \epsilon^{\alpha \beta} q_{MN} ( D_\beta Y^N - V_\beta^N) \right) \,.
\end{split} 
\ee
To proceed, we need the following identity:
\be
T^2 \partial_M = q_{MN} \gM^{NP} q_{PQ} \gM^{QK} \partial_K \,.
\label{T2magic}
\ee
To prove this, use the constraint \eqref{magic} to write
\be
\begin{split}
T^2 \partial_M & = \frac{1}{2d} \gM^{PQ} \gM^{KL} q_{PK} q_{QL} \partial_M 
\\ & = \frac{1}{2d} \gM^{PQ} \gM^{KL} Y^{RS}{}_{QL} q_{PK} q_{MR} \partial_S 
\\ & = \frac{1}{2d} \gM^{RL} \gM^{SN} Y^{PK}{}_{LN} q_{PK} q_{MR} \partial_S 
\\ & = \gM^{RL} \gM^{SN} q_{LN} q_{MR} \partial_S \,.
\end{split}
\ee
In going from the second to the third line, we used the fact that the generalised metric is a group element and so preserves the Y-tensor. Then in going to the final line we used the fact that $q_{MN} \in R_2$.
This (rather unexpected) identity allows us to rewrite the anomalous variation as:
\begin{align}
\hat \Delta_\xi \mathcal{L} 
& =  \frac{\gamma_{\alpha \gamma} \epsilon^{\gamma \delta}}{ T \sqrt{-\gamma}} 
\gM^{MP} q_{PQ} \gM^{QK} \partial_K \xi^\mu g_{\mu \nu} \partial_\delta X^\nu
\Big(
 T \sqrt{-\gamma} \gamma^{\alpha \beta} \gM_{MN} D_\beta Y^N 
- 2 (x-y) \epsilon^{\alpha \beta} q_{MN} ( D_\beta Y^N - V_\beta^N )
\Big)
\nonumber
\\ & 
\qquad + \delta'_\xi V_\alpha{}^M \left( 
2 y  T \sqrt{-\gamma} \gamma^{\alpha \beta}  \gM_{MN} D_\beta Y^N - \epsilon^{\alpha \beta} q_{MN} ( D_\beta Y^N - V_\beta^N) \right) \,.
\label{tidyanomvar}
\end{align} 
Say we take
\be
\delta'_\xi \gV_\alpha^M =  -\frac{z}{T \sqrt{-\gamma}} \gamma_{\alpha \beta} \epsilon^{\beta \gamma} \gM^{MP} q_{PQ} \gM^{QK} \partial_K \xi^\mu g_{\mu\nu} \partial_\gamma X^\nu 
\label{tildedeltaV}
\ee
for some constant $z$. To preserve $\gV_\alpha^M \partial_M = 0$ we need:
\be
\label{magic3}
\gM^{MP} q_{PQ} \gM^{QK} \partial_M \otimes \partial_K = 0 \,,
\ee
To show this, we note that in general one can write 
\be
q_{MN} = \eta_{MN \mathfrak{M} } q^{\mathfrak{M}}
\ee
where $\mathfrak{M}$ is an $\bar R_2$ index, and $\eta_{MN \mathfrak{M}}$ is an invariant tensor(an ``eta-symbol'' in the language of \cite{Sakatani:2017xcn}), which is basically just the projector of a pair of symmetric $R_1$ indices into $R_2$. Invariance of this tensor implies that
\be
\gM^{MP} \gM^{NQ} \eta_{PQ \mathfrak{M}} = \eta^{MN \mathfrak{N} } \gM_{ \mathfrak{M} \mathfrak{N} }
\ee
where $\gM_{\mathfrak{M} \mathfrak{N}}$ is the generalised metric in the $R_2$ representation. The section condition is expressible as $\eta^{MN \mathfrak{M} } \partial_M \otimes \partial_N = 0$. 
As a result,
\be
\gM^{MP} q_{PQ} \gM^{QK} \partial_M \otimes \partial_K 
= \eta^{MN \mathfrak{M} } \gM_{\mathfrak{M} \mathfrak{N} } q^{\mathfrak{N}} \partial_M \otimes \partial_N = 0 
\ee
by the section condition. For instance, for $\Gsix$, we have $q_{MN} = d_{MNP} q^P$, as $R_2 = \bar R_1$, and the section condition is given in terms of the other cubic invariant as $d^{MNP} \partial_M \otimes \partial_P = 0$. The generalised metric obeys $\gM^{MP} \gM^{NQ} \gM^{KL} d_{MNP} = d^{PQL}$ and so $\gM^{MP} q_{PQ} \gM^{QK} = d^{MKL} \gM_{LN} q^N$, confirming the general result in this case.

The transformation \eqref{tildedeltaV} cancels the anomalous variation if $2yz = 1$ and $2(x-y) = z$. 
This implies $\frac{1}{2y} = 2(x-y)$, which also means that one can view the anomalous variation (first line of \ref{tidyanomvar}) as being cancelled on-shell by the equation of motion of $\gV_\alpha^M$ (which can be read off the second line of \ref{tidyanomvar}). Technically, it is the only the equations of motion of the non-zero components of $\gV_\alpha^M$ which can be used, but as both $\gV_\alpha^M$ and $\gM^{MP}q_{PQ}\gM^{QK} \partial_K \xi^\mu$ are zero when contracted with $\partial_M$, this is consistent. 

We can not directly fix the coefficient $z$ using the symmetry requirements considered above.
However, a more in-depth study of the constraint imposed by $\gV_\alpha^M$ allows us to set $z=1$, as we now explain.

\subsection{Fixing the action: twisted self-duality} 

Naively, the equation of motion of $\gV_\alpha^M$ suggests that we are using it to impose the following ``twisted self-duality'' constraint:
\be
\label{covariantisedVEOM}
D_\alpha Y^M = \frac{z\gamma_{\alpha\beta} \epsilon^{\beta \gamma} }{T\sqrt{-\gamma}} \gM^{MN} q_{NP} D_\gamma Y^P \,,
\ee
where now $D_\alpha Y^M \equiv \partial_\alpha Y^M + \partial_\alpha X^\mu A_\mu{}^M$, 
which implies that
\be
D_\alpha Y^M = \frac{z^2}{T^2} \gM^{MN} q_{NP} \gM^{PQ} q_{QK} D_\alpha Y^K \,.
\ee
Contracting with $\partial_M$ and using \eqref{T2magic} we find that we must have $z^2 = 1$ otherwise $D_\alpha Y^M \partial_M = 0$.

Actually, $\gV_\alpha^M \partial_M = 0$, so that not all components of $\gV_\alpha^M$ and its equation of motion are present. 
To analyse the equation of motion of $\gV_\alpha^M$, 
we need to first analyse  the constraint \eqref{magic}. Splitting $M=(i,A)$ and solving the section condition as
$\partial_i \neq 0$, $\partial_A = 0$, we can write \eqref{magic} as
\be
q_{M B} Y^{B k}{}_{PQ} \partial_k = q_{PQ} \partial_M 
\ee
where we know that $Y^{ij}{}_{PQ} = 0$ as otherwise $\partial_i \neq 0$ would not solve the section condition. This also means $Y^{PQ}{}_{ij} = 0$.
Letting $M=A$ we find
\be
q_{AB} Y^{Bk}{}_{PQ} \partial_k = 0 
\Rightarrow
q_{AB} = 0\,.
\ee 
We also have
\be
q_{iC} Y^{Cj}{}_{AB} \partial_j = 0 \quad,\quad
q_{iC} Y^{Cj}{}_{Ak} \partial_j = q_{kA} \partial_i 
\quad,\quad
q_{iC} Y^{Cj}{}_{kl} \partial_j = q_{kl} \partial_i \,.
\ee
As in general $Y^{A i}{}_{jk} = 0$, the last of these implies $q_{ij} = 0$.
We conclude the only non-zero components of $q_{MN}$ are $q_{iA} = q_{Ai}$. This can also be obtained on a case-by-case basis for each $D$ and follows from the explicit formulas for $q_{MN}$ in the appendix.

Then, we can obtain from the $T^2$ identity \eqref{T2magic} that
\be
q_{A i} q_{B j} \left( \gM^{i B} \gM^{j k} + \gM^{ij} \gM^{B k} \right) = 0 \quad,\quad
q_{Ai} q_{B j} \left( \gM^{AB} \gM^{jk} + \gM^{Aj} \gM^{Ak} \right) = T^2 \delta_i^k \,,
\label{T2magic2}
\ee
while from \eqref{magic3} we learn that
\be
q_{Ak} \gM^{k(i} \gM^{j)A} = 0  \,.
\label{magic5}
\ee
Now, the general generalised metric takes a sort of Kaluza-Klein-esque form given the splitting $M=(i,A)$, and as studied in appendix \ref{gmfacts} can be parametrised as:
\be
\gM_{MN} = \begin{pmatrix}
\bar \gM_{ij} + \gM_{CD} U_i{}^C U_j^D & \gM_{BC} U_i{}^C \\
\gM_{AC} U_j{}^C & \gM_{AB} 
\end{pmatrix} 
\label{param1}
\ee
where
\be
U_i{}^A \equiv ( \gM_{AB})^{-1} \gM_{iB} \quad,\quad
\bar{\gM}_{ij} \equiv\gM_{ij}- \gM_{CD} U_i{}^C U_j{}^D\,.
\ee
The inverse, assuming that $\gM_{ij}$ and $\gM_{AB}$ are invertible, is
\be
\gM^{MN} = \begin{pmatrix} 
\bar{\gM}^{ij} & - \bar{\gM}^{ik} U_k{}^B \\
- \bar{\gM}^{jk} U_k{}^A & ( \gM_{AB} )^{-1} + \bar{\gM}^{kl} U_k{}^A U_l{}^B
\end{pmatrix} \,,
\label{param2}
\ee
The equation of motion for the non-zero components $\gV_\alpha^A$ is:
\be
\gM_{A M} DY^M = \frac{1}{zT} q_{Ai} \star DY^i \,,
\ee
using $\star$ the worldsheet Hodge star for simplicity ($\star^2=1$ for Lorentzian worldsheets). This implies
\be
DY^A = ( \gM_{AB} )^{-1} \left( \frac{1}{zT} q_{B j} \star DY^j - \gM_{Bj } DY^j \right)\,.
\label{DY}
\ee
In order to preserve formal $\ED$ covariance, we want this equation of motion to imply the remaining components of the constraint \eqref{covariantisedVEOM}.
This means we need
\be
\gM_{ij} DY^j = \frac{1}{zT} q_{iA}  \star DY^A - \gM_{iA} DY^A \,.
\ee
Substituting in \eqref{DY} and then making use of the identities \eqref{T2magic2} and \eqref{magic5} alongside the parametrisations \eqref{param1} and \eqref{param2} we find that this, and hence \eqref{covariantisedVEOM}, holds, provided that $z^2 =1$.

We can now without loss of generality take $z=1$ as changing the sign of $z$ amounts to changing the sign of the Wess-Zumino term, which is equivalent to $q\to-q$.

Therefore requiring that all components of the twisted self-duality relation \eqref{covariantisedVEOM} follow from the equation of motion of $V$ fixes the action completely.

\section{Reductions} 
\label{chargered}

\subsection{Reduction to 10-dimensions}
\label{reductionsection}

Consider the action \eqref{actionnice}.
Split $Y^M = (Y^i , Y^A)$ such that $\partial_i \neq 0$ and $\partial_A = 0$ defines a solution to the section condition.
Let $S = S_0 + S_V$ where
\be
\begin{split}
 S_0 & = - \frac{1}{2} \int d^2\sigma \, \Big( T \sqrt{-\gamma} \gamma^{\alpha \beta}  \left( g_{\alpha \beta} + \frac{1}{2} \gM_{ij} D_\alpha Y^i D_\beta Y^j \right) 
  \\ & \qquad\qquad \qquad
  + \epsilon^{\alpha \beta}  \left( q_{MN} \left( B_{\alpha \beta}{}^{MN}  + A_\alpha{}^M \partial_\beta Y^N \right)  + q_{iM} D_\alpha Y^M D_\beta Y^i \right) \Big) \,.  
\end{split} 
\ee
and
\be
\begin{split}
 S_{\gV}  = - \frac{1}{2} \int d^2\sigma \, \Big( T \sqrt{-\gamma} \gamma^{\alpha \beta}  \left( \frac{1}{2} \gM_{AB} D_\alpha Y^A D_\beta Y^B  + \gM_{Ai} D_\alpha Y^A D_\beta Y^i\right) 
  - \epsilon^{\alpha \beta} q_{AM} D_\alpha Y^A \bar D_\beta Y^M  \Big) \,.  
\end{split} 
\ee
Here 
\be
\bar D_\alpha Y^M = \partial_\alpha Y^M + A_\alpha{}^M \,,
\ee
and we have shortened $g_{\alpha \beta} \equiv g_{\mu\nu} \partial_\alpha X^\mu \partial_\beta X^\nu$, and similarly for $B_{\alpha \beta}{}^{MN}$ and $A_\alpha{}^M$.

Now, the non-zero components $\gV^A$ appear only in $S_V$. To integrate them out, we can complete the square, resulting in
\be
\begin{split}
S_V & \rightarrow - \frac{1}{2} \int d^2\sigma \Big( T \sqrt{-\gamma} \gamma^{\alpha \beta} \big( - \frac{1}{2} \gM_{iA} ( \gM_{AB} )^{-1} \gM_{Bj} D_\alpha Y^i D_\beta Y^j 
\\ & \qquad\qquad\qquad\qquad\qquad\qquad
+ \frac{1}{2T^2} (\gM_{AB})^{-1} q_{AM} q_{BN} \bar D_\alpha Y^M \bar D_\beta Y^N \big) 
\\ & \qquad\qquad\qquad
+ \epsilon^{\alpha \beta}  (\gM_{AB})^{-1} \gM_{Bi} q_{AM} D_\alpha Y^i \bar D_\beta Y^N \Big) \,.
\end{split} 
\ee
Here we assume that we can invert the components $\gM_{AB}$ of the generalised matrix carrying dual indices only. 
Next, we substitute in the result (see the appendix \ref{gmfacts})
\be
\gM_{ij} -  \gM_{iA} ( \gM_{AB} )^{-1} \gM_{Bj} = \Omega^{-1} \phi_{ij} \,,
\ee
where the conformal factor $\Omega = \phi^{\omega}$ for M-theory and IIB sections starting from Einstein frame, and $\Omega = \phi^{\omega} e^{-4\omega \Phi}$ for IIA sections starting from string frame.
Then,
using that $q_{iA} = q_{Ai}$ are the only non-zero components of the charge, one has
\be
\begin{split} 
 S & = - \frac{1}{2} \int d^2\sigma \, \bigg( T \sqrt{-\gamma} \gamma^{\alpha \beta}  \big( g_{\alpha \beta}
+ \frac{1}{2} \left[
 \Omega^{-1} \phi_{ij} + \frac{1}{T^2} (\gM_{AB})^{-1} q_{Ai} q_{Bj} 
\right] D_\alpha Y^i D_\beta Y^j
\big)
  \\ & \qquad\qquad \qquad
+ q_{Ai}  \epsilon^{\alpha \beta} \Big( 
2 B_{\alpha \beta}{}^{Ai} 
+   2 A_\alpha{}^A \partial_\beta Y^i + A_\alpha{}^A A_\beta{}^i
\\ & 
\qquad\qquad\qquad\qquad\qquad\qquad 
 + ( \gM_{AB} )^{-1} \gM_{Bj} D_\alpha Y^j D_\beta Y^i 
+ \partial_\alpha Y^A \partial_\beta Y^j 
\Big)
\bigg) \,.
\end{split} 
\label{reductionresult} 
\ee
This can be further simplified in general terms: we note that from \eqref{param1} as well as the combination of \eqref{param2} and \eqref{magic5} that
\be
q_{iA} (\gM_{AB})^{-1} \gM_{Bj} = q_{iA} U_j{}^A = q_{[i| A|} U_{j]}{}^A \,.
\ee
We also have from \eqref{T2magic2} and \eqref{param2} that
\be
\frac{1}{T^2} (\gM_{AB})^{-1} q_{Ai} q_{Bj}  = \Omega^{-1} \phi_{ij} \,.
\ee
Hence, with $X^{\hmu} = ( X^\mu , Y^i)$ and $\hat g_{\hmu \hnu}$ the 10-dimensional metric decomposed as in \eqref{metricdecomp} we have found
\be
\begin{split} 
 S & = - \frac{1}{2} \int d^2\sigma \, \bigg( \Omega^{-1} T \sqrt{-\gamma} \gamma^{\alpha \beta} 
 \hat g_{\hmu \hnu} \partial_\alpha X^{\hmu} \partial_\beta X^{\hnu} 
  \\ & \qquad\qquad \qquad
+ q_{Ai}  \epsilon^{\alpha \beta} \Big( 
2 B_{\alpha \beta}{}^{Ai} 
+   2 A_\alpha{}^A \partial_\beta Y^i + A_\alpha{}^A A_\beta{}^i
 +  U_j{}^A D_\alpha Y^j D_\beta Y^i 
+ \partial_\alpha Y^A \partial_\beta Y^j 
\Big)
\bigg) \,.
\end{split} 
\label{reductionresult2} 
\ee
Note that the only place the dual coordinates appear is
\be
S \supset - \frac{1}{2} \int d^2 \sigma \epsilon^{\alpha \beta} q_{A i} \partial_\alpha Y^A \partial_\beta Y^i \,.
\ee
This is a total derivative. Recall that a similar term appears in the reduction of the doubled sigma model, and is cancelled there by adding to the action a topological term \eqref{topterm} based on an antisymmetric tensor $\Omega_{MN}$.
Here we could define such an object to have non-vanishing components $\Omega_{Ai} = - \Omega_{iA} = q_{iA}$.
This is a bit different to the $O(d,d)$ case, as such an $\Omega_{MN}$ depends on the charge $q$, and so would be different in IIA and IIB sections.
It would be interesting to explore the uses and consequences of such an exceptional topological term and symplectic form, either on the worldsheet or more speculatively in spacetime \cite{Freidel:2015pka,Freidel:2017yuv}.

To complete the reduction, we must show that for explicit choices of $q$ and parametrisations of the EFT fields that we obtain known 1-brane actions.
This can be done group by group.
In section \ref{exE6}, we will focus on the example of $\Gsix$ in detail.

\subsection{Reduction to the doubled sigma model} 
\label{redtodouble}

An alternative reduction one can do is to reduce from our exceptional sigma model to the doubled sigma model itself. 
The Kaluza-Klein reduction of exceptional field theory to double field theory has been examined in the internal sector in \cite{Thompson:2011uw}. 
Let us write the EFT generalised metric as $\gM_{\hat M \hat N}$ and split the index $\hat M = (M,A)$ where $M$ is an $O(d,d)$ doubled index (here again $d=D-1$ after starting with $\ED$). 
We write
\be
\gM_{\hat M \hat N} = 
\begin{pmatrix} 
e^{4 \omega \Phi_d} \mathcal{H}_{MN} + \phi_{AB} A_M{}^A A_N{}^B & \phi_{BC} A_M{}^C \\
\phi_{AC} A_N{}^C & \phi_{AB} 
\end{pmatrix} 
\ee
where $\mathcal{H}_{MN}$ is the usual DFT generalised metric, $\Phi_d$ is the doubled dilaton, $A_M{}^A$ is generically a vector-spinor containing the RR fields, and $\det \phi_{AB} = e^{-8 \omega d \Phi_d}$ (in requiring $\det \gM_{\hat M \hat N} = 1$ we implicitly exclude the $\G$ EFT from our general analysis -- this case can be treated separately). We assume that $\partial_A = 0$. 
The Y-tensor components that are non-zero are $Y^{MN}{}_{PQ}$, $Y^{MA}{}_{NB}$ and $Y^{AB}{}_{CD}$ (generally built from $\eta_{MN}$ and the gamma matrices $\gamma_M{}^{AB}$, $\gamma^M{}_{AB}$) and the constraint \eqref{magic} then implies that $q_{AB} = q_{AM} = 0$. The only non-zero components of the charge are $q_{MN} = T_{F1} \eta_{MN}$.
Then $T(\gM,q) = T_{F1} e^{-4\omega \Phi_d}$. 

We now seek to integrate out the components $V_\alpha^A$. The Wess-Zumino term only involves the coordinates $Y^M$, so we only need to look at the kinetic term, which contains
\be
\begin{split}
S & \supset - \frac{T_{F1}}{4} \int d^2 \sigma \sqrt{-\gamma} \gamma^{\alpha \beta}
\Big(
 \mathcal{H}_{MN} D_\alpha Y^M D_\beta Y^N 
\\ & \qquad\qquad\qquad\qquad\qquad\qquad + e^{-4\omega \Phi_d} \phi_{AB} 
(D_\alpha Y^A + A_M{}^A D_\alpha Y^M )(
D_\beta Y^B + A_N{}^B D_\beta Y^N )
\Big)
\end{split}
\ee
from which one trivially uses the $V_\alpha{}^A$ equation of motion to eliminate the second line entire, leaving one with the standard double sigma model, identifying the EFT components $B_{\mu\nu}{}^{MN}$ and $A_\mu{}^M$ with their DFT counterparts. (Note that choosing $\partial_A = 0$ and then integrating out $\gV_\alpha^A$ means this reduction can only give the F1, and not the D1.)

\section{Example: the $\Gsix$ exceptional sigma model}
\label{exE6} 

\subsection{The charge constraint}

Here we have $Y^M$ in the $\mathbf{27}$ of $\Gsix$. The charge in $\bar R_2 = \mathbf{27}$ is $q^M$. The constraint \eqref{magic} is:
\be
10 d_{MNP} d^{PQR} q^N \partial_R = q^Q \partial_M \,.
\ee
Contracting the free indices implies that $q^P \partial_P = 0$. 

We again check the unsurprising solutions:
\begin{itemize}
\item IIB section: here we decompose $M = ( i, {}_{i\alpha}, {ij}, {}_\alpha)$, where $i=1,\dots,5$ and $\alpha$ the usual $\mathrm{SL}(2)$ index. We have $\partial_i \neq 0$ and $\partial^{i\alpha} = \partial_{ij} = \partial^\alpha = 0$. One finds
\be
q^k \partial_i = 0
\quad,\quad
q_{[i|\alpha} \partial_{|j]} = 0 
\quad,\quad
\delta_i^{[k} q^{l] j} \partial_j = 0
\quad,\quad
\epsilon^{ijklm} q_{l \alpha} \partial_m = 0
\quad,\quad
q^{ij} \partial_j = 0 
\ee 
and the only solution is again $q^\alpha \neq 0$ and the others zero. The non-zero components of the charge $q_{MN}$ are:
\be
q_{i}{}^{j a} = \frac{1}{\sqrt{10}} \delta^j_i \epsilon^{ab} q_b \,.
\label{IIBq} 
\ee

\item M-theory section: let $M=(i , {}_{ij} , \bar i)$ with $i, \bar i = 1,\dots,6$ and $ij$ antisymmetric. The section choice is $\partial_i \neq 0$, $\partial^{ij} = 0 = \partial_{\bar i}$. We see immediately that $q^i = 0$. One finds the constraints:
\be
q^{\bar i} \partial_i = 0 \quad,\quad \epsilon^{ijklmn}q_{lm} \partial_n=0
\quad,\quad
q^{[k} \delta^{l]}_{[i} \partial_{j]} = 0\,.
\ee
There are no solutions.

\item IIA section: we now take one of the M-theory directions $i=1$ (say) to be an isometry, $\partial_1 = 0$. This allows for $q^{\bar 1} \neq 0$, giving the F1 string, with all other charge components zero. 
The non-zero components of the charge $q_{MN}$ are:
\be
q_{i}{}^{\mk \ml} = \frac{q^{\bar 1}}{\sqrt{5}} \delta^{[\mk}_i \delta^{\ml]}_1 
\Rightarrow q_i{}^{j 1} = \frac{q^{\bar 1}}{2 \sqrt{5}} \delta_i^j\,.
\label{IIAq}
\ee
where $\mk = ( k , 1)$ includes the 5 IIA directions labelled by $i$ and the M-theory direction labelled by $1$.

\end{itemize}
The tension can be written as
$T = \frac{1}{\sqrt{10}} \sqrt{ q^M q^N \gM_{MN} }$.

\subsection{Reduction to IIA F1}
\label{IIAred}

In appendix \ref{E6Mdict}, the dictionary relating the EFT fields to those of 11-dimensional supergravity is given.
It is convenient to continue to use the 11-dimensional variables for a time. 
Let $\hat i = ( i , 1)$ denote the 5-dimensional internal IIA index $i$ along with the single index ``1'' corresponding to the M-theory direction.
We write the 27-dimensional $R_1$ index as $M = (i,A)$ where now $Y^A = ( Y^{\bar{\hat i}} , Y_{\hat i \hat j} , Y^1)$.
The IIA string corresponds to $q^{\bar 1} \neq 0$. We let $q^{\bar 1} =q $.
The symmetric charge $q_{MN} = d_{MNP} q^P$ has non-zero components as in \eqref{IIAq}.
Using appendix \ref{E6Mdict}, we can extract the dual components of the generalised metric, finding
\be
\gM_{AB} = U_A{}^C U_B{}^D \bar \gM_{CD} 
\ee
with
\be
U_A{}^C = \begin{pmatrix}
\delta_{\mi}^{\mm} & 0 & 0 \\
+ \frac{1}{\sqrt{2}} \tilde A^{\mi \mj \mm} & \delta^{[\mi \mj]}_{\mm \mn} & 0 \\
- \delta_1^{\mm} \varphi + \frac{1}{4} \tilde A^{\mm\mp\mq} A_{1\mp\mq} & + \frac{1}{\sqrt{2}} A_{1\mm\mn} & 1 
\end{pmatrix} 
\,,\,
\bar \gM_{CD} =
\begin{pmatrix}
\hat\phi^{-2/3} \hat\phi_{\mm\mp} & 0 & 0 \\
0 & \hat\phi^{1/3} \hat\phi^{\mp[\mm} \phi^{\mn]\mq} & 0 \\
0 & 0 & \hat\phi^{1/3} \hat\phi_{11}
\end{pmatrix} \,.
\label{UA}
\ee
Here $\hat \phi_{\mi\mj}$ denote the internal components of the \emph{11-dimensional} metric, while $A_{\mi \mj \mk}$ are the internal components of the \emph{11-dimensional} three-form.
We have also let $\tilde A^{\mi \mj \mm} = \frac{1}{6} \epsilon^{\mi\mj\mk \mm \mn \mp} A_{\mm\mn\mp}$.
The inverse may be straightforwardly calculated: $U^{-1}$ is obtained by flipping the signs of $\varphi$ and $C_{ijk}$ in $U$. Let us call $\hat \gM^{AB} \equiv (\gM_{AB})^{-1}$.

We also need to know the components of $\gM_{A k} = ( \gM_{\bar{\mi} k} , \gM^{\mi\mj}{}_k, \gM_{1 k})$. We find that 
\be
\gM_{Ak} = U_A{}^B \bar \gM_{Bk} 
\quad,\quad
\bar \gM_{Ak} = \begin{pmatrix}
- \hat \phi^{-2/3} \hat \phi_{\mi k} \varphi + \frac{1}{4}\hat  \phi^{-2/3} \hat \phi_{\mi \mp} \tilde A^{\mp \mq \mr} A_{k \mq \mr} \\
+ \frac{1}{\sqrt{2}} \hat \phi^{1/3} \hat \phi^{\mi\mp} \hat \phi^{\mj\mq} A_{k \mp\mq} \\
\hat \phi^{1/3} \hat\phi_{1 k} 
\end{pmatrix} 
\ee
From this one finds that
\be
(\gM_{AB})^{-1} \gM_{Bi} = (U^{-T})^A{}_B ( \bar \gM_{BC} )^{-1} \bar \gM_{Ci} \,,
\label{bit1}
\ee
where
\be
( \bar \gM_{AB} )^{-1} \bar \gM_{Bk} 
=\begin{pmatrix}
- \varphi \delta^{\mi}_k + \frac{1}{4} \tilde A^{\mi\mp\mq} A_{k\mp\mq} \\
+ \frac{1}{\sqrt{2}} A_{k \mi \mj} \\ 
\hat \phi_{11}^{-1} \hat \phi_{1 k} 
\end{pmatrix} \,.
\label{bit2}
\ee

\subsubsection*{The kinetic term}

The tension gives 
\be
T =  \frac{1}{\sqrt{10}} \sqrt{ \gM_{MN} q^M q^N } =  \frac{1}{\sqrt{10}} q \sqrt{ \gM_{\bar 1 \bar 1} } =\frac{1}{\sqrt{10}} q \sqrt{ \hat \phi^{-2/3} \hat \phi_{ 1  1} }\,,
\ee
Recall that $\hat \phi$ is still the M-theory internal metric. 
Let us denote the IIA one by $\phi_{ij}$. We have
\be
\hat \phi_{ij} - \frac{\hat \phi_{i1}\hat \phi_{j1}}{\hat \phi_{11}} = e^{-2\Phi/3} \phi_{ij} 
 \quad,\quad
\det \hat \phi = e^{-2 \Phi} \det  \phi \quad,\quad
\hat \phi_{11} = e^{4\Phi/3}\,,
\ee
where $\Phi$ is the IIA dilaton. As a result, we find
\be
T =  \frac{1}{\sqrt{10}} q \phi^{-1/3} e^{+4\Phi/3}\,.
\ee
Note that here the conformal factor $\Omega = \phi^{-1/3} e^{+4\Phi/3}$ appears.
We identify $q = \sqrt{10} T_{F1}$.  We can verify the rest of the kinetic term works out explicitly in this case. Consider the quantity
\be
\frac{1}{2T^2} ( \gM_{AB} )^{-1} q_{Ai} q_{Bj} D_\alpha Y^i D_\beta Y^j
= \frac{1}{\hat \phi^{-2/3}\hat \phi_{ 1  1} } (\hat\gM_{i1,j1}) D_\alpha Y^i D_\beta Y^j\,,
\ee
appearing in \eqref{reductionresult}.
Using the parametrisations above (recall $\hat \gM^{AB} \equiv (\gM_{AB})^{-1}$), we find 
\be
\hat \gM_{\mi \mj, \mk \ml} = \hat \phi^{-1/3} \hat \phi_{\mk[\mi} \hat \phi_{\mj]\ml} + \frac{1}{2 \hat \phi^{1/3} \hat \phi_{11} } A_{1\mi\mj} A_{1\mk\ml}  
\Rightarrow \hat \gM_{i1, j1} = \frac{1}{2} \hat \phi^{-1/3} (\hat \phi_{ij } \hat \phi_{11} - \hat \phi_{i 1} \hat \phi_{j1}) \,,
\ee
and so
\be
\frac{1}{2T^2} ( \gM_{AB} )^{-1} q_{Ai} q_{Bj} D_\alpha Y^i D_\beta Y^j
 = \frac{1}{2} \phi^{1/3} e^{-4\Phi/3}   \phi_{ij} D_\alpha Y^i D_\beta Y^j \,.
\ee
Then as expected the kinetic term of \eqref{reductionresult2} becomes
\be 
\begin{split} 
 S_{kin} 
 = - \frac{T_{F1}}{2} \int d^2\sigma \,   \sqrt{-\gamma} \gamma^{\alpha \beta} 
\hat g_{\hmu \hnu} \partial_\alpha X^{\hmu} \partial_\beta X^{\hnu} 
\end{split}
\ee
where $X^{\hmu} = (X^\mu, Y^i)$ are the usual ten-dimensional coordinates.

\subsubsection*{The Wess-Zumino term}

The Wess-Zumino term from \eqref{reductionresult} is found on using the result for the charge \eqref{IIAq} to be:
\be
\mathcal{L}_{WZ} = 
\frac{q}{\sqrt{5}} \epsilon^{\alpha \beta} \left( \sqrt{5} B_{\alpha \beta \bar{1} } + 2 A_{\alpha i 1} \partial_\beta Y^i + A_{\alpha i 1} A_\beta{}^i 
- ( \gM_{i1 , B})^{-1} \gM_{Bj} D_\alpha Y^i D_\beta Y^j \right)\,.
\ee
From \eqref{UA}, \eqref{bit1} and \eqref{bit2}, one finds that 
\be
( \gM_{\mk\ml, B} )^{-1}\gM_{B j} = + \frac{1}{\sqrt{2}} A_{j \mk \ml} + \frac{1}{\sqrt{2}} \frac{\hat \phi_{j1}}{\hat\phi_{11}} A_{\mk \ml1}  
\Rightarrow ( \gM_{i1 , B})^{-1} \gM_{Bj} =  -\frac{1}{\sqrt{2}} A_{i  j 1} \,.
\ee
The EFT dictionary of appendix \ref{E6Mdict} provides us with the information that:
\be
\begin{split} 
A_{\mu i 1}  = \frac{1}{\sqrt{2}} ( \hat C_{\mu i 1} +  A_\mu{}^j \hat C_{ij 1} )\,, \qquad  
\sqrt{5} B_{\mu \nu \bar 1} = \frac{1}{\sqrt{2}} ( \hat C_{\mu \nu 1} + A_{[\mu}{}^i \hat C_{\nu] i 1} )\,, 
\end{split}
\ee
where $\hat C_{\hmu\hnu\hrho}$ denotes the 11-dimensional three-form with kinetic term $- \frac{1}{48} F^2$. We therefore find
\be
\mathcal{L}_{WZ} = \frac{q }{\sqrt{10}} \epsilon^{\alpha \beta}\left( \hat C_{\alpha \beta 1} + 2 \hat C_{\alpha i 1} \partial_\beta Y^i + \hat C_{ij 1} \partial_\alpha Y^i \partial_\beta Y^j \right) \,.
\ee
Identifying as usual $\hat C_{\hmu \hnu 1} = \hat B_{\hmu \hnu}$ with $\hat B_{\hmu \hnu}$ the 10-dimensional B-field with kinetic term $- \frac{1}{12} H^2$, and $q \equiv T_{F1} \sqrt{10}$ we find the standard Wess-Zumino term for the fundamental string.

\subsection{Reduction to IIB $(m,n)$ string}
\label{IIBred}

We now turn to the IIB section. We have $Y^M = (Y^i, Y_{i a}, Y^{ij} , Y_{a} )$ where $i=1,\dots,5$ and $a=1,2$.
The charges allowed are $q_a$. 
The non-zero components of the charge $q_{MN}$ are given by \eqref{IIBq}.
We can turn directly to the papers \cite{Hohm:2013vpa, Baguet:2015xha} to find the generalised metric. 
The dual components can be written succintly as 
\be
\gM_{AB} = U_A{}^C U_B{}^D \bar \gM_{CD} \,,
\ee
with
\be
\bar \gM_{AB} = \begin{pmatrix} 
\phi^{1/3} \phi^{ij} m^{a b} & 0 & 0 \\ 
0 & \phi^{-2/3} \phi_{i[k} \phi_{l]j} & 0 \\
0 & 0 & \phi^{-2/3} m^{a b} 
\end{pmatrix} \,,
\ee
\be
U_A{}^C = \begin{pmatrix} 
\delta^i_m \delta^a_c & \frac{1}{\sqrt{2}} \epsilon^{imnpq} b_{pq}{}^a & \frac{1}{2} \epsilon^{ipqrs}( \epsilon_{b d} b_{pq}{}^a b_{rs}{}^d - \frac{1}{12} \delta^a_b C_{pqrs} ) \\
0 & \delta^{[mn]}_{ij} & \sqrt{2} \epsilon_{ c d} b_{ij}{}^d  \\
0 & 0 & \delta^a_c
\end{pmatrix} \,.
\ee
Here $\phi_{ij}$ are the internal components of the 10-dimensional \emph{Einstein frame} metric, $m^{ab}$ is an $\mathrm{SL}(2) / \mathrm{SO}(2)$ matrix, $b_{ij}{}^a = - 2 \hat C_{ij}{}^a$, where $\hat C_{ij}{}^a$ are the internal components of the two-form doublet, and $C_{ijkl}$ are related to the internal components of the RR four-form. 
In addition we have
\be
\gM_{Ai} = U_A{}^B \bar \gM_{Bi} \,,
\ee
with\footnote{The expression for $\gM^{ka}{}_i$ is not provided in \cite{Hohm:2013vpa} or \cite{Baguet:2015xha} and we are grateful to Henning Samtleben for providing us with the missing details. One can also verify this component by studying its transformation under generalised diffeomorphisms.}
\be
\bar \gM_{Ai} = \begin{pmatrix} 
- 2 b_{ij}{}^c \epsilon_{b c} \phi^{1/3} \phi^{jk} m^{a b} 
\\
- \frac{1}{6\sqrt{2}} \phi^{-2/3} \phi_{kp} \phi_{lq} \epsilon^{pqrst} ( C_{irst} - 6 \epsilon_{c d} b_{ir}{}^c b_{st}{}^d ) \\
\frac{2}{3} \phi^{-2/3} m_{b c} \epsilon^{kpqrs} ( b_{ik}{}^{[a} b_{pq}{}^{b]} b_{rs}{}^c + \frac{1}{8} \epsilon^{a b} b_{ik}{}^c C_{pqrs} )
\end{pmatrix} \,.
\ee 

\subsubsection*{Kinetic term}

Using the charge \eqref{IIBq} and the above expressions for the generalised metric parametrisation, we find that: 
\be
T^2 = \frac{1}{10} \phi^{-2/3} m^{ab} q_a q_b \quad,\quad
( \gM_{AB})^{-1} q_{Ai} q_{Bj} = \frac{1}{10} \phi^{-1/3}  \phi_{ij} m^{ab} q_a q_b \,.
\ee
Hence,
\be
\begin{split} 
S_{kin} 
= - \frac{1}{2} \int d^2\sigma
\sqrt{\frac{ q_a q_b m^{ab}}{10}  }  \sqrt{-\gamma} \gamma^{\alpha \beta}   \hat g_{\hmu \hnu} \partial_\alpha X^{\hmu} \partial_\beta X^{\hnu} \,,
\end{split}
\ee
where $\hat g_{\hmu \hnu}$ is the 10-dimensional Einstein frame metric, $X^{\hmu} = (X^\mu , Y^i)$.
If we write 
\be
q_a = \sqrt{10} T_{F1} ( m , n ) \,,
\ee
and parametrise the $\mathrm{SL}(2)/\mathrm{SO}(2)$ coset matrix $m^{ab}$ as 
\be
m^{ab} = e^\Phi \begin{pmatrix} 1 & C_{(0)} \\ C_{(0)} & ( C_{(0)})^2 + e^{-2\Phi} \end{pmatrix}\,,
\ee
then, in terms of the string frame metric $\hat g^{str}_{\hmu \hnu} = e^{\Phi/2} \hat g_{\hmu \hnu}$, the action takes the form  
\be
S_{kin} =
- \frac{T_{F1}}{2} \int d^2 \sigma
\sqrt{  e^{- 2 \Phi}  n^2 + ( m+ C_{(0)} n )^2} \sqrt{-\gamma}\gamma^{\alpha \beta} \hat g^{str}_{\hmu \hnu}  \partial_\alpha X^{\hmu} \partial_\beta X^{\hnu} \,.
\ee
This is a form of the action for an $(m,n)$ string. It is related to the F1 action by an S-duality transformation from $(1,0)$ to $(m,n)$, and can be obtained from the usual D1 action by integrating out the Born-Infeld vector \cite{Schmidhuber:1996fy}.
We see that for $(m,n) = (1,0)$ we immediately get the F1 action, while for $(m,n) = (0,1)$ we get an action with tension $T_{D1} = g_s^{-1} T_{F1}$. 

\subsubsection*{Wess-Zumino term} 

We find that
\be
q_{Ai} ( \gM_{AB})^{-1} \gM_{Bj} =
 \frac{1}{\sqrt{10}} 2 q_a b_{ij}{}^a \,,
\ee
and hence the Wess-Zumino term from \eqref{reductionresult} is found to be:
\be
\mathcal{L}_{WZ} 
= \frac{1}{\sqrt{10}} q_a \epsilon^{\alpha \beta} (
\sqrt{10} B_{\alpha \beta}{}^a 
- \epsilon^{ab} 2 A_{\alpha i b} \partial_\beta Y^i - \epsilon^{ab}  A_{\alpha i b} A_\beta{}^i 
 -2 b_{ij}{}^a D_\alpha Y^i D_\beta Y^j ) 
)\,.
\ee
The EFT dictionary of appendix \ref{E6Bdict} tells us that
\be
\begin{split}
- 2 b_{ij}{}^a  = \hat C_{ij}{}^a \,,\qquad
\sqrt{10} B_{\mu\nu}{}^a  = \hat C_{\mu\nu}{}^a + A_{[\mu}{}^k \hat C_{\nu] k}{}^a \,,\qquad
A_{\mu i a}  = \epsilon_{ab} ( \hat C_{\mu i}{}^b + A_\mu{}^j \hat C_{ij}{}^b )\,,
\end{split} 
\ee
from which one gets
\be
\mathcal{L}_{WZ} =
 \frac{1}{\sqrt{10}} q_a \left( 
\hat C_{\alpha \beta}{}^a + 2 \hat C_{\alpha i} \partial_\beta Y^i + \hat C_{ij}{}^a \partial_\alpha Y^i \partial_\beta Y^j 
\right) 
\ee
so with $q_a = \sqrt{10} \tilde q_a = \sqrt{10} T_{F1} (m,n)$ we find
\be
\mathcal{L}_{WZ} = \epsilon^{\alpha\beta} \tilde q_a \hat C_{\hmu \hnu}{}^a \partial_\alpha X^{\hmu} \partial_\beta X^{\hnu} \,,
\ee
which is the expected result.

\section{Fradkin-Tseytlin term}
\label{FTterm}

The doubled sigma model can be extended with the addition of a Fradkin-Tseytlin term using the doubled dilaton $\Phi_d$, which is related to the usual dilaton $\Phi$ by $\Phi_d = \Phi - \frac{1}{4} \log \det \phi$. This doubled FT term \cite{Hull:2006va} is just:
\be
S_{FT} = \frac{1}{4\pi} \int d^2\sigma \sqrt{-\gamma} \Phi_d R 
\label{ddil}
\ee 
where $R$ is the worldsheet Ricci scalar. Integrating out the gauge fields $V_{\alpha i}$ from the action
\be
S \supset \frac{T_{F1}}{4} \int d^2 \sigma \sqrt{-\gamma}\gamma^{\alpha \beta} \gM^{ij} V_{\alpha i} V_{\beta j} 
\ee
generates a shift of the doubled dilaton
\be
\Phi_d \rightarrow \Phi_d - \frac{1}{4} \log \det ( \gM^{ij} ) = \Phi_d + \frac{1}{4} \log \det \phi = \Phi
\ee
which turns the doubled FT term into the ordinary FT term with the conventional normalisation, thereby fixing its coefficient relative to the rest of the doubled sigma model.

We can consider something similar for our exceptional sigma model.
Although there is no exceptional dilaton, we propose to use $T(\gM,q) / T_{F1}$ to write down a scalar. Effectively, the charge $q_{MN}$ allows us to construct a scalar. 
One can check that:
\be
T^2 ( \gM,q)/T_{F1}^2 = \begin{cases}
( \phi^{-2} e^{8 \Phi} )^{1/(n-2)} & \text{IIA} \\
 \phi^{-2 / (n-2)} ( e^{-\Phi} n^2 + e^{\Phi} ( m  + C_{(0)} n)^2 ) & \text{IIB} 
\end{cases} \,,
\ee
while
\be
\det \gM_{AB} = \begin{cases} ( \phi^{-2} e^{(10-n)\Phi} )^{4/(n-2)} & \text{IIA} \\ 
( \phi^{-2} )^{4/(n-2)} & \text{IIB}
\end{cases} \,.
\ee
Hence, the combination 
\be
\log\Big[\left(\frac{T^2}{T^2_{F1}}\right)(\det M)^{-1/4}\Big]=\begin{cases} \Phi &\text{IIA}\\ \log ( e^{-\Phi} n^2 + e^{\Phi} (m + C_{(0)} n)^2  ) &\text{IIB}\,.\end{cases}
\ee
which therefore recovers the string dilaton in IIA and in IIB for $m=1,n=0$.

We integrate out the vector fields $V_\alpha^A$ from 
\be
S \supset \frac{T_{F1}}{4} \int d^2\sigma \sqrt{-\gamma} \gamma^{\alpha \beta} \left(T/T_{F1} \right) \gM_{AB} V_\alpha^A V_\beta^B
\ee
which produces a term 
\be
\frac{1}{4\pi} \int \sqrt{-\gamma} \left( - \frac{1}{4} \log \det ( \left(T/T_{F1} \right) \gM_{AB} )\right) R 
\ee
As $\det (T \gM_{AB} ) = T^{\dim R_1 - d} \det \gM_{AB}$, we combine this with the exceptional sigma model Fradkin-Tseytlin term:
\be
\frac{1}{4\pi} \int \sqrt{-\gamma} \frac{1}{4} ( \dim R_1 -d + 8) \log \left(T/T_{F1} \right)  R
\label{EFradT}
\ee
so that the integration out of the $V_\alpha^A$ produces the conventional FT term, at least for the IIA string and $(1,0)$ IIB string.

It may seem strange that the coefficient depends on $\dim R_1$ and $d$ and so differs from group to group, while the rest of the exceptional sigma model action took a universal form. However, this is in fact natural and consistent with the fact that one could reduce from $\ED$ to $\mathrm{E}_{D-1(D-1)}$ by integrating out a subset of the dual coordinates, thereby altering the term \eqref{EFradT}.

It is also possible to obtain the doubled dilaton Fradkin-Tseytlin term for the doubled sigma model directly.
The reduction from the exceptional sigma model to the doubled sigma model was explained in section \ref{redtodouble}.
The tension $T/T_{F1} = e^{-4\omega \Phi_d}$. 
Hence \eqref{EFradT} is initially
\be
\frac{1}{4\pi} \int \sqrt{-\gamma}
( -\dim R_1 +d - 8) \omega \Phi_d R
\label{this}
\ee
to which we add after integrating out
\be
\frac{1}{4\pi} \int \sqrt{-\gamma} \left( -\frac{1}{4}\right) \log ( \det ( e^{-4\omega \Phi_d}  \phi_{AB}) ) R\,.
\label{that}
\ee
As $\det ( e^{-4\omega \Phi_d}\phi_{AB} ) =  e^{-4\omega ( \dim R_1 - 2d ) \Phi_d}e^{-8\omega d \Phi_d}=e^{-4\omega \dim R_1 \Phi_d}$
and $\omega = - \frac{1}{n-2}= \frac{1}{d-8}$ the combination of \eqref{this} and \eqref{that} gives exactly \eqref{ddil}.

\section{Quasi-tensionless uplift} 
\label{tless}

In \cite{Blair:2017gwn}, the following action for the $R_1$ multiplet of particle states in $n$ dimensions:
\be
S = \int d\tau \left( - \sqrt{p_Mp_N \gM^{MN}} \sqrt{ - \det g_{\mu\nu} \dot{X}^\mu \dot{X}^\nu } + p_M \dot{X}^\mu A_\mu{}^M \right) 
\label{particle1}
\ee
was shown to uplift to the action for a massless particle on the extended spacetime of DFT/EFT:
\be
S = \int d\tau \frac{\lambda}{2} \left( g_{\mu\nu} \dot{X}^\mu \dot{X}^\nu + \gM_{MN}
 ( \dot{Y}^M + \gV^M + \dot{X}^\mu A_\mu{}^M )
 ( \dot{Y}^N + \gV^N + \dot{X}^\mu A_\mu{}^N )
\right)\,,
\label{particle2}
\ee
with the momenta in the extended directions $Y^M$ corresponding to the masses/charges $p_M$. 

One might wonder whether our string action could be interpreted as that of a tensionless string in the extended spacetime.
However, although massless particles reduce to massive particles on reduction (which underlies the relationship between the two forms of the particle action), tensionless strings do not reduce to tensionful strings.
Early explorations of this concept \cite
{deAzcarraga:1991px, 
Townsend:1992fa, 
Bergshoeff:1992gq}
found that one can instead replace the string tension with a dynamical one-form, which may have some (unclear) geometrical interpretation.
These ideas led, by combining the tension one-form with the Born-Infeld one-form to the $\mathrm{SL}(2)$ covariant description of the F1 and D1
\cite{
Townsend:1997kr, 
Cederwall:1997ts
}. The actions of \cite{Bergshoeff:1992gq, Townsend:1997kr, Cederwall:1997ts} can be termed ``quasi-tensionless'' in that they take the form $S \sim \int d^2\sigma \lambda ( \det g + (\star F_2)^2)$, where $\lambda$ is a Lagrange multiplier and $F_2$ is the field strength for some worldsheet one-forms. If the $(\star F_2)^2$ term was not present, then this would be a tensionless action. 

In general, one would encode the tension of $p$-brane in terms of a worldvolume $p$ form. 
For the particle case, this is a worldline scalar. In the action \eqref{particle2}, these scalars are the $Y^M$, which can be interpreted as target space coordinates.
The action evidently takes the form we have just mentioned, where the field strength of the $Y^M$ is $\dot{Y}^M + \dots$. However, the target space interpretation is indeed that it is tensionless (i.e. massless). 

Let us see how one can take a similar approach to our exceptional sigma model action, \eqref{ESintro}. 
Integrating out the worldsheet metric, we can write the action as:
\be
S = 
 \int d^2\sigma \left( 
-T \sqrt{ - \det \left( g+ \frac{1}{2} \gM \right) }
- \frac{1}{2} \epsilon^{\alpha \beta} q_{MN} \mathcal{B}_{\alpha \beta}{}^{MN}
\right) 
\label{ESNG}
\ee
where $\mathcal{B}_{\alpha \beta}{}^{MN} = B_{\mu\nu}{}^{MN} \partial_\alpha X^\mu \partial_\beta X^\nu + \dots$ denotes the full Wess-Zumino term, and $g+\frac{1}{2} \gM$ is shorthand for $g_{\mu\nu} \partial_\alpha X^\mu \partial_\beta X^\nu + \frac{1}{2} \gM_{MN} D_\alpha Y^M D_\beta Y^N$.

The action \eqref{ESNG} can be obtained from the following quasi-tensionless action:
\be
\begin{split}
S & =
\int d^2\sigma \frac{1}{2} \lambda 
\Big(
\det \left( g + \frac{1}{2} \gM \right)
\\ & \qquad\qquad\qquad\quad
+ 2d \gM_{MP} \gM_{NQ} \epsilon^{\alpha \beta} \epsilon^{\gamma \delta} \!
\left( \partial_\alpha Z_\beta^{MN} + \frac{1}{2} \mathcal{B}_{\alpha \beta}{}^{MN} \right)\!\!
\left( \partial_\gamma Z_\delta^{PQ} + \frac{1}{2} \mathcal{B}_{\gamma \delta}{}^{PQ} \right)
\Big) \,.
\end{split} 
\label{uplift1}
\ee
To demonstate this, we remove the new worldsheet one-form $Z_\alpha \in R_2$ following \cite{Townsend:1997kr}.
Define the momentum (we have $\epsilon^{01} = -1$)
\be
Q_{MN} = \frac{\partial \mathcal{L}}{\partial \dot{Z}_1^{MN} } = -\lambda 2 d \gM_{MP} \gM_{NQ} \epsilon^{\alpha \beta} 
\left( \partial_\alpha Z_\beta^{PQ} + \frac{1}{2} \mathcal{B}_{\alpha \beta}{}^{PQ} \right)\,.
\ee
The action in Hamiltonian form is:
\be
\begin{split}
S & = \int d^2 \sigma \left( \dot{Z}_1^{MN} Q_{MN} - \mathcal{H} \right)
\\ & = 
\int d^2 \sigma \Big(
\dot{Z}_1^{MN} Q_{MN} 
- \frac{1}{2 \lambda} \frac{1}{2d} \gM^{MP} \gM^{NQ} Q_{MP} Q_{NQ} 
+ \frac{\lambda}{2} \det \left( g+\frac{1}{2} \gM\right) 
\\ & \qquad\qquad\qquad- Q_{MN} \left( Z_0^{\prime MN} + \frac{1}{2} \epsilon^{\alpha \beta} \mathcal{B}_{\alpha \beta}{}^{MN} \right) 
\Big) \,.
\end{split}
\ee
Now $Z_0$ and $Z_1$ play the role of Lagrange multipliers, setting $Q_{MN} = q_{MN}$ to be constant. 
Replacing it in the action correspondingly, we get
\be
S = \int d^2\sigma \left(
- \frac{1}{2 \lambda}T^2 
+ \frac{\lambda}{2} \det \left( g+\frac{1}{2} \gM\right) 
- \frac{1}{2} q_{MN} \epsilon^{\alpha \beta} \mathcal{B}_{\alpha \beta}{}^{MN} 
\right) 
\label{downlift}
\ee
and integrating out $\lambda$ leads to \eqref{ESNG}.

We must then confront the issue that, just as in \cite{Blair:2017gwn} the initial particle action involved the field strength $\dot{Y}^M + \dot{X}^\mu A_\mu{}^M$ which was not covariant under generalised diffeomorphisms, the naive field strength we have used here, $\partial_\alpha Z_\beta^{MN} + \frac{1}{2} \mathcal{B}_{\alpha \beta}{}^{MN}$, will not be invariant under gauge transformations: we know that the Wess-Zumino term only transforms as a total derivative (which can be cancelled by assigning $Z_\alpha{}^{MN}$ the approriate transformation) when contracted with an appropriately constrained $q_{MN}$. 
The solution in the particle case was to introduce $\gV^M$, a worldline one-form. The generalisation of this is to introduce a worldsheet two-form, $W_{\alpha \beta}$. 
This gives the beginnings of a worldvolume mirroring of the tensor hierarchy for $p$-branes: 
\be
\begin{array}{cccc}
p & \mathrm{Charge} & \mathrm{Coordinate} & \mathrm{Gauge}\,\,\mathrm{field} \\
0 & p_M & Y^M & V_\alpha{}^M \\
1 & q_{MN} & Z_\alpha{}^{MN} & W_{\alpha \beta}{}^{MN} \\
\vdots & & &
\end{array}
\ee
It is interesting to compare the situation here with the suggestions in \cite{Aldazabal:2013via} that one could introduce extended ``coordinates'' associated to each gauge transformation parameter in the tensor hierarchy, leading to a notion of an extended ``mega-space'' beyond that already used involving just $Y^M$. (The dual coordinates contained in the latter are of course already associated to the purely internal gauge transformation parameters.) This may therefore be ``natural'' from the point of view of branes in EFT, though it is not clear that one should really view for instance $Z_\alpha{}^{MN}$, which is a worldvolume one-form, as something geometric in extended spacetime.

Under a gauge transformation,
\be
\delta_\lambda B_{\alpha \beta}{}^{MN} = 2 \partial_{[\alpha} ( \lambda_{\mu}{}^{MN} \partial_{\beta]} X^\mu ) + \Delta_{\alpha \beta}{}^{MN} \,,
\ee
where $\Delta_{\alpha \beta}{}^{MN}$ denotes the anomalous transformation. This vanishes for $\partial_M = 0$ and obeys $q_{MN} \Delta_{\alpha \beta}{}^{MN}=0$ for appropriately constrained $q_{MN}$,
and so we require
\be
\begin{split} 
\delta_\lambda Z_{\alpha}{}^{MN} & = - \lambda_\mu{}^{MN} \partial_\alpha X^\mu \,, \\
\delta_\lambda W_{\alpha \beta}^{MN} & = -\frac{1}{2} \Delta_{\alpha \beta}{}^{MN} \,.
\end{split} 
\ee
Evidently, we need to impose constraints on $W_{\alpha \beta}^{MN}$ which are dual to those \eqref{magic} on $q_{MN}$. These take a cumbersome form which we will not display here.
Then, the full action
\be
S - \int d^2 \sigma \frac{1}{2} \lambda \left( \det \left( g+ \frac{1}{2} \gM \right)
+ 2 d \gM_{MP} \gM_{NQ} 
F^{MN} F^{PQ} 
 \right)
\label{uplift2} 
\ee
with
\be
F^{MN} \equiv \epsilon^{\alpha \beta} \left( 
 \partial_\alpha Z_\beta^{MN} + W_{\alpha \beta}^{MN}  + \frac{1}{2} \mathcal{B}_{\alpha \beta}{}^{MN} \right) 
\ee
leads to \eqref{ESNG} plus the extra term:
\be
\int d^2\sigma \epsilon^{\alpha \beta} q_{MN} W_{\alpha \beta}^{MN} \,.
\label{extraterm}
\ee
The equation of motion for the non-zero components of the constrained $W_{\alpha \beta}^{MN}$ then sets to zero exactly the components of $q_{MN}$ which must vanish by the constraint \eqref{magic}. With the constraint on $W$ understood we propose that the action \eqref{uplift2} represents the quasi-tensionless uplift of the exceptional sigma model. 

We may also wonder about the checks of how \eqref{uplift2} respects generalised diffeomorphism and external diffeomorphism invariance.
Consider the transformation of the Wess-Zumino term, $\mathcal{B}_{\alpha \beta}{}^{MN}$.
We know that under any of the transformations that we care about, this transforms into a sum of the following pieces: total derivatives, anomalous terms that vanish on contraction with $q_{MN}$, and the anomalous terms arising in the case of external diffeomorphisms that were cancelled using contributions from the transformations of the kinetic terms. The total derivative type transformations can be cancelled by appropriate transformations of $Z_\alpha^{MN}$, while those that vanish against $q_{MN}$ can be safely absorbed into transformations of $W_{\alpha \beta}{}^{MN}$. 
The sole subtlety here is that one must take the Lagrange multiplier, $\lambda$, to transform under generalised diffeomorphisms and external diffeomorphisms in order to cancel the extra weight terms, similarly to the particle case \cite{Blair:2017gwn}. 

As a result, the only danger appears to lie in the final anomalous part of the transformation under external diffeomorphisms. In this case, we can write the total potentially anomalous variation as, using \eqref{anomvars} without the $T$,
\be
\begin{split}
\hat \Delta_\xi S & = 
\int d^2\sigma \frac{1}{2} \lambda \Bigg(
g_{\mu\nu} \partial_\alpha X^\mu \partial_P \xi^\nu 
\left(
\det \mathcal{G} \,\mathcal{G}^{\alpha \beta} D_\beta Y^P 
- 2 d \epsilon^{\alpha \beta} \gM_{MN} F^{NP} D_\beta Y^M 
\right)
\\ & 
\qquad\qquad\qquad\qquad
+ \tilde \delta_\xi \gV_\alpha^M \gM_{MN} \left( 
\det \mathcal{G}\, \mathcal{G}^{\alpha \beta} D_\beta Y^N
- 2 d \epsilon^{\alpha \beta} \gM_{KP} F^{NP} \bar D_\beta Y^K
\right) 
\\ & 
\qquad\qquad\qquad\qquad
+ \delta'_\xi W_{\alpha \beta}^{MN}  4 d \gM_{MP} \gM_{NQ} F^{PQ} \epsilon^{\alpha \beta} 
\Bigg)\,,
\end{split}
\ee
where $\mathcal{G}_{\alpha \beta} \equiv g_{\alpha \beta} + \frac{1}{2} \gM_{\alpha \beta}$, 
$\bar D_\alpha Y^M = D_\alpha Y^M - \gV_\alpha^M$.
The following extra tranformations of $V$ and $W$:
\be
\begin{split}
\tilde \delta_\xi V_\alpha^M & = - 2d \frac{\mathcal{G}_{\alpha \beta} \epsilon^{\beta \gamma}}{\det \mathcal{G}} F^{MP} \partial_P \xi^\mu g_{\mu\nu} \partial_\gamma X^\nu\,,\\
\delta'_\xi W_{\alpha \beta}^{MN} & = \frac{1}{2} \epsilon_{[\alpha | \gamma|} \mathcal{G}^{\gamma \delta} \partial_{\beta]} X^\mu g_{\mu\nu} \partial_P \xi^\nu
\left( F^{MN} \bar D_\delta Y^P - 2 d F^{P(M} \bar D_\delta Y^{N)} \right)\,,
\end{split} 
\label{deltaVW}
\ee
lead to 
\be
\hat \Delta_\xi S =
\int d^2\sigma \frac{1}{2} \lambda 
\mathcal{G}^{\alpha \beta} 
g_{\mu\nu} \partial_\alpha X^\mu  \partial_K \xi^\nu 
\bar D_\beta Y^K \left(
\det \mathcal{G} + 2d \gM_{MP} \gM_{NQ} F^{MN} F^{PQ} 
\right) \,.
\ee
The quantity in brackets is zero by the quasi-tensionless condition enforced by the equation of motion of $\lambda$: evidently we can cancel it off-shell by additionally taking 
\be
\tilde \delta_\xi \lambda = - 
\mathcal{G}^{\alpha \beta} 
g_{\mu\nu} \partial_\alpha X^\mu  \partial_K \xi^\nu 
\bar D_\beta Y^K \,.
\ee
We conclude that the uplifted action \eqref{uplift2} shares the same features as our original action \eqref{ESintro} with respect to the invariances of EFT.

Note that the transformation $\delta'_\xi V_\alpha^M$ in \ref{deltaVW} is equal to the transformation \eqref{tildedeltaV} on making use of the equations of motion for $Z_\alpha^{MN}$ and $\lambda$ used in the reduction, and that for the worldsheet metric $\gamma_{\alpha \beta}$. 

Finally, we must check that the transformation $\tilde \delta_\xi W_{\alpha \beta}^{MN}$ does not break the constraints we wanted to impose on $W_{\alpha\beta}^{MN}$. 
One way to confirm this is to use the fact that as 
$F^{MN} \in R_2$ we have $F^{MN} = \frac{1}{2d} Y^{MN}{}_{KL} F^{KL}$, so that $\tilde \delta_\xi W_{\alpha\beta}^{MN}$ is proportional to
\be
\bar D Y^Q F^{KL} \left( \delta^M_K \delta^N_L \partial_Q - Y^{P(M}{}_{KL} \delta^{N)}_Q \partial_P \right) 
\ee
which when contracted with $q_{MN}$ gives $q_{KL} \partial_Q - q_{MQ} Y^{MP}{}_{KL} \partial_P$ i.e. it leads to the constraint \eqref{magic}.

\section{Comments on branes} 
\label{mem} 

\subsection{Exceptional democracy (and why 10-dimensional sections are special)} 

In this paper, we taken a route towards a reformulation of one-brane actions that began with exceptional field theory. 
The latter is a reformulation of supergravity; there is much more to life than supergravity, and so our work forms part of the bigger picture of attempting to describe all the usual interesting braney features of string theory or M-theory in $\ED$ covariant language. The motivation here is to view EFT as a new organising principle for string and M-theory. This organising principle \emph{knows} something about the dualities that appear on toroidal reduction, but beyond that it provides access to a formulation that underlies different limits of the duality web - the same EFT structure elegantly describes 11-dimensional and 10-dimensional supergravities in one systematic fashion. 

Our approach towards branes suggests we should attempt to construct $p$-brane actions coupling to the generalised $p$-form fields of EFT. (One might envisage some difficulties here. Magnetic branes will couple to dualisations of the usual gauge potentials, which in EFT will ultimately involve exotic duals of various sorts. The description of branes whose worldvolume dimension exceeds the number of external directions in the EFT is also not immediately clear, but presumably involves coupling to generalised forms of such a type. Of course, the doubled string is known to work when there are \emph{no} external directions, while a doubled five-brane action \cite{Blair:2017hhy} can also be constructed, which indeed involves a WZ coupling to an unusual generalised $O(D,D)$ four-form, which at least linearly can be viewed as the dual of the generalised metric \cite{Bergshoeff:2016ncb}).
Supposing this is possible, we have a picture of an $\ED$ covariant theory in which $p$-branes for all $p$ are present, coupling to an extended tensor hierarchy of generalised form fields and to the generalised metric.

This amounts to a reorganisation of the description of branes in the usual type II and M-theory pictures. We can view this as an exceptional brane democracy.  
Solving the section condition to give the standard 11-dimensional, 10-dimensional type IIA and 10-dimensional type IIB sections, these brane actions -- which will be characterised by charges including and generalising our $q_{MN}$, obeying particular constraints -- collapse down to the usual ones. 
This leads to the observation, starting from the point of view of the extended theory, that 10-dimensional sections are special, because these contain a fundamental brane - the F1.

\subsection{Membranes and topological terms} 

Let us look ahead to membranes in particular.
These couple to the EFT field $C_{\mu\nu\rho} \in R_3$. 
As before, to describe these, we will be led to introduce a constant charge $q \in \bar R_3$. 
We expect there to be a constraint on $q$, which we would anticipate to follow from requiring $\mathcal{L}_\Lambda q = 0$. 
Then we would build up the Wess-Zumino term as before. 

An interesting question here would be to see if there is a topological term for membranes.
We commented already on the fact that one could perhaps define a symplectic form $\Omega_{MN}$ using the total derivative that is left after integrating out the dual components $V_\alpha^A$. 
In EFT, which, if we view it as a glimpse into the exceptional geometry of M-theory, has no brane more fundamental than any other, it seems that one should expect to collect a collection of symplectic $p$-forms for each brane worlvolume. Whether these play any role in the spacetime theory, as $\Omega_{MN}$ does in some approaches to doubled geometry based on a doubled sigma model \cite{Freidel:2015pka,Freidel:2017yuv}, is then an interesting question. 

As an example, let us consider the group $\Gfour$. Here the three-form $C_{\mu\nu\rho}{}^a$ is in the five-dimensional fundamental representation, for which we use the indices $a,b,c,\dots$. The charge $q_a$ should satisfy $q_{[a} \partial_{bc]} = 0$ (which has appeared in a similar setting in \cite{CederwallKorea, MalekTalk}). The M-theory section is defined by splitting $a=(i,5)$ and taking $\partial_{i5} \neq 0$, $\partial_{ij} = 0$. 
A natural guess, based on the available index contractions, for a topological term would be something like:
\be
S \supset - \frac{1}{2} \int d^3 \sigma \epsilon^{\alpha \beta \gamma} q_5 \epsilon_{i jkl 5} \partial_\alpha Y^{ij} \partial_\beta Y^{k5} \partial_\gamma Y^{l5}\,,
\ee
leading to a totally antisymmetric $\Omega_{MNP}$ with $\Omega_{ij, k5, l5} \sim \epsilon_{ijkl}$.
We may recall that T-duality of strings is basically a canonical transformation in phase space: the generalisation from symplectic two-forms to symplectic $p$-forms leads from Hamiltonian to Nambu mechanics. In the simplest generalisation, with $p=3$, phase space is $3N$ dimensional with canonical triples of phase space coordinates (rather than coordinate/momenta pairs). It looks likely that one can view the set of coordinates $Y^{ij}$, $Y^{k5}, Y^{l5}$ as a set of ``Nambu triples'', with for fixed $i \neq j$, $Y^{i5}, Y^{j5}$ and $\frac{1}{2}\epsilon^{ijkl} Y_{kl}$ a Nambu triple.
This may shed light on the full structure of the exceptional geometry, and we will continue this investigation in future work.
(Of course, beyond $\Gfour$, coordinates that can be associated to M5 windings will appear also.)

\section{Conclusions}
\label{concl}
In this paper, we have thoroughly investigated the exceptional sigma model, whose form we wrote down in \cite{Arvanitakis:2017hwb}.
This is the action for a string coupled to a background of exceptional field theory, a unified reformulation of the 11-dimensional and type II supergravities.
It generalises the doubled sigma model based on $O(d,d)$ to the exceptional groups $\ED$.

One can view the exceptional sigma model action in four ways:
\begin{itemize}
\item as an action for a multiplet of charged strings in an ``extended spacetime'', with extra worldsheet scalars corresponding to dual directions,
\item as an action for the usual 1-brane states in $10$ dimensions on integrating out these dual coordinates,
\item as an action for strings corresponding to wrapped branes in $n$-dimensions, on further reduction,
\item and also, by encoding the charges $q$ as the momenta for further worldsheet one-forms, the action \eqref{ESintro} becomes that of a quasi-tensionless string \eqref{uplift1} generalising the $\mathrm{SL}(2)$ covariant string \cite{Townsend:1997kr, Cederwall:1997ts}.
\end{itemize}

Our approach was grounded in respecting the local symmetries of exceptional field theory. 
We were able to construct the Wess-Zumino term by requiring invariance under gauge transformations of the EFT generalised two-form, and showed that requiring a form of covariance under the generalised and external diffeomorphism transformations allowed us to essentially fix the whole Polyakov-style Weyl invariant action.

We should note that one limitation of the present paper was that we effectively restricted ourselves to $D \leq 6$, where one can use a common general description of the $\ED$ EFTs. For $D=7$, the Y-tensor is no longer symmetric, and one has in addition to the standard generalised two-form an additional covariantly constrained two-form in $\bar R_1$. 
However, it is likely that the exceptional sigma model can be constructed in this case (for the standard generalised two-form), and we should do this.

It would be interesting to further explore our action in the case $\partial_M = 0$, corresponding to a (torodial) reduction to $n$-dimensions, where the $\ED$ symmetry of the action becomes the standard U-duality. Perhaps one can study, or define, U-fold backgrounds from the sigma model, just as the double sigma model was introduced in order to better understand T-folds. 

There are some other obvious problems that should be addressed in the future. We should supersymmetrise the exceptional sigma model.
The beta functional equations should provide some truncation of the full EFT field equations. 
We know already that the background field equations following from the doubled sigma model lead to those of DFT \cite{Berman:2007xn, Berman:2007yf, Copland:2011wx}. One might therefore expect to obtain the same equations, with the replacement $B_{\mu\nu} \rightarrow q_{MN} B_{\mu\nu}{}^{MN}$. There are question marks here about how exactly one makes sense of the (possible) appearance of the charge $q$, and how much information one can really extract from this procedure. 

We have assumed throughout that the generalised metric admits a standard parametrisation on section in terms of the usual spacetime fields. This was important in integrating out the dual coordinates, for example, where we needed to assume invertibility of some block of the generalised metric. 
This assumption ignores the possibility of alternative ``non-Riemannian'' parametrisations which have been explored for the doubled sigma model in \cite{Lee:2013hma, Morand:2017fnv}. 
This would also be interesting to consider in the context of our exceptional sigma model. 

Also, in this paper we generalised the doubled string action of Hull \cite{Hull:2004in, Hull:2006va}.
An alternative (equivalent) approach is that of Tseytlin \cite{Tseytlin:1990nb, Tseytlin:1990va} where the chirality constraint --- what we called twisted self-duality earlier --- follows from the equations of motion \emph{without} the introduction of a gauge field $V^M_\alpha$ at the cost of losing manifest Lorentz invariance (which can be recovered using a PST-style approach as for instance in \cite{Driezen:2016tnz}).
It would be intriguing to see if these methods apply to the exceptional sigma model, perhaps using an action of the form
\be
S = -\frac{1}{2} \int d^2 \sigma \left( q_{MN} D_0 Y^M D_1 Y^M - T \gM_{MN} D_1 Y^M D_1 Y^N + \dots \right) \,.
\ee
This involves $q_{MN}$ playing the role of the $O(d,d)$ metric $\eta_{MN}$.
A Hamiltonian study of doubled string actions of this form shows that worldsheet diffeomorphism invariance leads to the section condition on the background fields
\cite{Blair:2013noa}, and this would be intriguing to see here. Such methods could also be applied to quantise the Hull-type exceptional sigma model we have used \cite{HackettJones:2006bp}.

Finally, as we have already mentioned, our approach could in principle lead to the actions for branes of higher worldvolume dimension coupled to the EFT fields, and thus to ``U-duality-covariant'' branes. 

\vspace{1em}

{\bf Acknowledgements:} 
The authors are grateful to David Berman, Martin Cederwall, Chris Hull, Emanuel Malek, Jeong-Hyuck Park, Henning Samtleben, Daniel Thompson, and Arkady Tseytlin for useful discussions,
and to the organisers of the conference ``String Dualities and Geometry'' held at the Centro Atomico Bariloche, Argentina, for hospitality while we completed this work. 
ASA is supported by the EPSRC programme grant ``New Geometric Structures from String Theory'' (EP/K034456/1). 
CB is supported by an FWO-Vlaanderen postdoctoral fellowship, by the Belgian Federal Science Policy Office through the Interuniversity Attraction Pole P7/37 ``Fundamental Interactions'', by the FWO-Vlaanderen through the project G.0207.14N, by the Vrije Universiteit Brussel through the Strategic Research Program ``High-Energy Physics''. 

\appendix 

\section{EFT dictionaries}
\label{appA}

\subsection{Supergravity, decomposed}
\label{SUGRAdecomp} 

Let $\hat g_{\hmu \hnu}$ denote the metric of 10- or 11-dimensional supergravity.
We split the coordinates $X^{\hmu} = (X^\mu, Y^i)$ and take the metric to be given by
\be
\hat g_{\hmu \hnu} 
= \begin{pmatrix} 
\Omega g_{\mu\nu} + \phi_{kl} A_\mu{}^k A_\nu{}^l & \phi_{jk} A_\mu{}^k \\
\phi_{ik} A_\nu{}^k & \phi_{ij} 
\end{pmatrix} \,.
\label{metricdecomp}
\ee
This amounts to a Lorentz gauge fixing that breaks $\mathrm{SO}(1,10) \rightarrow \mathrm{SO}(1,n-1) \times \mathrm{SO}(D)$. 
If $\hat g_{\hmu\hnu}$ is the 10- or 11-dimensional Einstein frame metric, the conformal factor is $\Omega = \phi^\omega$, while if it is a 10-dimensional string frame metric, then $\Omega = \phi^\omega e^{-4\Phi\omega}$. Here $\omega =0$ in DFT and $\omega = - \frac{1}{n-2}$ in EFT (and $\phi \equiv |\det \phi|$). Note that this guarantees that EFT reduces to Einstein frame in $n$ dimensions, while DFT reduces from 10-dimensional string frame to $n$-dimensional string frame with the generalised dilaton $e^{-2\Phi_d} = e^{-2\Phi} \phi^{1/2}$.

Now consider the form fields, $\hat C_{\hmu_1 \dots \hmu_p}$.
To define covariant $n$-dimensional $p$-forms, it is convenient to treat these to the standard field redefinition, whereby
\be
A_{\mu_1 \dots \mu_p i_1 \dots i_q} =
 \hat e_{\mu_1}{}^{\bar a_1} \hat e_{\bar a_1}{}^{\hmu_1} 
 \dots \hat e_{\mu_p}{}^{\bar a_p}\hat  e_{\bar a_p}{}^{\hmu_p} \hat C_{\hmu_1 \dots \hmu_p i_1 \dots i_q}
\ee 
with $\hat e_{\hmu}{}^{\hat a}$ the vielbein for the metric $\hat g_{\hmu \hnu}$, and $\bar a$ the flat $n$-dimensional index.

Let us apply this to the 11-dimensional three-form, $\hat C_{\hmu \hnu \hrho}$. This procedure leads to:
\be
\begin{split}
A_{mnp} & = \hat C_{mnp} \,,\\
A_{\mu mn} & = \hat C_{\mu mn} - A_\mu{}^k \hat C_{k m n}\,, \\ 
A_{\mu \nu m} & = \hat C_{\mu \nu m} - 2A_{[\mu}{}^k \hat C_{\nu] m k} 
+ A_\mu{}^k A_\nu{}^l \hat C_{m k l } \,, \\ 
A_{\mu\nu\rho} & = \hat  C_{\mu\nu\rho} - 3 A_{[\mu}{}^k \hat  C_{\nu\rho]k} 
+ 3 A_{[\mu}{}^k A_\nu{}^l \hat  C_{\rho]kl} 
- A_\mu{}^k A_\nu{}^l A_\rho{}^m \hat  C_{klm} \,,
\end{split} 
\label{AC}
\ee
where $A_\mu{}^k$ is the KK vector. The inverse relations are
\be
\begin{split}
\hat  C_{mnp} & = A_{mnp} \,, \\
\hat C_{\mu mn} & = A_{\mu mn} + A_\mu{}^k A_{k m n} \,, \\ 
\hat C_{\mu \nu m} & = A_{\mu \nu m} + 2A_{[\mu}{}^k A_{\nu] m k} 
+ A_\mu{}^k A_\nu{}^l A_{m k l } \,, \\ 
\hat C_{\mu\nu\rho} & = A_{\mu\nu\rho} + 3 A_{[\mu}{}^k A_{\nu\rho]k} 
+ 3 A_{[\mu}{}^k A_\nu{}^l A_{\rho]kl} 
+ A_\mu{}^k A_\nu{}^l A_\rho{}^m A_{klm} \,,.
\end{split} 
\label{CA}
\ee
Similarly, for type IIB, let $\hat C_{\hmu \hnu}{}^\alpha$ denote the doublet of two-forms. We define
\be
\begin{split} 
A_{ij}{}^\alpha & = \hat C_{ij}{}^\alpha \,,\\
A_{\mu i}{}^\alpha & = \hat C_{\mu i}{}^\alpha - A_\mu{}^j \hat C_{ji}{}^\alpha \,, \\
A_{\mu \nu}{}^\alpha & = \hat C_{\mu \nu}{}^\alpha  - 2 A_{[\mu}{}^j C_{|j| \nu]}{}^\alpha + A_\mu{}^i A_\nu{}^j \hat C_{ij}{}^\alpha \,.
\end{split} 
\ee
The four-form redefinitions are not needed in this paper. 

Unfortunately, the above fields will not immediately correspond to components of the generalised form fields appearing in EFT, which may in fact be further redefinitions of the above.
This must be checked on a case-by-case basis. We will provide the details for $\Gsix$ below.

\subsection{Properties of the generalised metric}
\label{gmfacts}

Here we record some useful and very general facts about the generalised metric, $\gM_{MN}$.
Let us split the extended coordinates as $Y^M = (Y^i,Y^A)$, where the $Y^i$ are physical and the $Y^A$ are dual.
Then the generalised metric can be decomposed as
\be
\gM_{MN} = \begin{pmatrix}
\bar \gM_{ij} + \gM_{CD} U_i{}^C U_j^D & \gM_{BC} U_i{}^C \\
\gM_{AC} U_j{}^C & \gM_{AB} 
\end{pmatrix} \quad,\quad
U_i{}^A \equiv ( \gM_{AB})^{-1} \gM_{iB} 
\ee
and where
\be
\bar{\gM}_{ij}=\gM_{ij}- \gM_{CD} U_i{}^C U_j{}^D\,.
\ee
The inverse, assuming that $\gM_{ij}$ and $\gM_{AB}$ are invertible, is
\be
\gM^{MN} = \begin{pmatrix} 
\bar{\gM}^{ij} & - \bar{\gM}^{ik} U_k{}^B \\
- \bar{\gM}^{jk} U_k{}^A & ( \gM_{AB} )^{-1} + \bar{\gM}^{kl} U_k{}^A U_l{}^B
\end{pmatrix} \,,
\ee
where $\bar{\gM}^{ij} \equiv ( \bar\gM_{ij})^{-1}$.

For $\Lambda^M = (\Lambda^i, \lambda^A )$ one has
\be
\begin{split} 
\mathcal{L}_\Lambda \gM_{ij} & = L_\Lambda \gM_{ij} + 2 \partial_{(i} \lambda^C \gM_{j) C} - 2 Y^{Ck}{}_{D(i|} \partial_k \lambda^D \gM_{|j)C}\,,\\
\mathcal{L}_\Lambda \gM_{iA} & = L_\Lambda \gM_{iA} + \partial_i \lambda^C \gM_{AB} - 2 Y^{Ck}{}_{D(i|} \partial_k \lambda^D \gM_{|A) C} \,,\\
\mathcal{L}_\Lambda \gM_{AB} & = L_\Lambda \gM_{AB} - 2 Y^{Ck}{}_{D(A|} \partial_k \Lambda^D \gM_{|B)C} \,,\\
\mathcal{L}_\Lambda (\gM_{AB})^{-1} & = L_\Lambda (\gM_{AB})^{-1} + 2 Y^{(A|k}{}_{DC} \partial_k \Lambda^D (\gM_{B)C})^{-1} \,,\\
\end{split}
\ee
where the ordinary Lie derivative $L_\Lambda$ includes weights of $-2\omega$ (and $+2\omega$ for $(\gM_{AB})^{-1}$). 
Hence one can show that
\be
\mathcal{L}_\Lambda \bar\gM_{ij}  = L_\Lambda \bar\gM_{ij}\,,
\ee
so we conclude this is invariant under form field gauge transformations and so up to possible dilaton factors should take the form
$\bar \gM_{ij} = \phi^{-\omega} \phi_{ij}$. 
Thus, we will always have $\bar \gM_{ij} = \Omega^{-1} \phi_{ij}$ and $\gM^{ij} = \Omega \phi_{ij}$. 

In addition, we record the gauge transformations of the object $U_i{}^A$:
\be
\mathcal{L}_\Lambda U_i{}^A = L_\Lambda U_i{}^A + \partial_i \lambda^A - Y^{Ak}{}_{Di} \partial_k \lambda^D 
+ Y^{Ak}{}_{CB} \partial_k \lambda^C U_j{}^B\,.
\ee

\subsection{The $\Gsix$ dictionary: M-theory section}
\label{E6Mdict}

The EFT fields are the generalised metric $\gM_{MN}$, and the two-form fields $\mathcal{A}_\mu{}^M$ and $\mathcal{B}_{\mu\nu M}$ (we have denoted these calligraphically to prevent confusion with the components $A$ of the supergravity form fields). 

Note that in the conventions of \cite{Hohm:2013vpa}, the kinetic term for the three-form is normalised as $-\frac{1}{12} F_4^2$. 
Our conventions will be that the kinetic term is normalised as $-\frac{1}{48} F_4^2$.
Hence, $\hat C_{\text{there}} = \frac{1}{2} \hat C_{ \text{here} }$ and similarly for the redefined components.
Bearing this factor of two in mind, we can read off from \cite{Hohm:2013vpa} the relationship between EFT tensor hierarchy fields and the redefined three-form components $A_{\mu \dots}$ of  \eqref{AC} as follows:
\be
\begin{split}
\mathcal{A}_{\mu}{}^m & = A_\mu{}^m \,, \\
\mathcal{A}_{\mu mn} & = \frac{1}{\sqrt{2}} A_{\mu mn} \,, \\
\tilde{\mathcal{B}}_{\mu \nu \bar m} & \equiv \sqrt{5} \mathcal{B}_{\mu \nu \bar m } + A_{[\mu}{}^n \mathcal{A}_{\nu] nm}  
\\ & =\frac{1}{\sqrt{2}}A_{\mu \nu m} \,. \\ 
\end{split} 
\ee
The degrees of freedom appearing in the generalised metric are $( \phi_{ij} , A_{ijk}, \varphi )$.\footnote{In \cite{Hohm:2013vpa}, the parametrisation is in terms of the unit determinant matrix $m_{ij} = e^{\Phi/3} \phi_{ij}$ and $e^{\Phi} = \phi^{-1/2}$, so that $e^\Phi m_{ij} = \phi^{-2/3} \phi_{ij}$, $e^{-\Phi} m_{ij} = \phi^{1/3} \phi_{ij}$.}.

The components $\phi_{ij}$ are those appearing in the decomposition of the 11-dimensional metric \eqref{metricdecomp}, while the components $A_{ijk}$ are the internal components of the three-form.
The scalar $\varphi$ is the dualisation of the external three-form (and so can be regarded as the internal components of the six-form dual to the three-form, i.e. $\varphi \equiv \frac{1}{6!} \epsilon^{ijklmn} C_{ijklmn}$. 
To simplify the expressions, we define
\be
\tilde A^{ijk} \equiv \frac{1}{6} \epsilon^{ijklmn} A_{lmn} \,.
\ee
Then the paper \cite{Hohm:2013vpa} allows us to extract the components of the generalised metric $\gM_{MN}$:
\be
\begin{split}
\gM_{ i j} & = \phi^{1/3} \phi_{ij}
 + \phi^{1/3} \frac{1}{2} A_{ikl} A_{jpq} \phi^{kp} \phi^{ql}
- \frac{1}{2} \phi^{-2/3} \phi_{p(i} A_{j)qr} \tilde A^{pqr} \varphi 
\\ & \qquad\qquad\qquad+ \frac{1}{16} \phi^{-2/3} \phi_{kl} A_{ipq} A_{jrs} \tilde A^{kpq} \tilde A^{lrs} 
+ \phi^{-2/3} \phi_{ij} \varphi^2
\\
\gM_{i \bar j} & = \frac{1}{4} \phi^{-2/3} \phi_{jp} \tilde A^{pqr} A_{iqr} - \phi^{-2/3} \phi_{ij} \varphi \\
\gM_i{}^{jk} & =  + \frac{1}{\sqrt{2}} \left( A_{ipq} \phi^{1/3} \phi^{jp} \phi^{kq} + 
 \frac{1}{4} \phi^{-2/3} \phi_{pq} \tilde A^{pjk} \tilde A^{qrs} A_{irs}  - \phi^{-2/3} \phi_{ip} \tilde A^{pjk} \varphi \right)\\
\end{split}
\ee
and 
\be
\begin{split}
\gM_{ \bar i \bar j} & = \phi^{-2/3} \phi_{ij} \\
\gM^{ij,kl} & = \phi^{1/3} \phi^{i[k} \phi^{l]j} + \frac{1}{2}\phi^{1/3} \phi_{pq} \tilde A^{pij} \tilde A^{qkl} \\
\gM_{ \bar i}{}^{jk} & = +  \frac{1}{\sqrt{2}} \phi^{-2/3}\phi_{ip} \tilde A^{pkl} \\
\end{split} 
\ee
The easiest way to obtain the IIA relationships is to carry out the usual reduction on the 11-dimensional variables.

\subsection{The $\Gsix$ dictionary: IIB section}
\label{E6Bdict}

According to \cite{Baguet:2015xha}, we have the following relationships on the IIB section:
\be
\begin{split}
\mathcal{A}_\mu{}^i & = A_\mu{}^i \\
\mathcal{A}_{\mu i \alpha} & = \epsilon_{\alpha \beta} A_{\mu i}{}^\beta \\
\tilde{\mathcal{B}}_{\mu\nu}{}^\alpha & \equiv \sqrt{10} \mathcal{B}_{\mu \nu}{}^\alpha  - \epsilon^{\alpha \beta} A_{[\nu}{}^j \mathcal{A}_{\nu] j \beta}\\
 & = A_{\mu \nu}{}^\alpha \\
\end{split}
\ee
In addition, the generalised metric contains an internal field denoted in \cite{Baguet:2015xha} by $b_{ij}{}^\alpha$, and we have $b_{ij}{}^\alpha  = - \frac{1}{2} A_{ij}{}^\alpha$.
Note that in the conventions of \cite{Baguet:2015xha}, the kinetic term for the two-forms is normalised as $-\frac{1}{12} F_3^2$. We use the same convention. 

The full expressions for the generalised metric are lengthy and provided in more concise form in section \ref{IIBred} already, except for $\gM_{ij}$, for which we only need the general results in appendix \ref{gmfacts} anyway.

\section{General DFT and EFT conventions} 
\label{EFTconv}

We provide here a useful summary of the conventions for the fields of DFT and EFT, focusing on the generalised metric and those that appear in the $R_1$ and $R_2$ representations.
We provide the constraints on the generalised metric, and following \cite{Wang:2015hca} the expressions for the operations $\p: R_1 \otimes R_1 \rightarrow R_2$ and $\hat \partial : R_2 \rightarrow R_1$.

\subsection*{Summary of general $R_1$ and $R_2$ index conventions}

For $B_{\mu\nu} \in R_2$ we write $B_{\mu\nu}{}^{MN}$. In the construction of the exceptional sigma model, we introduce a charge $q \in \bar R_2$ which we will also write as $q_{MN}$, where the symmetrisation (and projection) is implicit. 
We require that:
\be
q_{MN} B_{\mu\nu}{}^{MN} = q \cdot B_{\mu\nu} 
\ee 
in order to determine the precise map between $q_{MN}$ and $q\in \bar{R}_2$. 
Note that as $Y^{MN}{}_{PQ}$ projects into $R_2$ or $\bar R_2$ we have 
\be
q_{MN} \frac{1}{2d}  Y^{MN}{}_{PQ} = q_{PQ} \,,
\ee
where $d= D-1$ in the $\ED$ case.

Following the conventions and notation of appendix \ref{EFTconv}, we can write:
\be
A^M = \begin{cases}
A^M & O(d,d) \\
( A^a, A^s ) & \G\\
A^{I a}   & \Gthree \\ 
A^{[ab]} & \Gfour \\
A^M  & \Gfive \\
A^M & \Gsix
\end{cases}
\quad,\quad
B^{MN} = \begin{cases} \frac{1}{2d} \eta^{MN} B & O(d,d) \\
\delta^{(M}_a\delta^{N)}_s B^{a s} & \G\\
\frac{1}{4} \epsilon^{IJK} \epsilon^{ab} B_K & \Gthree \\ 
\frac{1}{6} \epsilon^{abcd e} B_e & \Gfour \\
\frac{1}{8} \gamma_I^{MN} B^I & \Gfive \\
d^{MNK} B_K & \Gsix
\end{cases}
\,.
\ee
We in general write $M$ for an $R_1$ index, except in the cases of $\G$, where $M = (a,s)$ with $a$ an $\mathrm{SL}(2)$ fundamental index and $s$ a singlet index; $\Gthree$, where $M = (Ia)$ with $I$ an $\mathrm{SL}(3)$ fundamental and $a$ an $\mathrm{SL}(2)$ fundamental; and $\Gfour$, where $M = [ ab]$ is a pair of antisymmetric fundamental indices. 
Note as well that the $B$-field in the $\Gsix$ EFT is normalised differently to the $B$-field in the lower rank cases: a more consistent pattern could be achieved by redefining both the invariant tensor $d^{MNP}$ and $B_M$ by factors of $\sqrt{10}$, however we maintain full alignment with the EFT conventions of the original paper \cite{Hohm:2013vpa} here. 
In addition, the charges are given by
\be
q_{MN} =  \begin{cases} 
q\eta_{MN}  & O(d,d) \\
2 \delta_{(M}^a \delta_{N)}^s q_{\alpha s} & \G \\
\epsilon_{IJK} \epsilon_{ab} q^K & \Gthree \\
q^e \epsilon_{abcd e} & \Gfour \\
\frac{1}{2} q_I \gamma^I_{MN} & \Gfive \\ 
d_{MNK} q^K  & \Gsix
\end{cases}
\label{qdef}
\ee

\subsection*{$O(d,d)$ DFT}

The $R_1$ representation is the fundamental, and we write $A^M$ for a field in this representation. 
The Y-tensor is
\be
Y^{MN}{}_{PQ} = \eta^{MN} \eta_{PQ} \,,
\ee
so the section condition is 
\be
\eta^{MN} \partial_M \otimes \partial_N = 0 \,.
\ee
The generalised metric $\gM_{MN}$ obeys
\be
\eta^{MN} \gM_{MP} \gM_{NQ} = \eta_{PQ} \,. 
\ee
A field in the fundamental ($R_1$) representation is denoted $A^M$ while a field in the trivial $R_2$ representation is $B$. 
We have 
\be
A_1 \p A_2 = \eta_{MN} A_1^M A_2^N 
\quad,\quad
( \hat \partial B)^M = \eta^{MN} \partial_N B \,.
\ee

\subsection*{$\G$ EFT}

This EFT was constructed in \cite{Berman:2015rcc}.
The $R_1$ representation of the $\G$ EFT is reducible, being the $\mathbf{2}_1 \oplus \mathbf{1}_{-1}$.
We let $a=1,2$ be a fundamental $\mathrm{SL}(2)$ index.
Then we write $A^M = ( A^a, A^s)$ for a field in this representation (where the singlet index $s$ refers to the component in the $\mathbf{1}_{-1}$). 
The $R_2$ representation is the $\mathbf{2}_0$, and a field here is denoted $B^{a s}$. We have
\be
(A_1 \p A_2)^{a s} = A_1^a A_2^s + A_1^s A_2^a
\quad,\quad
(\hat \partial B)^a = \partial_s B^{as} \,,
\quad,\quad
(\hat \partial B)^s = \partial_a B^{as} \,.
\ee
The Y-tensor has components
\be
Y^{as}{}_{bs} = \delta^a{}_b
\ee
and others related by symmetry. The section condition is
\be
\partial_a \otimes \partial_s = 0 \,.
\label{sl2sec}
\ee
The generalised metric is also reducible, consisting of a two-by-two matrix $\gM_{ab}$ and a one-by-one matrix $\gM_{ss}$. 
These are not independent: one has $\det \gM_{ab} = (\gM_{ss})^{-3/2}$.

\subsection*{$\Gthree$ EFT}

This EFT was constructed in \cite{Hohm:2015xna}.
The $R_1$ representation of the $\Gthree$ EFT is the $( \mathbf{3} , \mathbf{2})$.
We let $I=1,2,3$ be a fundamental $\mathrm{SL}(3)$ index and $a=1,2$ be a fundamental $\mathrm{SL}(2)$ index.
Then we write $A^M = A^{Ia}$ for a field in this representation. 
The $R_2$ representation is the $(\mathbf{\bar 3}, \mathbf{1})$, and a field in this representation is denoted by $B_I$. 
We have
\be
( A_1 \p A_2 )_I = \epsilon_{IJK} \epsilon_{ab} A_1^{Ja} A_2^{Kb} 
\quad,\quad
( \hat \partial B)^{Ia} = \epsilon^{IJK} \epsilon^{ab} \partial_{Jb} B_K \,.
\ee
The Y-tensor is
\be
Y^{Ia,Jb}{}_{Kc,Ld} = \epsilon^{IJ K^\prime} \epsilon_{KL K^\prime} \epsilon^{ab} \epsilon_{cd} \,,
\ee
and the section condition is
\be
\epsilon^{IJK} \epsilon^{ab} \partial_{Ia} \otimes \partial_{Jb} = 0 \,.
\ee
The generalised metric is decomposable, with $\gM_{I a, Jb} = \gM_{IJ} \gM_{ab}$, where both $\gM_{IJ}$ and $\gM_{ab}$ have determinant one. 

\subsection*{$\Gfour$ EFT}

The full $\Gfour$ EFT was constructed in \cite{Musaev:2015ces}. The internal sector had been pioneered in \cite{Berman:2010is, Berman:2011cg}.
The $R_1$ representation of the $\Gfour$ EFT is the $\mathbf{10}$. This is the antisymmetric representation.
We let $a= 1,\dots,5$ be a fundamental $\mathrm{SL}(5)$ index.
Then we write $A^M = A^{ab} = A^{[ab]}$ for a field in $R_1$.
We define contraction to be accompanied by a factor of $1/2$, thus $A^M D_M \equiv \frac{1}{2} A^{ab} D_{ab}$. 
The $R_2$ representation is the $\mathbf{\bar 5}$ and is denoted $B_a$. We have
\be
(A_1 \p A_2)_a = \frac{1}{4} \epsilon_{a bcde} A_1^{bc} A_2^{de} 
\quad,\quad
(\hat \partial B)^{ab} = \frac{1}{2} \epsilon^{abcde} \partial_{cd} B_e \,.
\ee
The Y-tensor is
\be
Y^{aa^\prime, bb^\prime}{}_{cc^\prime,dd^\prime} = \epsilon^{aa^\prime bb^\prime e} \epsilon_{cc^\prime dd^\prime e} \,.
\ee
The section condition is that
\be
\partial_{[ab} \otimes \partial_{cd]} = 0 \,.
\label{sl5sec}
\ee
The generalised metric is decomposable, with $\gM_{ab} = \gM_{a[c} \gM_{d]b}$, where $\det \gM_{ab}  =1$. 

\subsection*{$\Gfive$ EFT}

The full $\Gfive$ EFT was constructed in \cite{Abzalov:2015ega}, the internal sector having earlier appeared in \cite{Berman:2011jh}.
The $R_1$ representation of the $\Gfive$ EFT is the $\mathbf{16}$, which is a Majorana-Weyl spinor representation. 
We let $A^M$, where $M= 1,\dots ,16$ denote a field in $R_1$. 
The $R_2$ representation is the fundamental, and we denote a field in this representation by $B_I$, where $I = 1,\dots ,10$ is a fundamental index. 
Note that we can raise and lower such indices using $\eta_{IJ}$, the $\Gfive$ structure. 
Let $\gamma_I{}^{MN}$ and $\gamma^I{}_{MN}$ denote the sixteen-by-sixteen off-diagonal components of the full gamma matrices, obeying
\be
 \gamma^I{}_{MN} \gamma^{J N P} +  \gamma^J{}_{MN} \gamma^{I N P} 
= 2 \eta^{IJ} \delta^P_M \,,
\ee
which implies $\gamma^I{}_{MN} \gamma_{J}{}^{MN} = 16 \delta^I_J$.
We have
\be
( A_1 \p A_2 )_I = \frac{1}{2} \gamma_{I M N } A_1^M A_2^N 
\quad,\quad
(\hat \partial B)^M = \gamma_I{}^{MN} \partial_N B^I \,.
\ee
The Y-tensor is
\be
Y^{MN}{}_{PQ} = \frac{1}{2} \gamma_I{}^{MN} \gamma^I{}_{PQ} \,,
\ee
and the section condition
\be
\gamma_I{}^{MN} \partial_M \otimes \partial_N = 0 \,.
\ee
The generalised metric is $\gM_{MN}$, and can also be written in the fundamental as $\gM_{IJ}$ with $\gM_{IJ} \gamma^{I}{}_{MN} \gM^{MP} \gM^{NQ} \sim \gamma_J{}^{PQ}$.

On an M-theory section, with $\Gfive \rightarrow \mathrm{SL}(5)$, we have $A^M = (A^i, A_{ij}, A^z)$, where $i$ is five-dimensional and $z$ is a singlet index. We also have $B^I = ( B^i, B_i )$. The invariant tensor $\eta_{IJ}$ has the usual components $\eta_i{}^j = \delta_i{}^j$ and the components of the gamma matrices can be taken to be
\be
\begin{array}{ccccc}
(\gamma_I)^{MN} & \rightarrow & ( \gamma^i)^j{}_{kl} = 2 \delta^{ij}_{kl} & (\gamma_i)^j{}_z = \sqrt{2} \delta_i^j & (\gamma_i)_{jklm} = \frac{1}{\sqrt{2}} \epsilon_{ijklm} \\
(\gamma^I)_{MN} & \rightarrow & ( \gamma_i)_j{}^{kl} = 2 \delta_{ij}^{kl} & (\gamma^i)_j{}^z = \sqrt{2} \delta^i_j & (\gamma^i)^{jklm} = \frac{1}{\sqrt{2}} \epsilon^{ijklm} 
\end{array} 
\ee
On a IIB section, with $\Gfive \rightarrow \mathrm{SL}(4) \times \mathrm{SL}(2)$, we have $A^M =  ( A^i, A_{i a} , A^{\bar i} )$, where $i,\bar i$ are four-dimensional indices and $a$ is the $\mathrm{SL}(2)$ index.
We also have $B^I = ( B_a , B^{ij} , B_{\bar a} )$. 
The invariant tensor $\eta_{IJ}$ has components $\eta^{ \bar a b } = \epsilon^{ab}$ and $\eta_{ij,kl} = \frac{1}{2} \epsilon_{ijkl}$.
The gamma matrix components can eb taken to be 
\be
\begin{array}{cccc}
(\gamma_I)^{MN} & \rightarrow &  ( \gamma^a )^{\bar i}{}_{j b} = - \sqrt{2} \delta^i_j \delta^a_b
 & (\gamma^{\bar a} )^i{}_{j b} = \sqrt{2} \delta^i_j \delta^a_b \\ 
 & & ( \gamma_{ij} )_{ k a , l b} = \epsilon_{ijkl} \epsilon_{ab} & (\gamma_{ij} )^{\bar k l} = - 2 \delta^{kl}_{ij} \\
(\gamma^I)_{MN} & \rightarrow &  ( \gamma_a )_{\bar i}{}^{j b} = - \sqrt{2} \delta_i^j \delta_a^b
 & (\gamma_{\bar a} )_i{}^{j b} = \sqrt{2} \delta_i^j \delta_a^b \\ 
 & & ( \gamma^{ij} )^{ k a , l b} = \epsilon^{ijkl} \epsilon^{ab} & (\gamma^{ij} )_{\bar k l} = - 2 \delta^{ij}_{kl} 
\end{array} 
\ee

\section{The charge constraint in general EFTs} 
\label{qconstraint}

Here we write down the specific form of the charge constraint \eqref{magic}, and solve it, relying on the conventions of appendix \ref{EFTconv}.

\subsection*{DFT} 

In this case, the constraint \eqref{magic} is in fact an identity, reducing to $q \eta_{PQ} \partial_M$ on both sides. Therefore there is always a doubled string, as we would expect.
The non-zero components of $q_{MN}$ are
\be
q_i{}^j = q^j{}_i = q \delta_i{}^j \,.
\ee
The tension simplifies to $T = q$.

\subsection*{$\G$ EFT} 

Here the extended coordinates are $Y^M = (Y^a, Y^s) \in \mathbf{2}_1 \oplus \mathbf{1}_{-1}$ of $\G$. The charge in $\bar R_2$ is $q_{as}$.
The constraint \eqref{magic} becomes
\be
\epsilon^{ab} q_{a s} \partial_b = 0 \,.
\label{sl2magic}
\ee
The solutions are as follows:
\begin{itemize}
\item IIB section: we have $\partial_s \neq 0$ and $\partial_a = 0$. The constraint \eqref{sl2magic} therefore does not constraint the charge $q_{a s}$ at all. This means the charge is a doublet, and the objects that couple to $B_{\mu\nu}{}^{a s}$ are $(p,q)$ strings. 
The non-zero components of $q_{MN}$ are $q_{s a} = q_{a s}$.
\item M-theory section: we have $\partial_s = 0$ and $\partial_a \neq 0$. The constraint \eqref{sl2magic} forces $q_{a s} = 0$: so there are no strings in eleven dimensions.
\item IIA section: now we have $\partial_s = 0$ and $\partial_1 = 0$, but $\partial_2 \neq 0$. We are allowed have $q_{2s} \neq 0$ and $q_{1s} = 0$. This gives the IIA fundamental string. 
The non-zero components of $q_{MN}$ are $q_{2s} = q_{s2}$.
\end{itemize}
In this case, the generalised metric is reducible, consisting of components $\gM_{ab}, \gM_{ss}$. 
The tension is $T = \sqrt{ q_{as}q_{bs} \gM^{ab} \gM^{ss}}$.

\subsection*{$\Gthree$ EFT} 

Here we have $Y^M = Y^{I a}$ in the $(\mathbf{3}, \mathbf{2})$ of $\Gthree$. The charge in $\bar R_2 = ( \mathbf{3},\mathbf{1})$ is $q^I$.
The constraint \eqref{magic} becomes
\be
q^I \partial_{I a} = 0 \,.
\ee
The solutions are as follows:
\begin{itemize}
\item IIB section: we split $I = (\alpha, 3)$, and the section condition solution is $\partial_{3 a} \neq 0$, $\partial_{\alpha a} = 0$. 
The allowed charges are $q^\alpha \neq 0$, while $q^3 = 0$. 
The non-zero components of $q_{MN}$ are 
\be
q_{3 a , \alpha b} = q_{\alpha b, 3a} = \epsilon_{\alpha \beta} q^\beta \epsilon_{ab} \,.
\ee
\item M-theory section: we have $\partial_{I 1} \neq 0$ and $\partial_{I 2} = 0$. This means that $q^I = 0$, so there are no solutions.
\item IIA section: let $I = (1, i)$ and suppose that $\partial_{1 1} = 0$, so $\partial_{i 1} \neq 0$. Then we can have $q^1 \neq 0$. The non-zero components of $q_{MN}$ are
\be
q_{i 1 , j 2} = q_{j 2 , i 1} = \epsilon_{ij} q^1 \,.
\ee
\end{itemize} 
The generalised metric is also reducible, with $\gM_{Ia,Jb} = \gM_{IJ} \gM_{ab}$.
The tension becomes
$T = \sqrt{ \gM_{IJ} q^I q^J }$.

\subsection*{$\Gfour$ EFT}

In this case the extended coordinates carry a pair of antisymmetric 5-dimensional indices, and are written $Y^{ab}$. The charge in $\bar R_2 = \mathbf{5}$ is $q^a$.
The constraint \eqref{magic} boils down to
\be
q^a \partial_{ab} = 0 \,.
\label{sl5magic}
\ee 
The solutions are as follows:
\begin{itemize}
\item IIB section: here we decompose $a= ({}_i,\alpha)$ where $i =1,2,3$ and $\alpha$ is a 2-dimensional $\mathrm{SL}(2)$ S-duality index. Then we have $\partial^{ij} \neq 0$ for $i,j=1,2,3$, and $\partial_\alpha{}^i = \partial_{\alpha \beta} = 0$. The constraint \eqref{sl5magic} imposes that $q_i=0$ and $q^\alpha \neq 0$.
The non-zero components of $q_{MN}$ are:
\be
q^{ij,}{}^k{}_{\alpha} = q^k{}_{\alpha,}{}^{ij} = \epsilon^{ijk} \epsilon_{\alpha \beta} q^\beta \,.
\ee
\item M-theory section: here $a=(i,5)$ where $i=1,2,3,4$. The section condition is solved by $\partial_{i5}\neq 0$ and $\partial_{ij} = 0$. The charge $q^a$ is forced to be zero. 
\item IIA section: now $a=(i,4,5)$ with $i=1,2,3$. The section condition is $\partial_{i5} \neq 0$, $\partial_{ij} = \partial_{i4} = \partial_{45} = 0$. One can have $q^4 \neq 0$, $q^i=0 = q^5$. 
The non-zero components of $q_{MN}$ are:
\be
q_{i 5, jk } = q_{jk, i5} = - \epsilon_{ijk} q^4 \,.
\ee
\end{itemize}
The generalised metric can be decomposed in terms of a symmetric unit determinnat little metric $m_{ab}$ as $\gM_{ab,cd} = m_{ac} m_{bd} - m_{ad} m_{bc}$, and the tension becomes $T =\sqrt{ m_{ab} q^a q^b}$.

\subsection*{$\Gfive$ EFT} 

Here we have $Y^M$ in the $\mathbf{16}$ of $\Gfive$. The charge in $\bar R_2 = \mathbf{10}$ is $q^I$.
The constraint \eqref{magic} becomes
\be
q^I \gamma_I{}^{MN} \partial_N  = 0 \,.
\ee
The solutions are as follows:
\begin{itemize}
\item IIB section: we decompose $q^I = ( q_a, q^{ij} , q_{\bar a} )$, where $a$ and $\bar a$ are separate $\mathrm{SL}(2)$ fundamental indices. One finds that
\be
q_{\bar a} \partial_i = 0 \quad,\quad q^{ij} \partial_j = 0 \,.
\ee
Hence we must have $q_{\bar a} = 0$ and $q^{ij} = 0$, but $q_a \neq 0$. The non-zero components of the charge $q_{MN}$ are
\be
q_i{}^{ja} = \frac{1}{\sqrt{2}} \delta_i^j \epsilon^{ab} q_b\,.
\ee
\item M-theory section: we decompose $q^I = (q^i , q_i )$, with $i=1,\dots,5$. One finds that
\be
q_{[i } \partial_{j]} = 0 \quad,\quad q^i \partial_i = 0 \,,
\ee
which has no solutions. 

\item IIA section: we now take one of the M-theory directions $i=1$ (say) to be an isometry, $\partial_1 = 0$. This allows for $q^{ 1} \neq 0$, giving the F1 string, with all other charge components zero. 
The non-zero components of the charge $q_{MN}$ are (for $i,j = 1,\dots 4$ IIA indices):
\be
q_{i }{}^{1 j} = \frac{1}{2} q^1 \delta_i^j
\ee
\end{itemize} 
The tension can be written as $T = \sqrt{ \gM_{IJ} q^I q^J }$ where $\gM_{IJ}$ is the generalised metric in the fundamental representation rather than the $R_1$ representation.

\providecommand{\href}[2]{#2}\begingroup\raggedright\endgroup

\end{document}